\documentclass[pre,twocolumn,showpacs,preprintnumbers,amsmath,amssymb,floatfix,nofootinbib]{revtex4}
\usepackage{graphicx}
\usepackage{subfigure}
\usepackage{color}
\usepackage{xspace}
\usepackage{cases}
\usepackage{tikz}
\usepackage{pgfplots}
\usepackage{afterpage}

\expandafter\ifx\csname package@font\endcsname\relax\else
  \expandafter\expandafter
  \expandafter\usepackage
  \expandafter\expandafter
  \expandafter{\csname package@font\endcsname}%
\fi

\newcommand{\gpvec}[1]{\mathbf{#1}}

\newcommand{\kvec}{\gpvec{k}}
\newcommand{\mvec}{\gpvec{m}}

\newcommand{\xvec}{\gpvec{x}}
\newcommand{\yvec}{\gpvec{y}}
\newcommand{\imag}{\mathring{\imath}}

\usepackage{dsfont}

\newcommand{\gpset}[1]{\mathds{#1}}

\newcommand{\canetset}[1]{{\mathchoice {\hbox{$\sf\textstyle #1\kern-0.4em #1$}}
{\hbox{$\sf\textstyle #1\kern-0.4em #1$}}
{\hbox{$\sf\scriptstyle #1\kern-0.3em #1$}}
{\hbox{$\sf\scriptscriptstyle #1\kern-0.2em #1$}}}}

\newcommand{\Nset}{\gpset{N}}
\newcommand{\Zset}{\gpset{Z}}

\newcommand{\Rset}{\gpset{R}}

\newcommand{\latin}[1]{{\it #1}}
\newcommand{\ie}{\latin{i.e.}\@\xspace}
\newcommand{\eg}{\latin{e.g.}\@\xspace}
\newcommand{\cf}{\latin{cf.}\@\xspace}
\newcommand{\etc}{\latin{etc}\@\xspace}
\newcommand{\etal}{\latin{et al}\@\xspace}

\makeatletter
\newcommand{\supmarker}[1]{{\@ifempty{#1}{}{\text{(#1)}}}}
\makeatother

\newcommand{\slabel}[1]{\label{sec:#1}}

\newcommand{\Sref}[1]{Sec.~\ref{sec:#1}}

\newcommand{\Aref}[1]{Appendix~\ref{sec:#1}}

\newcommand{\elabel}[1]{\label{eq:#1}}
\newcommand{\eref}[1]{(\ref{eq:#1})}

\newcommand{\Eref}[1]{Eq.~(\ref{eq:#1})}
\newcommand{\Erefs}[1]{Eqs.~(\ref{eq:#1})}

\newcommand{\flabel}[1]{\label{fig:#1}}
\newcommand{\fref}[1]{Fig.~\ref{fig:#1}}
\newcommand{\Fref}[1]{Fig.~\ref{fig:#1}}
\newcommand{\Frefs}[1]{Figs.~\ref{fig:#1}}
\newcommand{\subfref}[1]{\subref{fig:#1}}

\newcommand{\CC}{\mathcal{C}}
\newcommand{\EC}{\mathcal{E}}
\newcommand{\FC}{\mathcal{F}}

\newcommand{\FChat}{\hat{\FC}}
\newcommand{\FCtilde}{\tilde{\FC}}
\newcommand{\GC}{\mathcal{G}}

\newcommand{\plaind}{\mathrm{d}}
\newcommand{\dint}[1]{\mathchoice{\!\plaind#1\,}{\!\plaind#1\,}{\!\plaind#1\,}{\!\plaind#1\,}}
\newcommand{\ddint}[1]{\mathchoice{\!\plaind^d#1\,}{\!\plaind^d#1\,}{\!\plaind^d#1\,}{\!\plaind^d#1\,}}
\newcommand{\dXint}[2]{\mathchoice{\!\plaind^{#1}#2\,}{\!\plaind^{#1}#2\,}{\!\plaind^{#1}#2\,}{\!\plaind^{#1}#2\,}}

\newcommand{\ave}[1]{\left\langle #1 \right\rangle}

\newcommand{\ket}[1]{\left|#1\right\rangle}
\newcommand{\ident}{\mathbf{1}}

\renewcommand{\exp}[1]{\mathchoice{\mathrm{e}^{#1}}{\operatorname{exp}\left(#1\right)}{\operatorname{exp}\left(#1\right)}{\operatorname{exp}\left(#1\right)}}

\newcommand{\half}{\mathchoice{\frac{1}{2}}{(1/2)}{\frac{1}{2}}{(1/2)}}

\newcommand{\fourth}{\mathchoice{\frac{1}{4}}{(1/4)}{\frac{1}{4}}{(1/4)}}

\newcommand{\quarter}{\fourth}

\newcommand{\Activity}{G}
\newcommand{\msd}{\Delta^2}
\newcommand{\ActivityTimeAve}{\overline{\Activity}}
\newcommand{\ActivityTimeAveSpave}{\widetilde{\Activity}}
\newcommand{\TimeIntegratedActivity}{\widehat{\Activity}}
\newcommand{\ActAct}{C}
\newcommand{\ActActTimeAve}{\overline{\ActAct}}
\newcommand{\ActActTimeAveSpave}{\widetilde{\ActAct}}
\newcommand{\ShapeAvaAbs}{R}
\newcommand{\ShapeAvaRel}{\widehat{\ShapeAvaAbs}}
\newcommand{\MFT}{\text{\tiny MFT}}
\newcommand{\AvaAvaCorr}{K}

\newcommand{\nhat}{\hat{n}}

\newcommand{\Kbar}{\overline{K}}
\newcommand{\Ds}{D_s}
\newcommand{\Dsbar}{\overline{\Ds}}

\newcommand{\roughness}{\chi}

\newcommand{\phihat}{\hat{\phi}}

\newcommand{\atilde}{\tilde{a}}
\newcommand{\btilde}{\tilde{b}}

\begin{document}
\title{Spatio-temporal Correlations in the Manna Model in one, three and
five dimensions}
\author{Gary Willis}
\affiliation{Department of Mathematics,
Imperial College London,
180 Queen's Gate,
London SW7 2AZ, United Kingdom}
\author{Gunnar Pruessner}
\email{g.pruessner@imperial.ac.uk}
\homepage{http://www.ma.ic.ac.uk/~pruess}
\affiliation{Department of Mathematics,
Imperial College London,
180 Queen's Gate,
London SW7 2AZ, United Kingdom}

\begin{abstract}
Although the paradigm of criticality is centred around spatial
correlations and their anomalous scaling, 
not many studies of
Self-Organised Criticality (SOC) 
focus on  spatial
correlations.  
Often, 
integrated observables, such as avalanche size
and duration, are 
used, not least as to avoid complications due
to the unavoidable lack of translational invariance. 
The present work is a survey of spatio-temporal correlation functions in
the Manna Model of SOC,
measured numerically in detail 
in $d=1,3$ and $5$ dimensions and compared to theoretical
results, in particular relating them to ``integrated'' observables such as
avalanche size and duration scaling, that measure them indirectly.
Contrary to the notion held by some of SOC models organising into
a critical state by re-arranging their spatial structure avalanche by avalanche, 
which may be expected to result in large, non-trivial, system-spanning 
spatial correlations in the quiescent
state (between avalanches),
correlations of inactive particles in the quiescent state have a small
amplitude that does not increase with the system size, although they display
(noisy) power law scaling over a range linear in the system size.
Self-organisation, however, does take place as the (one-point) density of
inactive particles organises into a particular profile that is asymptotically
independent of the
driving location, also demonstrated analytically in one dimension.  
Activity
and its correlations, on the other hand, display non-trivial long-ranged
spatio-temporal scaling with exponents that can be related to
established results, in particular avalanche size and duration
exponents. 
The correlation length and amplitude are set by the system size (confirmed
analytically for some observables), as expected in systems displaying
finite size scaling. 
In one dimension, we find some surprising
inconsistencies of the dynamical exponent. A (spatially extended) mean
field theory is recovered, with some corrections, in five dimensions.
\end{abstract}

\pacs{
05.65.+b, %
05.70.Jk %
}

\keywords{Self-organised criticality,
Universality,
Correlation functions,
Finite-size scaling}

\maketitle

\newcommand{\gcomment}[1]{}

\section{Introduction}
\slabel{introduction}
Correlations functions are at the heart of critical phenomena
\cite{Stanley:1971}. They capture spatio-temporal scaling in microscopic
variables (position and time) and, via integrals, also on the large
scale (system size and duration). In the form of propagators or response
functions, they govern most of our theoretical understanding of critical
phenomena, certainly all of field theory \cite{Taeuber:2014}. In fact,
originally, temporal correlation functions were the key-motivation of
Self-Organised Criticality (SOC)
\cite{BakTangWiesenfeld:1987,Pruessner:2012:Book}, namely to develop a
theory of $1/f$ noise \cite{vanderZiel:1950}. However, for a range of
reasons interest in correlation functions in SOC systems ceased very
quickly \cite{WatkinsETAL:2016}: Firstly, $1/f$ noise in the
Bak-Tang-Wiesenfeld Model was quickly repudiated
\cite{JensenChristensenFogedby:1989}, secondly spatial analogues were
difficult to come by numerically 
(because necessary boundary and initial
conditions 
spoil translational invariance thereby
making it impossible to improve estimates by taking spatial averages)
and thirdly, spatio-temporal integrals
were
very easily determined and linked very nicely with established theories
and systems, in particular via correlation functions
\cite{Stanley:1971,StaufferAharony:1994,Luebeck:2004}. Given modern
computing resources, most of the technical difficulties are fairly easily
overcome, except maybe for the effort needed to carefully implement 
the observables, so that they can be measured efficiently.

To make further theoretical progress, indeed a more complete
understanding of correlations in SOC systems is needed.  Does the
``substrate'', \ie the lattice occupied by immobile particles these models ``live on'', self-organise
in any form? Does it develop (long-ranged, clearly visible) correlations? Those questions are part of
the narrative of an SOC model developing into its critical state
\cite{BrokerGrassberger:1997,ChristensenMoloney:2005}. In the active
state, what does the response function look like, \ie where, when and
how much
activity is seen in a system after it is being perturbed (activated) somewhere? 
What is left of the old claim of $1/f$ noise? Which
correlations display non-trivial scaling and how is that related to the
known scaling of avalanches \cite{Pruessner:2012:Book}, or, in fact,
to growth models \cite{PaczuskiBoettcher:1996,Pruessner:2003}? How does the
behaviour above the upper critical dimension relate to mean field
theory?

The scaling that we are primarily concerned with is finite size scaling,
and more specifically scaling of amplitudes and characteristic (correlation)
lengths with the system size.
This has two key reasons: Firstly, in SOC systems the finite extent should
be the only finite (large length) scale.
Secondly, 
we expect that it is 
difficult to identify the scaling of an observable (in particular in natural
systems),
if its amplitude is fixed and cannot be increased by studying bigger systems.
If the (effective) lattice spacing is very small compared to the range of
observations, such a feature might be indiscernible in measurements.
Conversely, it is easier to measure scaling of an observable, when its amplitude
scales with the system size.

The aim of this work is
twofold: On the one hand, we want to confirm that many of the features
to be expected in a self-organised critical system are actually present,
\ie correlation functions really display what is expected according to
the paradigm
\cite{BakTangWiesenfeld:1987,BakTangWiesenfeld:1988,TangBak:1988a,TangBak:1988b,Pruessner:2012:Book,WatkinsETAL:2016,McAteerETAL:2016},
such as
the spatial correlation length scaling linearly in the system size and
the self-organisation being independent of details such as the driving.
A key-objective is to provide an overview of correlation
functions that is broad in scope as far as different correlations are concerned.
To keep the size of this work within reason we therefore focus on
a single model, namely on the Abelian Manna Model which to us seems particularly
well behaved \cite{HuynhPruessnerChew:2011,HuynhPruessner:2012b}. 
To our knowledge, the present work is nevertheless one of the most comprehensive
surveys
of correlation
functions in an SOC model to date. 
However, we are by far not the first to study correlation functions in SOC,
which have received prominent attention in the past
\cite{Dhar:1990a}, most notably in the form
of important exact results  
\cite{DharMajumdar:1990,MajumdarDhar:1991,MajumdarDhar:1992,Priezzhev:1994,Ivashkevich:1994,MahieuRuelle:2001,Ruelle:2002,Jeng:2005a,Jeng:2005b,JengPirouxRuelle:2006,SaberiETAL:2009,Azimi-TafreshiETAL:2010}
for the Abelian Sandpile Model
\cite{BakTangWiesenfeld:1987,Dhar:1990a} and its directed variant \cite{DharRamaswamy:1989},
but also as a key-feature in SOC
models generally \cite{Grinstein:1995,DickmanVespignaniZapperi:1998,Lise:2002,McAteerETAL:2016}.
Moreover, correlation functions have
recently been studied at the interface between 
absorbing state phase transitions and SOC
\cite{BasuETAL:2012,daCunhaETAL:2014,DickmandaCunha:2015,GrassbergerDharMohanty:2016}.

On the other hand, we want to make contact with theoretical and in
particular field-theoretical work, where response and correlation
functions play a central r{\^o}le. Some theories have been or still are being
developed, which we can compare our numerical findings to
\cite{LeDoussalWiese:2015,Pruessner:2017:FT}.  We can also verify
standard scaling forms \cite{Taeuber:2014} and relate the scaling of
correlation functions to those of observables normally investigated in
SOC, such as avalanche sizes and duration
\cite{Luebeck:2004,Pruessner:2012:Book}.  Many of our findings can also
be compared to mean-field theories, which serve as a first reference
point and which (normally) become exact above the upper critical
dimension $d_c=4$
\cite{LuebeckHucht:2002,LuebeckHeger:2003b,Luebeck:2004}. 
Mean-field theories are unable to capture non-trivial scaling and (most
of) the non-trivial physics and in the past they have often been equated
with a lack of spatial structure
\cite{ChristensenFlyvbjergOlami:1993,FlyvbjergSneppenBak:1993,JanowskyLaberge:1993,ZapperiLauritsenStanley:1995,LauritsenZapperiStanley:1996,VergelesMaritanBanavar:1997,BrokerGrassberger:1997,VespignaniZapperi:1997,VespignaniZapperi:1998,Luebeck:2003b,Yang:2004}.
However, there is no need for a mean field theory to do away with
dissipation at boundaries, which has been suggested to be so crucial to
SOC \cite{HwaKardar:1989a,nonotePaczuskiBassler:2000} and implement their
average effect by way of a global dissipation rate
\cite{VespignaniZapperi:1998,ChessaMarinariVespignani:1998,BarratVespignaniZapperi:1999,Pastor-SatorrasVespignani:2000e}.
Although we will briefly introduce the relevant features of the
mean-field theory in \Sref{mft},
we will not dwell on the details and
intricacies of mean-field theories in general and instead leave that for separate,
future work \cite{LeDoussalWiesePruessner:2016:MFT_unpublished}.

In the following, we will first introduce the Manna Model and our
observables in some detail. We will then present our findings for the
various one, two and three-point correlation functions, with a focus on
qualitative results, such as where scaling is found and whether it is
quantitatively consistent with the exponents reported in the literature.
We will not, however, attempt to extract very high accuracy estimates of
exponents, but rather explore different observables and probe for
consistency mostly using data collapses.  The core of this work has been
performed in $d=1$ dimensions, but we have also carried out extensive
simulations in $d=3$ and $d=5$ dimensions, the latter in order to make
contact with mean-field theory. 
In one dimension, the system sizes considered (linear extent up to $L=4095$
depending on the observable) are small in comparison to past studies 
\cite{HuynhPruessnerChew:2011}, but limited by the CPU-time needed to calculate
some correlation functions. 
Given that transients in some versions of the Manna Model can be extremely
long
\cite{BasuETAL:2012} and show significant scaling in $L$, large system sizes
become computationally prohibitively expensive. This is a trade-off between
small finite size corrections on the one hand and short transients and large
statistics on the other.
In dimensions $d=3$ and $d=5$, where memory requirements
become a limiting factor too, system sizes were commensurate with the literature
\cite{HuynhPruessner:2012b}.
In the last section, we will summarise
and discuss the numerical finding in particular in the light of recent
theoretical progress.

\section{Model, Observables and Methods}
\slabel{model}
The Abelian variant \cite{Dhar:1999a} of the Manna Model
\cite{Manna:1991a}, used throughout the present work, is defined as follows: Sites $\xvec$ of a lattice are
occupied by $z_{\xvec}$ particles. A site $\xvec$ occupied by not more than one particle, 
$z_{\xvec}\le1$, is said to be stable.
If all sites are stable, $\forall_{\xvec} z_{\xvec}\le1$, the configuration is
said to quiescent, otherwise active. The system is ``driven'' at times when it is
quiescent by adding a particle to a site, say $\xvec_0$, which may be
fixed or selected at random, say, with uniform probability, which is
then called
``uniform driving''.
The present study, however, focuses almost exclusively at centre driving, where $\xvec_0$ 
is fixed and chosen to be in the middle of the system.
The fact that driving never takes place while the system is active is known
as a separation of time scales.
The succession of such particle additions, ``drivings'', is said to
occur on the \emph{macroscopic} time scale. 

If the particle number at a site
exceeds the threshold of unity, \ie $z_{\xvec}>1$, then \emph{two}
particles are removed from the site and each placed randomly,
independently and with uniform probability among its nearest neighbours.
Such a redistribution is called a ``toppling'' and the arrival of a particle
at a nearest neighbour a ``charge''. The toppling of a driven
site and the subsequent charge of a nearest neighbour may give rise to the latter exceeding the threshold. In
each \emph{microscopic} time step, every site $\xvec$ that exceeds the
threshold at the beginning of the microscopic time step redistributes
each of two
particles randomly, independently and uniformly among its nearest
neighbours, until $z_{\xvec}\le1$, \ie it topples $\lfloor
z_{\xvec}/2\rfloor$ times in that time step. This may include some sites toppling more than once in that time step. 
For example, $z_{\xvec}=2$
topples to $z_{\xvec}=0$ via one toppling and $z_{\xvec}=5$ to
$z_{\xvec}=1$ via two topplings.
The microscopic
time step lasts until all sites initially exceeding the threshold have
completed their toppling. Only then sites that subsequently exceed the
threshold are considered at the beginning of the next microscopic
time step. This \emph{parallel} updating scheme in ``sweeps'' provides an integer-valued microscopic 
time $t$ which furnishes a convenient, well-defined and well justified way to estimate
time-resolved observables. This original scheme
is sometimes replaced by
random sequential or proper Poissonian updating with random, exponentially distributed waiting times, which lends itself more naturally to a
theoretical description.

For our purposes it proves most convenient for the microscopic time to
be reset to $t=0$ at each driving.  In that sense we will regard
microscopic time as the time passed since the last driving.

In the literature it is not always stated explicitly 
whether particles are
redistributed all at once (non-Abelian, original definition
\cite{Manna:1991a}) or in pairs
(Abelian definition \cite{Dhar:1999a}). The latter is more commonly implemented and has a
number of theoretical advantages \cite{Pruessner:2012:Book}. In
particular, for a fixed (random) sequence of directions of particle
redistributions, the final configuration of the Abelian Manna Model is
independent of the order in which sites exceeding the threshold topple.
This is not the case in the non-Abelian definition, as the number of
particles leaving a site equals the number of charges it has received
since the last toppling and up to the time of its toppling.

On a lattice that naturally divides into sublattices of sites that are
mutual nearest neighbours, in particular on hypercubic lattices with
suitable boundary conditions (see below), the scheme above has the
additional advantage that all sites that are active at the beginning of
a time step reside on the same sublattice.  The number of active
sites can therefore never exceed the size of the largest sublattice
(which are equally or almost equally large) and given that no two sites within the
same sublattice are nearest neighbours, they cannot charge each other
while toppling. More importantly, if a site toppling during a particular
time step were able become active again during the same time step, a trail of
computational implementation problems arises, namely of keeping track of the
number of
particles to topple in the present and in the future time steps.  Without
such safeguards, the number of topplings occurring at a site \emph{during
a particular microscopic time step} would be a function of the order of
updates of sites, as the Abelian property states only that the
distribution of \emph{final states} is invariant under changes of updating
order (strictly, changes of the order of initial charges), but says nothing about observables on the microscopic time scale
(or, strictly, any other observables). To avoid any form of bias against
certain sequences of events, one would
probably resort to random sequential updating. Rather than doing that,
we use hypercubic lattices that naturally divide into sublattices as
described above.

In the description above, there is no loss of particles anywhere, in
fact, only gain of particles added by the external driving. However,
particle loss occurs at the boundaries of the lattice. For the
following discussion it is easiest to think of such boundary sites as
being situated adjacent to \emph{sink sites}, where particles may
accumulate without that site ever exceeding the threshold. The
particles are effectively lost at those sites and such sink sites are
not subject to the lattice dynamics. The only lattices we have studied
are those where each and every regular (\ie non-sink) site has the same
number of nearest neighbours, \ie every boundary site is surrounded by
suitable number of sink sites.  In one dimension, a system consisting of
$L$ sites has sites $x\in\{1,2,\ldots,L\}$ with boundary sites $x=1$
and $x=L$ adjacent to sink sites at $x=0$ and $x=L+1$. In two dimensions, a
``frame'' of sink sites may be thought of surrounding the lattice.
However, in dimensions $d>1$ we have applied periodic boundary
conditions in all but one direction (which we refer to as
$x$-direction), \ie a site at $y=1$ is adjacent to a site at $y=L'$,
with $L'$ the length of the lattice in the periodic direction.  In the
following, this is referred to as hyper-cylindrical boundary conditions.
Coordinates in the periodic directions will be denoted by
$y_2,y_3,\ldots,y_d$ (or just $y$ where unambiguous), so that
the full lattice vector
$\xvec=(x,y_2,y_3,\ldots,y_d)$, has components $x\in\{1,2,\ldots,L\}$ with
open boundary conditions and
$y_2,\ldots,y_d\in\{1,2,\ldots,L'\}$ with periodic boundary conditions.

In order to implement centre driving, we have chosen $L$ odd and the
driving position $x_0=(L+1)/2$. However, to maintain the segregation of
toppling exclusively on either the even or the odd sublattice in
dimensions $d>1$, we had to choose an even length for the ``perimeter''
$L'$ of the $d-1$ periodic directions, which we took as $L'=L+1$. The
total number of sites in these systems is thus $N=L(L+1)^{d-1}$.  Below,
we will discuss the Manna Model in one dimension at length, presenting results
for a large
variety of observables. Only for a selected set of these observables we
will present results from simulations in higher dimensions, namely $d=3$
(just below the upper critical dimension $d_c=4$,
\cite{LuebeckHucht:2002,Luebeck:2004,HuynhPruessner:2012b}) and $d=5$ (above the upper
critical dimension, where mean field theory should apply).

According to the updating rules above, particle trajectories are those
of random walkers. The difference between the particles in the Manna
Model and a random walker is the fact that the former may get stuck
occasionally, when they arrive at an empty site. If each particle had
attached to it a local clock that ticks only whenever the particle
moves, then the particle's trajectory with time labels of the local
clock would be indistinguishable from that of a random walk. In the
following, we will call moving particles ``active'', the local (per-site) number of
\emph{topplings} ``activity density'' and their totality ``activity''.
Strictly, for any given time $t$ activity density is the \emph{number}
of topplings $n(\xvec,t;\xvec_0,L)=\lfloor z_{\xvec}(t)/2 \rfloor$ that
take place during parallel update $t$ to $t+1$ at site $\xvec$ after driving
at
site $\xvec_0$. Because time is reset to $t=0$ at driving,
$n(\xvec,t=0;\xvec_0,L)$ vanishes everywhere except possibly but not
necessarily (as the driving is not bound to result in activation) at
$\xvec=\xvec_0$, \ie $n(\xvec,t=0;\xvec_0,L)\propto
\delta_{\xvec,\xvec_0}$.  It is rare that a large fraction of sites is
active in parallel, so $n(\xvec,t;\xvec_0,L)$ is generally sparse in
$\xvec$. To distinguish different histories $n(\xvec,t;\xvec_0,L)$
after driving the system for the $i$th time, we use the notation
$n_i(\xvec,t;\xvec_0,L)$.  The average of $n_i(\xvec,t;\xvec_0,L)$ over
many realisations $i=1,2,\ldots,M$ is referred to as the
(time-dependent) activity density (the count per site) or response
propagator $\Activity(\xvec, t; \xvec_0,L)$ for a system of size $L$,
see \Eref{def_activity_estimator} below.

Particles that are not moving are part of the ``substrate'', \ie the
``backdrop'' in front of which activity 
unfolds. These inactive, immobile particles may be referred to as ``substrate particles'' and their
density, \ie fraction of singly occupied sites, as the substrate density
either spatially resolved as $D_s(\xvec;\xvec_0,L)$ or averaged as
$\zeta_L$, \Eref{def_zeta}, with $\lim_{L\to\infty}\zeta_L=\zeta_{\infty}$.

In each toppling, \emph{two} particle moves occur.
Each particle contributes (with weight $1/2$) to the activity density
(only while moving away). A time integral over the activity density hides the
complicated relationship between local time and actual microscopic time
and the resulting space-dependent density is that of a collection of
random walkers with a diffusion constant $D$ corresponding to one lattice
spacing squared per local time step (as moves occur only when local
time ticks away), $D=1/(2d)$ \cite{Pruessner_aves:2013}. Strictly, the
activity density is \emph{half} the density of random
walker trajectories emanating from the driving site.

The totality of all topplings triggered by driving a quiescent system is
called an avalanche. If the driving occurs at an empty site,
$z_{\xvec}=0$, so that it remains stable and no toppling takes place at
all, an avalanche size of $0$ is recorded. Otherwise, the avalanche size
$s$ is the total number of topplings that are triggered by adding a
particle as the system is driven. To indicate the instantaneous avalanche
size in an individual realisation, indexed by $i$, we may use the
notation $s_i$. Following from the discussion above, the expected
avalanche size is given by half the escape time\footnote{The escape time
of a random walker gives the number of its moves, 
which is half the number of topplings, as each toppling causes two moves.} of a random walker from
a lattice, which can (depending on boundary conditions) be calculated in
closed form \cite{Pruessner_aves:2013}. With hyper-cylindrical boundary
conditions as described above, the expected avalanche size is simply
\cite{Pruessner:2012:Book} 
\begin{equation}\elabel{ave_ava}
\ave{s}=\frac{x_0 (L+1-x_0)}{4D}
\end{equation}
in a $d-dimensional$ lattice (periodic in $d-1$ dimensions and open in
one of linear extent $L$) driven at distance $x_0$ away from the open boundary.

The number of parallel updates needed to make the system quiescent again
after driving it
defines the avalanche duration $T$, individually denoted by $T_i$ for
the duration of the $i$th avalanche.
If no toppling takes place we define $T_i=0$.
We may thus write
\begin{equation}
s_i=\sum_{t=0}^{T_i-1} n_i(\xvec,t;\xvec_0,L)  \ .
\end{equation}
As opposed to the Oslo Model \cite{ChristensenETAL:1996}, where
particles move deterministically as in the (BTW) Sandpile Model
\cite{BakTangWiesenfeld:1987} but with the difference that the threshold
is reset randomly, the Manna Model has no strict upper bound for the
avalanche size and the avalanche duration, as particles may keep
toppling back and forth indefinitely. As we know from the mapping to random
walker trajectories mentioned above, this has no serious implications as
far as the numerics are concerned (avalanches eventually terminate just
as walkers eventually dissipate), but algebraically, \eg in terms of
expressing the dynamics using Markov matrices
\cite{Pruessner:2012:Book}, a number of difficulties arise (see also
\Aref{appendix_generate_matrices}).

Avalanche size and avalanche duration are typical observables in the
study of SOC
models. Beyond a lower cutoff, the probability density
function of both observables displays scaling. Specifically, the
probability $P_s(s;L)$ of observing an avalanche of size $s$ large
compared to some lower cutoff in a system
of linear size $L$ displays \emph{simple (finite size) scaling}
\cite{BakTangWiesenfeld:1988}, that is, it scales like $a_s s^{-\tau} \GC_s(s/(b_s L^D))$ with
metric factors $a_s$ and $b_s$, scaling function $\GC_s$ and two universal
exponents $\tau$ and $D$. The former, $\tau$ is known as the avalanche
size exponent, the latter as the avalanche dimension $D$.
Correspondingly, the avalanche duration has probability distribution
$P_T(T;L)=a_T T^{-\alpha} \GC_T(T/(b_T L^z))$ with avalanche duration
exponent $\alpha$ and dynamical exponent $z$.

All measurements reported below are taken in the stationary (or steady)
state, which the Manna Model develops into over long times. 
Because of the two timescales (microscopic and macroscopic) and
an obvious dependence of observables on the microscopic time passed since
driving, stationarity 
may appear somewhat awkward
to define properly. To do
this, we introduce $P_i(\{z\})$, which denotes the probability of finding
the system in a certain quiescent configuration $\{z\}$ (the set $\{z\}$ denoting
all particle numbers on all sites) after driving it $i$ times and
allowing all avalanching to cease. The system is stationary if
$P_{i+1}(\{z\})=P_i(\{z\})$, \ie in case of invariance of $P_i(\{z\})$
under further drives. We will use ``stationary state'', ``steady state''
and the invariant joint probability $P_i(\{z\})$ synonymously in the
following. In the stationary state (or, equivalently, sampling initial
states from $P_i(\{z\})$), the expectation of any observable taken after
$t$ microscopic time steps after the $i$th drive, will be identical to
that after $t$ time steps after the $i+1$th drive. This state is what we
have attempted to characterise numerically below.

However, we do not aim to estimate or determine $P_i(\{z\})$ explicitly.
Rather, as in most Markov Chain Monte-Carlo procedures, we initialise
our system once (or a small number of times, letting multiple instances
run embarrassingly parallel) and trigger a large number of avalanches,
hoping that after passing (and dismissing as transient) many avalanches
the resulting configuration is far more representative (\ie likely) than
the initial configuration, so that all further evolution of the system may be
considered as an exploration of the stationary state, with each new
configuration being the initial configuration for the next avalanche.
Rare configurations may
still occur, but with suitably low frequencies. The present argument is
inherently quantitative, as 
\emph{every} configuration (modulo a certain conserved parity in certain
settings, see \Aref{appendix_unique_estat}) is recurrent, \ie configurations
transcended at (supposed) stationarity are rare ones,
not strictly transient ones.

It may appear natural to start the Manna Model from an empty lattice
\cite{Manna:1991a}, but because the substrate density is close to unity
at least in $d=1$ dimensions, it pays off to start from full occupation.
A similar approach has been proposed for the Oslo Model
\cite{Dhar:2004,Corral:2004c} (and a more sophisticated one recently \cite{GrassbergerDharMohanty:2016}); in that case, the distribution of
configurations after a single further charge is exactly the stationary
$P_i(\{z\})$. We are not aware of a similar proof for the Manna Model.
In some cases, we have initialised the lattice with bulk density (as
estimated in preliminary runs or as published
\cite{HuynhPruessnerChew:2011,HuynhPruessner:2012b}) throughout.  After
initialisation, we generally dismissed a generous number of typically
about $10^6$ avalanches as transient.

Measurements of most relevant observables were taken as averages over
$100$ or so ``chunks'' of about $10^5$ avalanches each.  By monitoring in particular (but not exclusively)
moments of the avalanche size we were able to determine whether the
transient was over, \ie satisfy ourselves that ``equilibration'' had
been achieved.  Although analytically known, the first moment is a
somewhat misleading indicator for that. We focused instead on higher moments,
taking as the end of the transient a small multiple of the number
avalanches from when on the estimate of the moment is no longer
monotonic in the number of avalanches since initialisation.  Increasing
the length of the transient beyond that has no noticeable effect on the
estimates (within the estimated error) and we are therefore confident
that the SOC Manna Model is not suffering from the same dependence on
the transient as recently reported for the fixed energy sandpile version
in one dimension
\cite{BasuETAL:2012}.
We further verified that our numerical findings are consistent with 
published data \cite{Luebeck:2000,Pastor-SatorrasVespignani:2001,LuebeckHeger:2003b,Bonachela:2008,HuynhPruessnerChew:2011,HuynhPruessner:2012b}.

After the transient, the chunks can always be merged
to create estimates based on bigger chunks, so that each chunk exceeds the correlation
time on the macroscopic time scale (see \Sref{ava_ava_corr}), which is
orders of magnitude shorter than the transient. Chunks of that size may
be treated as statistically independent. On the basis of about $100$
such
chunks statistical errors are easily calculated.  As a random number
generator we used the Mersenne Twister \cite{MatsumotoNishimura:1998a}.

All numerical results stated in the following are based on such
chunk-averages \cite{Pruessner:2012:Book}. To ease notation (and discussion) we will not
distinguish numerical estimators and exact population averages, which we
may
denote by $\ave{\cdot}$ (usually for avalanche sizes and duration
averaged across many avalanches).  Although we spend
much time on only one dimension, we will use vectors such as $\xvec$ and
$\xvec_0$ to denote positions, and use scalars such as $x$ and $x_0$
only if the result is either restricted to one dimension, or if the only
relevant component of the vectors in our hyper-cylindrical lattice is the
distance from the open boundary.

Many of the results below derive from observables that are defined for
the entire lattice, which suggests that expectation values are
calculated on the basis of scanning the entire lattice.  Because this is
computationally very costly, we made extensive use of stacks (that store
the list of sites active at a given time $t$) and ``integration by
parts'', as the time series $a_t$ for $t=0,\ldots,T-1$ obeys
\cite{Pruessner:2012:Book}
\begin{equation}\elabel{summation_by_parts}
\sum_{t=0}^{T-1} a_t = T a_{T} - \sum_{t=1}^T t (a_{t}-a_{t-1}) \ ,
\end{equation}
with arbitrary $a_T$, in particular $a_T=0$.  The right hand side is
computationally much less costly to calculate in cases where changes
$a_{t}-a_{t-1}$ of
$a_t$ are rare and naturally tracked (for example if $a_t$ represents
the occupation of a site in the quiescent state). The computational gain
may depend on the dimensionality of the lattice; if, for example, $a_{t}-a_{t-1}$ are
changes in the local occupation of the lattice, then on average $\propto
L^2$ changes have to be tracked between avalanches, compared to scanning
a lattice of size $L^d$.

\subsection{Observables}
\slabel{observables}
The main objective of the present work is to characterise
spatio-temporal correlation functions. In field theoretic terms, we are
considering both response functions and correlation functions, but also,
effectively, three-point functions. 

As far local degrees of freedom are concerned, the particle numbers (or
densities) observed at
a site naturally divide into two ``categories''.
Firstly,
there is the number of immobile particles (or ``substrate particles'')
residing at any site, unambiguously measured in the quiescent state, secondly the number of
particles moving, which is always a multiple of $2$, to be precise
$2\lfloor z_{\xvec}(t)/2 \rfloor$ for a site carrying $z_{\xvec}(t)$
particles. This latter observable is more elegantly expressed as the number
of
topplings or the ``activity''
occurring at any site at time $t$ over the course of
an avalanche, $n(\xvec,t;\xvec_0,L)=\lfloor z_{\xvec}(t)/2 \rfloor$ as
introduced above.
As each toppling involves two particles leaving the site,
the activity $n(\xvec,t;\xvec_0,L)$ is thus half the count of active
particles. 
These counts of immobile (substrate) particles and of the instantaneous
topplings (activity) can be correlated in time and space. We will not
mix them in the following, although that would give rise to very interesting
observables, such as the activity as a function of local particle
density. We will use the notion of ``count'' (per site) and ``density''
synonymously.

All of these observables (the ``counts'') must be considered as a
function of the driving position, $\xvec_0$. We will consider (almost)
exclusively centre driving (for the definition see above).  The driving position
makes,
effectively, any local count a two-point correlation function, namely
the count somewhere as a function of the driving somewhere else. In case
of the immobile particle count, it turns out that observables are
(under certain conditions) independent of the position of driving. As far as activity
is concerned, the opposite is the case, \ie location and time of
activity is quite obviously correlated to the position and time of
driving. Driving at a site results in activity at that site with a
probability equal to the probability of that site being occupied, which
displays a very shallow spatial profile, \ie in the bulk, the
probability of a site being occupied has very little dependence on its
position. Up to this pre-factor, the activity resulting from driving the
system may therefore be seen as the response (that is the activity
resulting from creating activity somewhere else).

We will also consider higher correlation functions, such as the immobile
particle count at two different points in space given the driving at the
centre. We will usually choose one of the two points to be the driving
site. Similarly, we will consider correlations in the activity, given
driving somewhere else and again, we will choose one site (probed for
activity possibly at a later time) to coincide
with the driving site. 

The decomposition of the system into an even and an odd sublattice, as
discussed above, results in certain correlation functions vanishing ---
if site $x$ topples at time $t$, site $y$ may topple at time $t'$ only
if $y+t'$ has the same parity as $x+t$.

Many of the results derived below will be based on collapses, which are
often not very sensitive as far as estimates of exponents are concerned.
Rather, they give qualitative results, indicating that scaling takes
place and whether exponents found are compatible with those in the
literature.

In the following we first present a mean field theory before discussing 
the results in detail, defining the various observables as we proceed,
first through the results in one dimension (\Sref{results_one_dim}) and
then in three and five dimensions (\Sref{higher_dimensions}). We will
conclude with a discussion of the results in \Sref{conclusion}.

Briefly summarising the key results, we will demonstrate that correlations in the substrate (\ie in the distribution of
immobile particles)  are quite faint and that the density profile of immobile particles that
the system adopts is (essentially) independent of the driving and very shallow. In one
dimension, we can qualify this statement further by providing an
analytical proof. In our interpretation, this result suggests that the
notion of ``self-organising to the critical state'' --- namely the one
and only critical state, given by the invariant ensemble $P(\{z\})$ and resulting in a particular particle density profile --- is indeed justified.
Further, we will show that
the activity profile (the response function) is essentially Gaussian in
space and
that its spreading is governed by the exponents as captured by the
avalanches normally analysed in SOC. Remarkably, this link seems somewhat
flawed in one dimension with inconsistencies occurring
within results presented in the following and in relation to the literature. We believe that validating the
relation between the scaling of avalanches and the scaling of
spatio-temporal correlation functions is crucial for the understanding of
SOC and the significance of avalanches as historically studied.
Notably, the correlation length of the activity as well as of the weak correlations in the substrate are linear in the
system size, as expected in a critical, finite system.

\subsection{Mean Field Theory}
\slabel{mft}
There is surprisingly little effort in the literature to devise any
spatially extended mean field theory (MFT). This is probably because
mean field theories are designed to ignore certain interactions and thus
correlations and fluctuations. If spatial correlations are neglected,
one may be tempted to disregard space as a whole. However, it has been
known for a long time that boundaries are fundamental to SOC, as
particles in bulk-conservative models can only leave via the boundary
\cite{HwaKardar:1989a,nonotePaczuskiBassler:2000}.  Some authors have
attempted to mimic their effect by introducing a bulk dissipation rate
\cite{VespignaniZapperi:1998,ChessaMarinariVespignani:1998,BarratVespignaniZapperi:1999,Pastor-SatorrasVespignani:2000e}. 

As far as the spreading of activity is concerned, one may think
of it as a spatially extended branching process, whereby activity
(similar to active particles) moves
on a lattice, ceases upon arrival at a site with probability $1/2$ or
doubles otherwise, with both ``offspring'' being redistributed randomly
and independently at nearest neighbours. This MFT model, a branching
random walk which arises as the tree level in a recent field-theoretic
study \cite{Pruessner:2017:FT}, differs from the Manna Model crucially
in the mechanism by which activity doubles --- in the Manna Model this
is dependent on the occupation number at the site, and that is in turn
dependent on whether or not activity has ceased at the site previously.
The particle number is conserved in the Manna Model, but noticeably
activity only in the sense that its time-integral is identical to that
of the density of (bulk-conserved) random walkers with the diffusion constant as stated
above. This last point is also captured by the MFT model.  We are
planning to publish a detailed analytical study of the MFT model soon
\cite{LeDoussalWiesePruessner:2016:MFT_unpublished}.  For comparison
with the numerics obtained in the present study, we will occasionally
draw on this MFT model, deriving some of its features in passing.

\section{Results in one dimension}
\slabel{results_one_dim}
\subsection{Quiescent state}
\slabel{quiescent_state_1D}
Because of the separation of time scales, avalanches are instantaneous on the time scale
of driving (the macroscopic time scale). 
Observing therefore the lattice at any given macroscopic time, it is
quiescent.\footnote{In the following we distinguish the ``quiescent state'',
which refers to the system being quiescent, ``quiescent configurations''
which is any of the $2^{N}$ configurations that are quiescent and the
``stationary state'' or ``steady state'' which means that the probability 
of any such configuration is invariant
under driving (at a certain site).} 
In fact, the connection between macroscopic and microscopic time scale
is sometimes made by devising an infinitely slow Poissonian rate
\cite{BakSneppen:1993}, so that the system is almost surely quiescent at
any randomly chosen microscopic time.
It
is thus natural to attempt to identify the signature of SOC in the
quiescent state. 
In the following, we will study the
one-point and two-point correlations of immobile (inactive) particles
that make up those quiescent configurations.

The one-point correlation $\Ds(\xvec;\xvec_0,L)$ is the expected count
(at site $\xvec$) or density of
inactive particles as a function of position $\xvec$ and the position
$\xvec_0$
where the driving takes place, in a system of linear extent
$L$. This function is below referred to as the ``density profile''. 
The measurements are taken at quiescence (when no avalanche is running) and in the stationary state.
Strictly $\Ds(\xvec;\xvec_0,L)$ is a response function of the density of inactive
particles at $\xvec$ in response to a driving taking place at $\xvec_0$
in the presence of an initial distribution of inactive particles and
in the long time limit (when the avalanche has ceased).
Because we are measuring in the stationary state, the density of
inactive particles is invariant under further external driving, \ie
there is no time-dependence (which does not
mean that the arrangement of inactive particles remains unchanged under
driving, but only that its \emph{joint probability distribution} is not
changing).
Since sites can be occupied by at most one particle, the expected particle
count at a site is the probability of finding a particle there at all.

\begin{figure*}
\subfigure[Density profile in relative coordinates.]{\includegraphics*[width=0.45\linewidth]{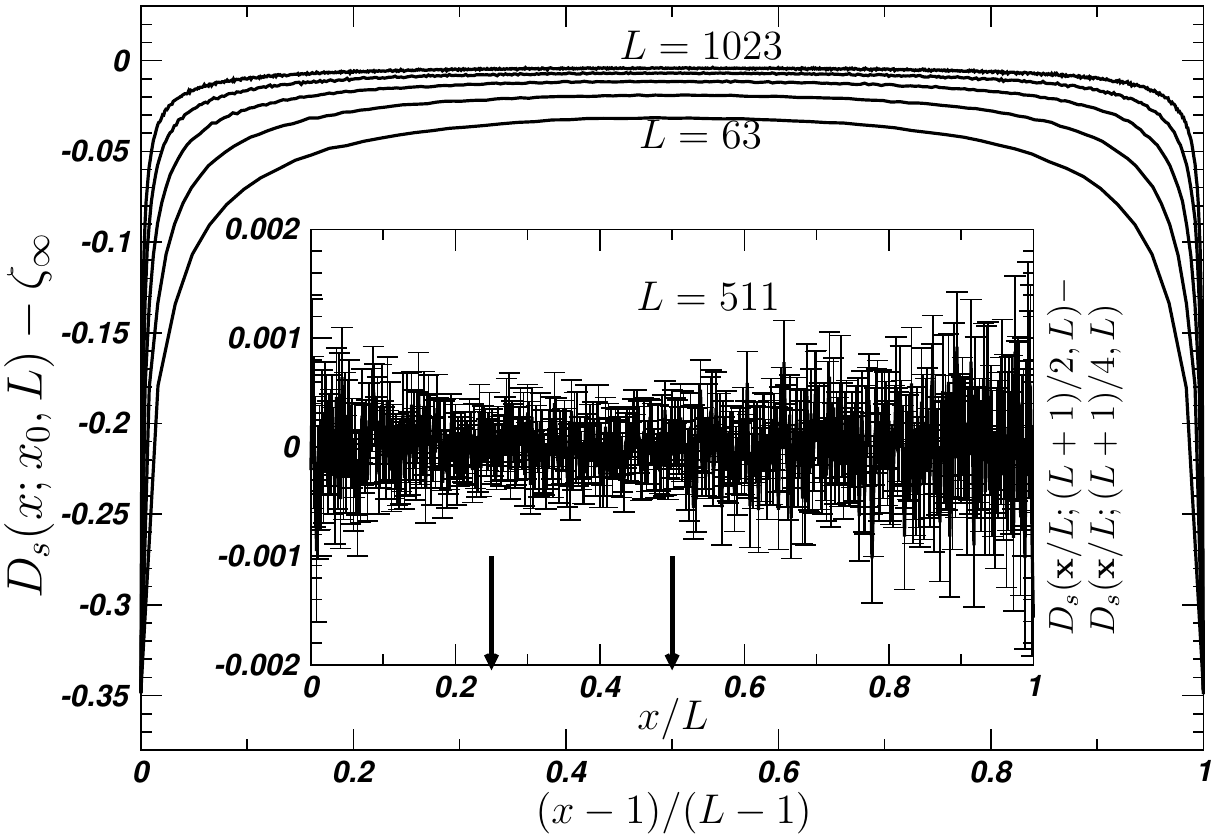}\flabel{substrate_density_1D_bulk}}
\subfigure[Density profile close to the boundary.]{\includegraphics*[width=0.45\linewidth]{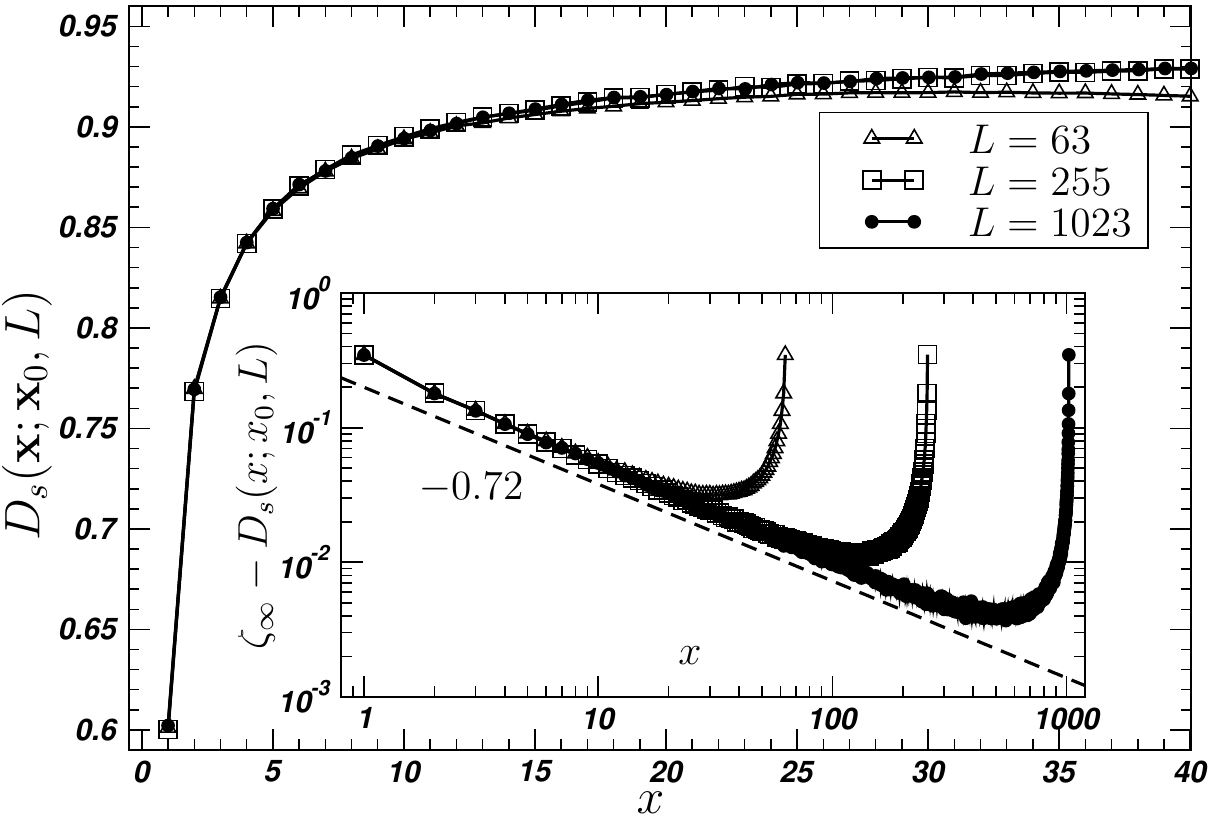}\flabel{substrate_density_1D_boundaries}}
\caption{\flabel{substrate_density_1D} The density of inactive
(immobile, substrate) particles during quiescence in the stationary
state, $\Ds(\xvec;\xvec_0,L)$, for different system sizes $L$ and as a
function of position $x$ (full lines to guide the eye, error bars
negligible unless shown).  \subfref{substrate_density_1D_bulk} Plotting
$\Ds(\xvec;\xvec_0,L)-\zeta_{\infty}$ versus the relative position
$(x-1)/(L-1)$ with $x\in\{1,\ldots,L\}$ reveals that deviations of the
density away from the asymptotic value $\zeta_{\infty}$ are confined to
the regions close to the boundary. With increasing system size, an
increasing fraction of sites is occupied with a probability that is
arbitrarily close to $\zeta_{\infty}$.  The data in the main panel is
for $x_0=(L+1)/2$ (centre driving).  The inset shows for $L=511$ the
difference between the density profile for different driving positions
(marked by arrows), $\Ds(\xvec/L;(L+1)/2,L)-\Ds(\xvec/L;(L+1)/4,L)$, \ie
driving at $x_0=(L+1)/2$ and at $x_0=(L+1)/4$. This deviation clearly remains within the
numerical error (\ie can be fitted against $0$ with almost unity as
goodness of fit).  Errors are smaller near the driven sites $x_0$ due
to better statistics.  \subfref{substrate_density_1D_boundaries} The
density profile close to the boundary quickly converges with increasing
$L$, as shown here for $L=63,255,1023$.  The ``shoulder'' in the profile
does not scale, in the sense that densities differ visibly among different
\emph{small} $L$ only
further in the bulk (as can be seen for $L=63$ and $L=255$, but for
$L=255$ and $L=1023$ only at bigger $x$ and then much less pronounced). 
The inset shows the scaling of the difference $\Ds(\xvec;\xvec_0,L)-\zeta_{\infty}$
as a function of $\xvec$, revealing an intermediate powerlaw dependence with
exponent of approximately $-0.72$ (dashed line).
}
\end{figure*}

The sum
\begin{equation}\elabel{def_zeta}
\zeta_L = N^{-1} \sum_{\xvec} \Ds(\xvec;\xvec_0,L)
\end{equation}
over all $N$ sites is the 
spatially averaged
(expected) density of particles in the system at stationarity (as
$\Ds(\xvec;\xvec_0,L)$ is taken at stationarity). As can be seen in \Fref{substrate_density_1D_bulk},
in large enough systems
$\Ds(\xvec,\xvec_0,L)$ shows little variation in the bulk, as boundary
effects decay, according to \Fref{substrate_density_1D_boundaries}, independent of the system size,
so that $\Ds(\xvec,\xvec_0,L)$ converges as $L\to\infty$ for fixed $\xvec$
and $\xvec_0$, \ie the shoulder of $\Ds(\xvec,\xvec_0,L)$, visible for small $x$, is
reproduced with increasing system size.

The inset of \Fref{substrate_density_1D_boundaries} suggests that the deviation of the density from the bulk value follows a power law as a function of the distance away from the boundary, 
$\Ds(\xvec;\xvec_0,L)-\zeta_{\infty} \propto |\xvec|^{-0.72}$, 
over a characteristic scale that is linear in the system size (probably related
to
the scaling of the correlations seen below in the inset of \Fref{substrate_correlations_1D_zoom}).
Because the amplitude of the deviation quickly converges with increasing $L$,
this is difficult to confirm in the bulk of large systems, as any deviations
eventually drown in noise. This is a common theme in substrate features: Amplitudes
do not display finite size scaling.
The inverse of the observed exponent, $1/0.72=1.388\ldots$ should be $\nu_\perp$ \cite{Luebeck:2004},
estimated below to be $1.395(3)$.
Bonachela and Munoz have studied a range of observables
in the Manna Model
as a function of the distance from the boundary \cite{BonachelaMunoz:2007}
(also \cite{BonachelaMunoz:2009b})
and Grassberger, Dhar and Mohanty \cite{GrassbergerDharMohanty:2016} recently found
the same scaling in the Oslo
Model
\cite{ChristensenETAL:1996}, which is thought to be in the same universality
class \cite{NakanishiSneppen:1997}. They found an exponent of approximately $0.75$,
which is
still compatible
with the present data. 

The density is bounded from above and from below, but nevertheless displays a small decrease with system size 
at the boundary sites ($x=1$ and $x=L$). Because of the lack of scaling of the amplitude of the deviation,
the drop in the 
rescaled plot
\Fref{substrate_density_1D_bulk} 
from the bulk 
density towards the lower density at the boundaries gets increasingly sharp with system size.
For very large system sizes, 
the density in the bulk may therefore 
be approximated nearly everywhere by $\zeta_L$, \Eref{def_zeta}. Numerically, the best
known estimate
for its value in the limit of $L\to\infty$ is $\zeta_{\infty}=0.9488(5)$
\citep{HuynhPruessnerChew:2011}, which our value of $0.94882(1)$ is compatible with.
We have extracted that from our data by fitting results for $L=31,63,\ldots,4095$ against 
\begin{equation}\elabel{fit_zeta}
\zeta_{\infty} + a_1 L^{\alpha} + a_2 L^{\alpha-1/2}  + a_3 L^{\alpha-1}
\end{equation}
which also produces a very good goodness of fit (about $0.64$) and
generates, in passing,
an estimate of $\nu_\perp=-1/\alpha$ of $1.395(3)$ which compares well with previous estimates
of $1.35(9)$ \citep{Luebeck:2004}.

That the density profile shows so little structure suggests that it does not
even reveal the position $\xvec_0$ where the driving takes place, \ie
that $\Ds(\xvec,\xvec_0,L)$ is independent of the driving position.
Numerically, this is indeed confirmed.  The inset of
\Fref{substrate_density_1D_bulk} shows the difference
$\Ds(\xvec/L;(L+1)/2,L)-\Ds(\xvec/L;(L+1)/4,L)$ between the density
profile of $L=511$ driven at $x_0=(L+1)/2$ and at $x_0=(L+1)/4$. This
data is fully compatible with the hypothesis that the profile does not
depend on the driving position.

The stationary state, \ie the invariant probability of finding a certain
profile of immobile particles, could in principle be dependent on the
way (where and how) the system is driven and also on the initialisation.
For the system sizes considered here ($L\ge63$), we do not find any such
dependence. The resulting profile is independent of $\xvec_0$ and other
details of the driving.

In \Aref{appendix_generate_matrices} and more particularly
\Aref{appendix_eigenstates}, we discuss an analytical approach to the stationary
state. There, we show that in one dimension 
\emph{the stationary state reached by driving the Manna Model at the first site, $x_0=1$, at the last 
site, $x_0=L$, or globally with positive probability at every site (such as uniform driving), is identical and unique}.
This stationary state is the 
 \emph{unique
invariant distribution of configurations that is common to all driving sites}. 
This is a remarkable feature that is in perfect agreement with
the notion of self-organisation in the Manna Model, namely that there is
one and only one stationary state (a distribution of configurations)
that the model evolves towards, irrespective of whether it is driven at a boundary site
or at all sites with
positive probability.

However, as discussed in \Aref{appendix_degeneracy}, it turns out that the
invariant probability is in fact two-fold degenerate (and can in
principle be even more degenerate), as there is a conserved quantity if
$L$ is odd and $x_0$ is even (as in the present case of odd $L$ and
centre driving). This degeneracy was not picked up in the
numerics mentioned above, because it is visible in the density profiles
$\Ds(\xvec/L;\xvec_0,L)$ only in very small systems (see
\Fref{degeneracy_lifting}). For systems of size $31$ and bigger it seems
numerically impossible to differentiate between the stationary states
resulting from these different initial conditions
and one will therefore arrive always at the
(numerically) same density profile $\Ds(\xvec/L;\xvec_0,L)$.

\subsection{Correlations}
\slabel{substrate_correlations}
\begin{figure*}
\subfigure[Correlations in the substrate occupation.]{\includegraphics*[width=0.45\linewidth]{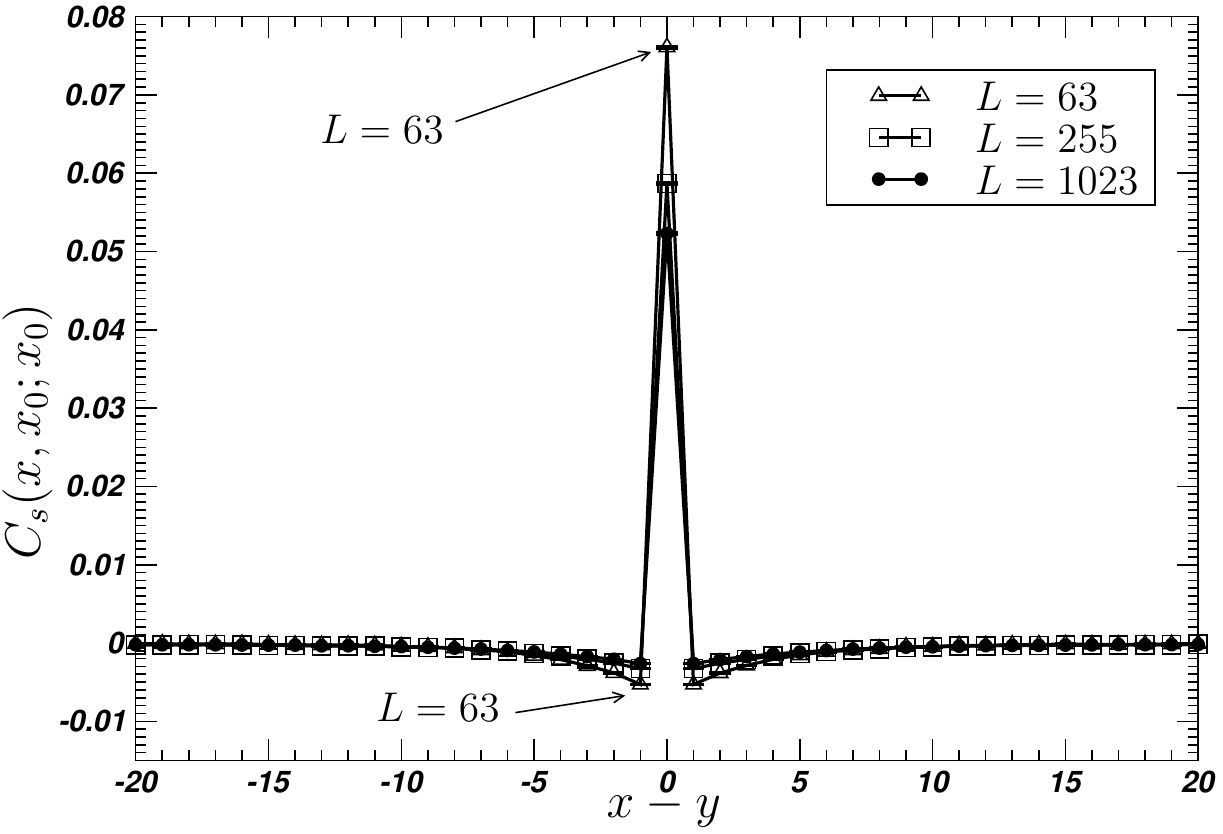}\flabel{substrate_correlations_full}}
\subfigure[Correlations in the substrate occupation (details).]{\includegraphics*[width=0.45\linewidth]{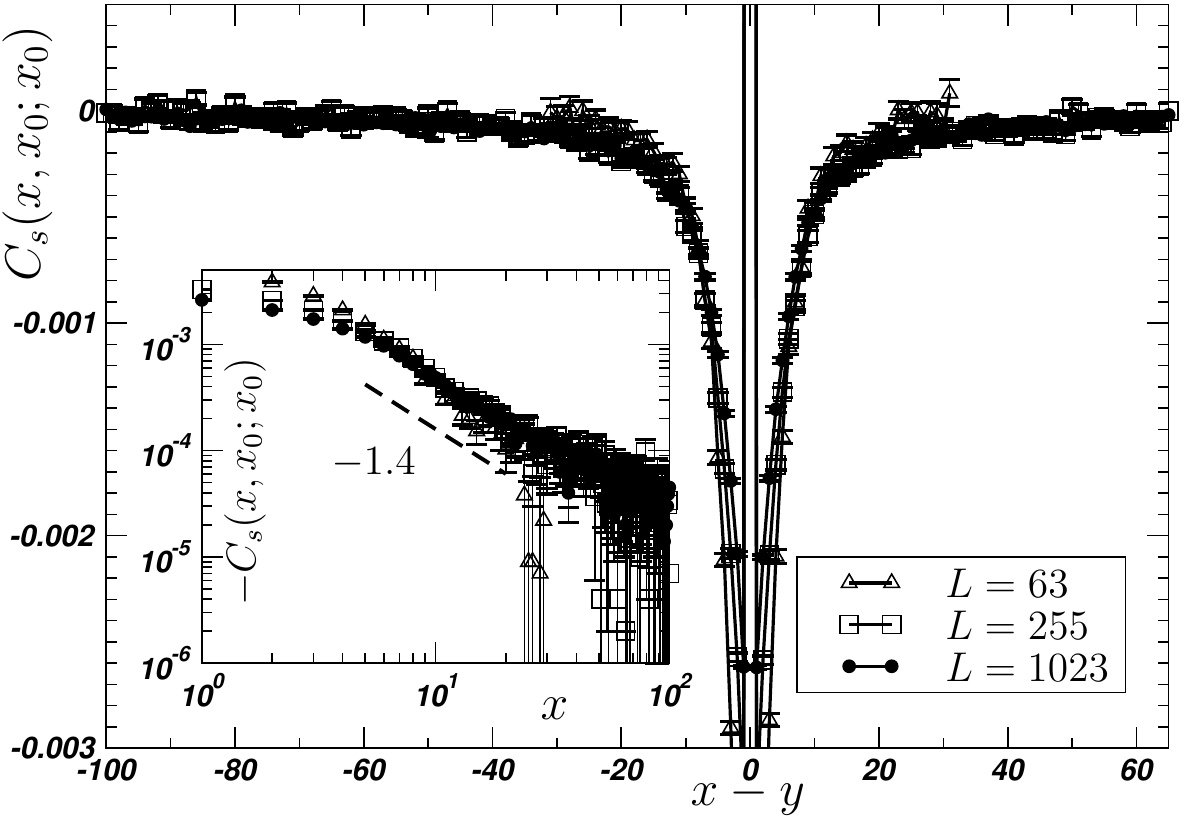}\flabel{substrate_correlations_1D_zoom}}
\caption{\flabel{substrate_correlations_1D}The correlations in
occupation (density)
by inactive particles, 
$C_s(x,y;\xvec_0,L)$,
\Eref{def_Cs}, versus the distance $x-y$,
measured during
quiescence and at stationarity.
The system is driven at the
centre site $x_0=(L+1)/2$, which is also the location where one of the
densities is taken, $y=x_0$. Correlations are in fact anticorrelations
(with an amplitude decreasing with increasing system size), because the
dynamics depletes sites of particles, depositing them at neighbouring
sites. \subfref{substrate_correlations_full} Correlations are slightly more
pronounced for the \emph{smaller}
systems, decay very quickly and are barely noticeable beyond $5$ to
$10$ sites. \subfref{substrate_correlations_1D_zoom} On a very fine
scale correlations are more discernible, and in fact might decay like a
powerlaw, as shown in the inset (the dashed line shows a powerlaw with
exponent $-1.4$).  As data for different system sizes $L$ collapse on the same plot
without rescaling, there is no non-trivial finite size scaling in these
correlations (other than possibly in the cutoff, see main text). The data
in \subfref{substrate_correlations_full} and \subfref{substrate_correlations_1D_zoom}
are identical but shown on different scales.}
\end{figure*}

In the stationary state, not only the one-point
density $\Ds(\xvec;\xvec_0,L)$ is invariant, but in fact the probability
of finding the system in any of its $2^N$ configurations.
Indeed, the
analytical results mentioned above (see
\Aref{appendix_eigenstates}) give access also to $n$-point
correlation functions (in principle even to the response function and the
``temporal shape of the avalanche'' discussed below) --- unfortunately, however, only for very
small system sizes. These results are therefore not shown.

Carrying on with numerical results for systems of size $L\ge 63$,
in the following we analyse two point correlations in the occupation by
inactive particles. 
If $P_s^{(2)}(\xvec_2,\xvec_1;\xvec_0,L)$ is the joint
probability of finding a particle at $\xvec_1$ and another one, at the same
(macroscopic, quiescent) time, at $\xvec_2$ after driving at $\xvec_0$, then
\begin{multline}\elabel{def_Cs}
C_s(\xvec_2,\xvec_1;\xvec_0,L) = 
P_s^{(2)}(\xvec_2,\xvec_1;\xvec_0,L) \\
- \Ds(\xvec_2;\xvec_0,L) \Ds(\xvec_1;\xvec_0,L)
\end{multline}
is the connected two-point correlation function. As shown in
\Fref{substrate_correlations_1D}, small anti-correlations are present in the
distribution of inactive particles. 
However, there is clearly no finite
size scaling of the amplitude of these correlations, which cannot possibly
increase indefinitely as the density is bounded everywhere. 
In fact, the anti-correlations die
off
very quickly in space. 
Despite being
slightly \emph{less} pronounced for large system sizes,
they seem to converge to a finite value for increasing $L$, \ie
they are not merely a finite size effect. 
Because
the spatial scale of the anti-correlation does not vary significantly with the size of the
system, this correlation function does not collapse under any
non-trivial rescaling. However, for distances $|x-y|$ of less than
about $50$ sites $C_s(x,y;\xvec_0,L)$ shows some noisy linear
behaviour in a double logarithmic plot, as shown in the inset of
\Fref{substrate_correlations_1D_zoom}. This may suggest a power law
dependence of $C_s(x,y) \propto |x-y|^{-1.4}$, 
albeit with a very small amplitude 
of about one third of the (local) variance 
$C_s(\xvec_0,\xvec_0;\xvec_0,L)$, which is itself  a small quantity (as discussed
below).
The
power law-like behaviour persists for $y\ne
x_0$ and, as a \emph{second} moment, may be related to the exponent of $-0.72$
found in the scaling of the
deviation of the density from the bulk value, away from the boundary 
\Fref{substrate_density_1D_boundaries} and thus expected to be $-2/\nu_\perp \approx 1.48(10)$ \cite{Luebeck:2004}.

Given the small amplitude of the anti-correlations,
which seems to converge
from above with
increasing system size, and the large relative statistical error, it is fair
to say that the anti-correlations are not very pronounced and difficult to measure. There is clearly
no scaling of the amplitude with system size and no rescaling needed to achieve
the (noisy) collapse of \Fref{substrate_correlations_1D_zoom}. This lack of
scaling is similarly found in the density profile shown in \Fref{substrate_density_1D_boundaries}, yet the exponent roughly characterising the scaling of the correlations is about twice that characterising the decay of the density difference from the bulk value away from the boundary. Similar power law scaling is observed in in three and five dimensions (\Frefs{substrate_density_3D_boundary} and \ref{fig:substrate_density_5D_boundary} respectively), but the data is obviously plagued by statistical noise.
Future studies, in particular
using more sophisticated observables and numerical techniques, may be more
successful in identifying features in the substrate whose amplitude scales up with increasing system size
and that (unlike, say, the shoulder in \Fref{substrate_density_1D_bulk} localised
close to boundary) remain visible even when the (apparent) lattice spacing
is very small compared to the range of observation. 

Within statistical
error the variance $C_s(\xvec_0,\xvec_0;\xvec_0,L)$ coincides with the Bernoullian
$\Ds(\xvec_0;\xvec_0,L)-\Ds(\xvec_0;\xvec_0,L)^2$ (which is a small quantity as
$\Ds(\xvec_0;\xvec_0,L)$ is close to unity, $\zeta_{\infty}=0.9488(5)$).
To explore the correlations apparent in \Fref{substrate_correlations_1D_zoom} further, 
we have also measured the distribution of distances between unoccupied sites, measured 
as the number $d_0$ of consecutively occupied sites between any two unoccupied
ones. If occupation is governed 
by a Bernoulli process, the frequency $P_0(d_0)$ of such distances $d_0$ should follow 
$(1-\zeta_{L})\zeta_{L}^{d_0}$. As shown in \Fref{zero_spacing}, a 
semi-logarithmic plot produces a mixed picture. On small scales (small $d_0$ or small system size)
significant deviations are apparent, but with increasing system size, the large scale
behaviour seems to approach the expected exponential, although with a higher density of
unoccupied sites (a steeper slope). Very long stretches of continually occupied sites are 
very rare and their statistics therefore subject to significant noise.
That barely any correlations are visible on the large scale is nevertheless consistent with the 
observation of hyperuniformity discussed in the following.

\begin{figure}
\includegraphics*[width=0.95\linewidth]{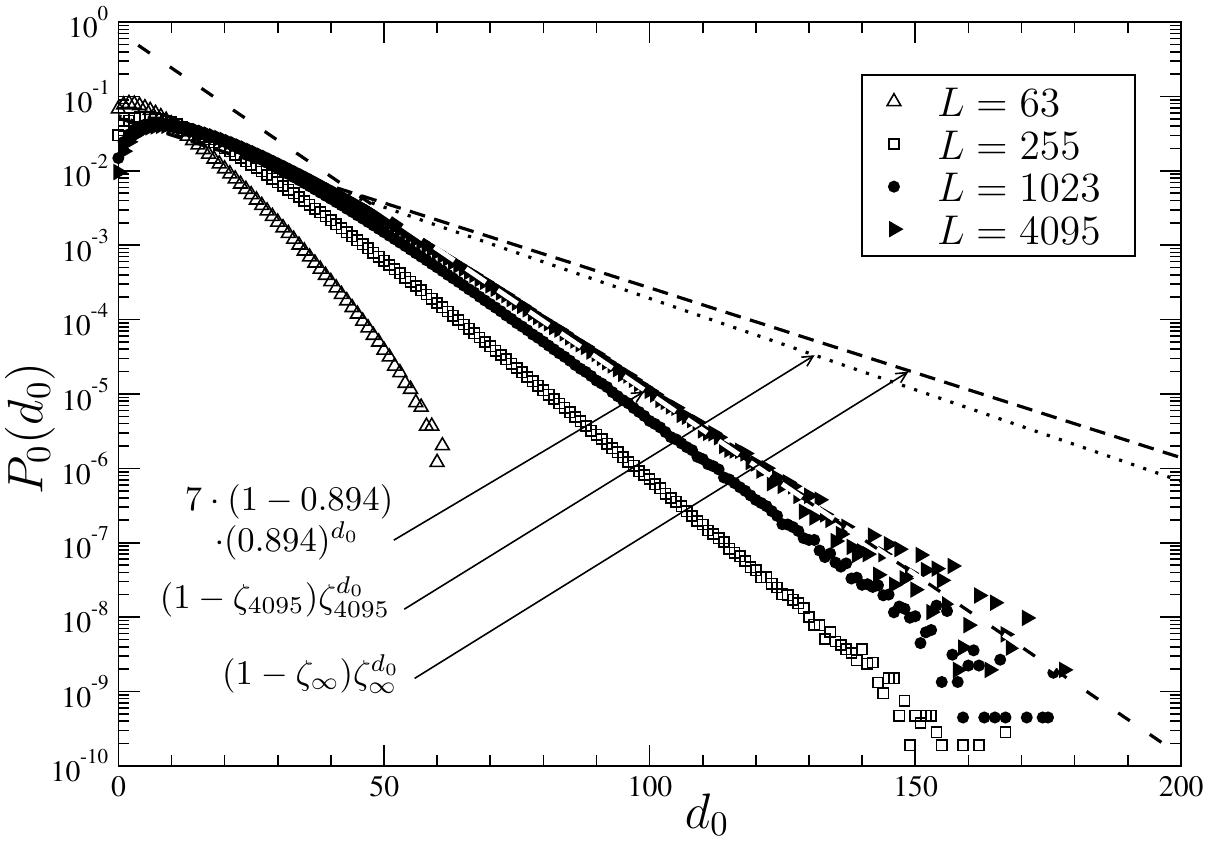}
\caption{\flabel{zero_spacing}
Semi-logarithmic plot of the distribution $P_0(d_0)$ of distances $d_0$ between consecutive unoccupied sites (separated by $d_0$ occupied ones) 
in one dimension for various system sizes. If sites are occupied independently, the data forms a straight line.
The thick (black and white) dashed line through the data for $L=4095$ is $7\cdot
(1-0.894)\cdot(0.894)^{d_0}$, 
the other (black) dashed line the Bernoullian $(1-\zeta_{\infty})\zeta_{\infty}^{d_0}$ using the asymptotic density
$\zeta_\infty=0.9488(5)$.
The dotted line shows $(1-\zeta_{4095})\zeta_{4095}^{d_0}$ with $\zeta_{4095}=0.945028(3)$.
}
\end{figure}
\begin{figure}
\includegraphics*[width=0.95\linewidth]{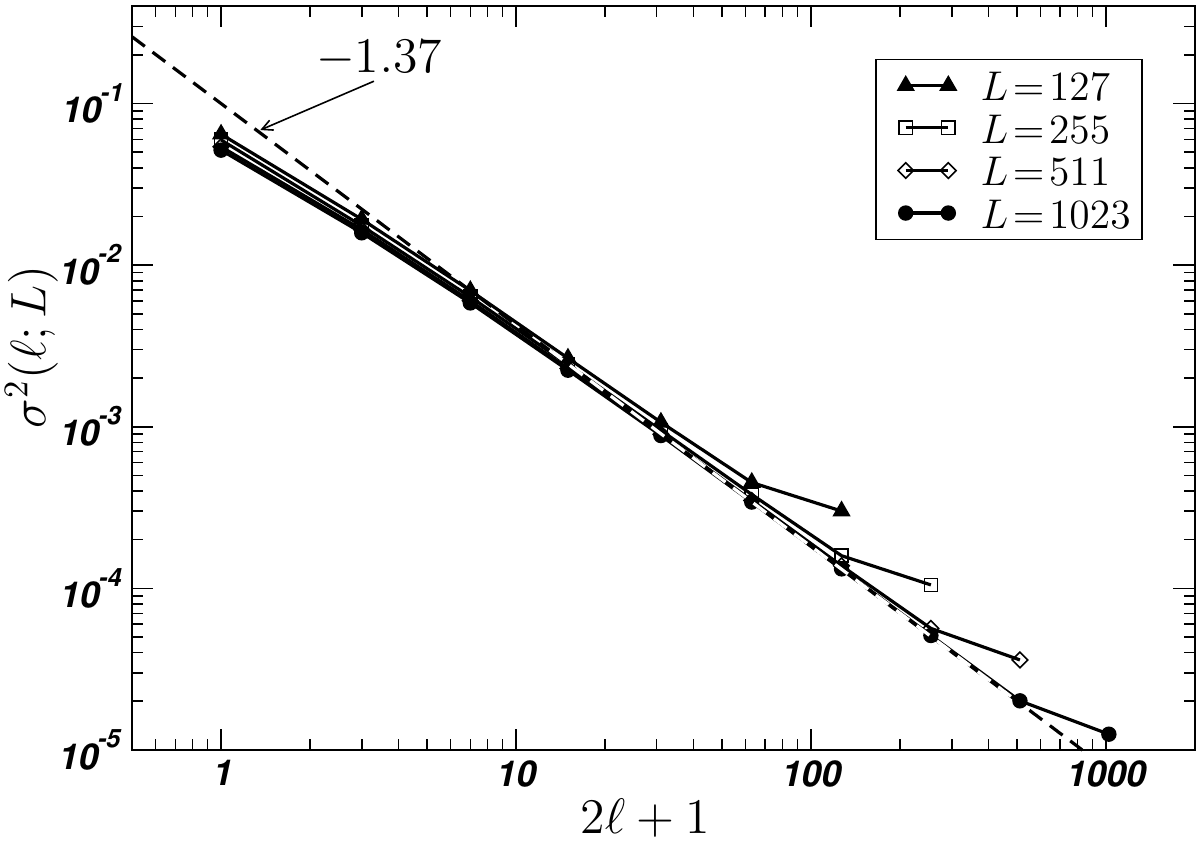}
\caption{\flabel{density_bracket_var}
Double logarithmic plot of the variance $\sigma^2(\ell;L)$ of the window-averaged (size $\ell$) substrate particle
density in one dimension for a range of system
sizes $L$. The dashed line shows the approximate scaling exponent $-1.37$,
which is steeper than $\sigma^2\propto\ell^{-1}$ expected when sites are independently
occupied.
}
\end{figure}

Through a different observable, there is already clear evidence for anti-correlations
in the fixed energy variant of the Manna Model, as Hexner and Levine 
\cite{HexnerLevine:2015} observed hyperuniformity
\cite{TorquatoStillinger:2003} in the substrate particle density and Basu \etal identified
``natural long-range correlations in the background''
\cite{BasuETAL:2012}.

Hyperuniformity refers to the (fast) scaling of the variance of the particle density with the volume over which this density is estimated. 
In one dimension, the instantaneous density might be measured as 
a window average
($\frac{1}{2\ell+1} \sum_{x=x_0-\ell}^{x_0+\ell} \ldots$) symmetrically around the driving site at the centre. Its variance is given by 
\begin{equation}\elabel{sigma2_from_C_s}
	\sigma^2(\ell;L) = \frac{1}{(2\ell+1)^2} \sum_{x_1,x_2=x_0-\ell}^{x_0+\ell}
	C_s(\xvec_2,\xvec_1;\xvec_0,L) \ge0\ ,
\end{equation}
and so $\sigma^2(\ell;L)\propto(2\ell+1)^{-1}$ if sites are independently occupied. In general, 
if $C_s(\xvec_2,\xvec_1;\xvec_0,L)$ was positive everywhere, $\sigma^2(\ell;L)$
could not decay faster than $\ell^{-1}$. If it does, this is referred to as hyperuniformity.
The variance $\sigma^2(\ell;L)$ can always be written as 
\begin{equation}\elabel{sigma2_behaviour}
\sigma^2(\ell;L)=\frac{\sigma^2(0;L)}{(2\ell+1)}+\sum_{\substack{x_1,x_2=x_0-\ell\\ x_1\ne x_2}}^{x_0+\ell}
C_s
(\xvec_2,\xvec_1;\xvec_0,L)
\end{equation} 
with $\sigma^2(0;L)=C_s(\xvec_1,\xvec_1;\xvec_0,L)$, the correlation
at $x_2=x_1$, which
is bound to be non-negative. At the heart of hyperuniformity is the behaviour
of the sum in \Eref{sigma2_behaviour}.
Even if $C_s(\xvec_2,\xvec_1;\xvec_0,L)$ is negative for $x_2\ne x_1$, it
might still be subleading, resuling in $\sigma^2(\ell;L)\propto\ell^{-1}$.
However, as illustrated in \Fref{density_bracket_var},
we found a scaling of $\sigma^2(\ell;L)\propto\ell^{-1.37}$, for $L=1023$ 
in an intermediate range of the width of about $31<2\ell+1\le 511$. We believe this value of the exponent 
is compatible with $-1.425(25)$ found by Hexner and Levine for the same quantity in the fixed energy version of the Manna Model. 

Ignoring the contributions from $C_s(\xvec_2,\xvec_1;\xvec_0,L)$ for small
$|x_2-x_1|$ or, equivalently, assuming that the positive contributions at
$x_2-x_1=0$, which scale like $\ell^{-1}$, are cancelled by negative ones
from small, positive $|x_2-x_1|$ (where it does not follow a power law), the
scaling of $\sigma^2(\ell;L)$ in large
$\ell$ is due to the (intermediate) asymptote of $C_s(\xvec_2,\xvec_1;\xvec_0,L)$,
which means that the exponent of $-1.4$ in the inset of \Fref{substrate_correlations_1D_zoom}
is to be compared to $-1.37$ and $-1.425(25)$, found for $\sigma^2(\ell;L)$ here and in \cite{HexnerLevine:2015}, respectively. 
Standard finite size scaling indeed suggests $\sigma^2(\ell;L)\propto\ell^{-2/\nu_\perp}$ \cite{GrassbergerDharMohanty:2016}.
Notably, the scaling of $C_s
(\xvec_2,\xvec_1;\xvec_0,L)$, which has to be cut off when $|x_2-x_1|$ exceeds
$L$ and the absence of finite size scaling of the amplitude are compatible
with hyperuniformity.
Our numerics indicate that the scaling of $\sigma^2(\ell;L)$ persists up to
$\ell\approx L/2$, which suggests that the scaling of $C_s(\xvec_2,\xvec_1;\xvec_0,L)$
is long-ranged, possibly of the form $|x_2-x_1|^{-1.4} \CC(|x_2-x_1|/L)$,
with a cutoff length linear in the system size.
Algebraic correlations of the substrate have first been observed analytically
in the seminal work by Majumdar and Dhar
\cite{MajumdarDhar:1991} on the paradigmatic Abelian Sandpile Model \cite{BakTangWiesenfeld:1987,Dhar:1990a}
and thus may be considered
the fingerprint of the critical state.

\subsection{Active state}
\slabel{active_state_1D}

\begin{figure*}
\subfigure[Activity as a function of position $x$ for different times
$t$ (and $L=511$ fixed).]{\includegraphics*[width=0.45\linewidth]{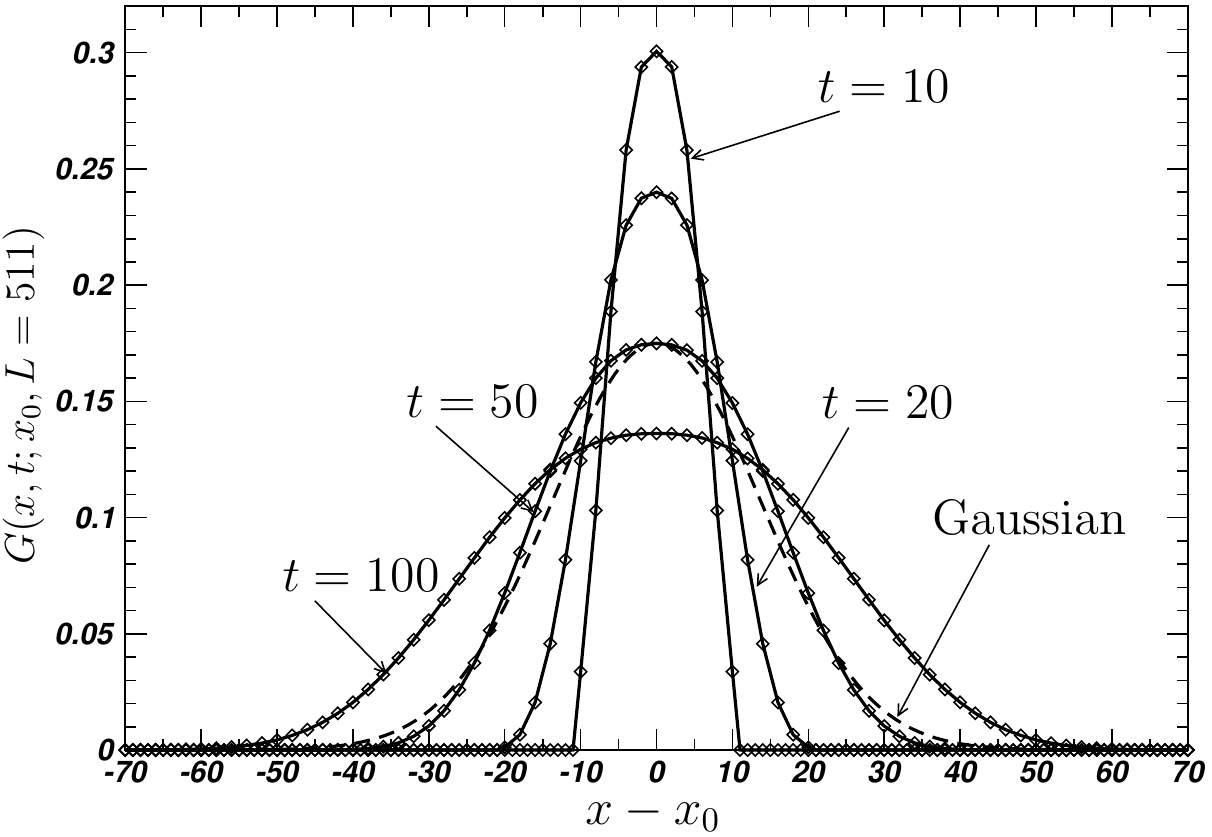}\flabel{response_t}}
\subfigure[Activity as a function of position $x$ for different system
sizes $L$ (and $t=10$ fixed).]{\includegraphics*[width=0.45\linewidth]{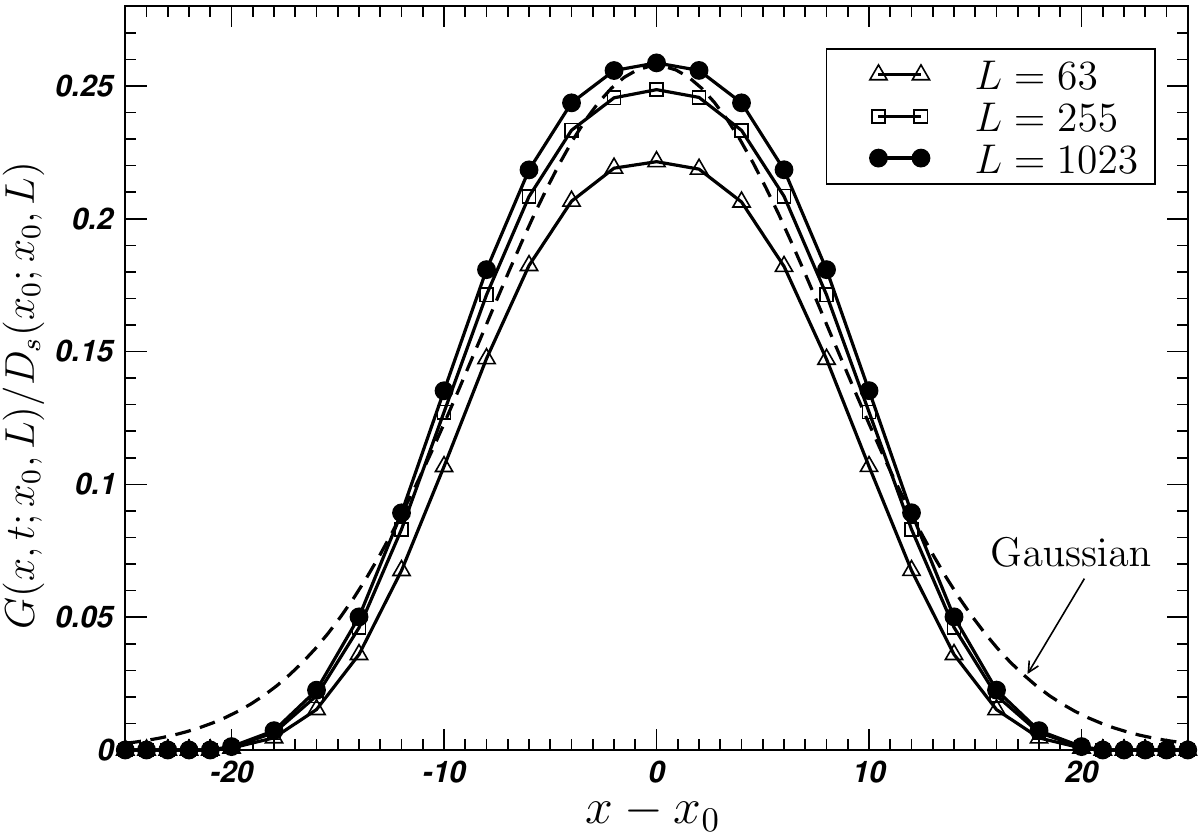}\flabel{response_L}}
\caption{\flabel{response_sliced}The activity $\Activity(x, t; x_0,L)$
in one dimension as a function of position $x$, for centre driving,
$x_0=(L+1)/2$, at various times and for different system sizes $L$ (for
a collapse see \Fref{full_collapse}).
Actual data are shown as symbols, which are connected by a line as guide
for the eye. Data points that necessarily vanish because of parity
conservation (see text) have been omitted.  \subfref{response_t} The
activity as a function of the distance from the driving site, $x-x_0$,
for fixed system size $L=511$ at various times $t$ as indicated. The
shape resembles a Gaussian (dashed line, for $t=50$), but deviates
clearly from it.  \subfref{response_L} The activity rescaled by the
particle density at the driving site as a function of the distance
$x-x_0$ for different system sizes $L$ and at fixed time $t=20$. The
difference in shape cannot be caused by the activity having reached the
boundary (as it cannot possibly given $t$), nor by the probability $\Ds(x_0;x_0,L)$
by which activity is
triggered. Again, a Gaussian (dashed) is shown for comparison.  }
\end{figure*}
The features in the active state, \ie during the course of an 
avalanche, are much richer not least due to the additional time-dependence. In
the following, we will follow roughly the order of observables above.
The one-point correlation function is, as above, really the response
function $\Activity(\xvec, t; \xvec_0,L)$, as defined below. It is the activity density at
$\xvec$ and (microscopic) time $t$ after the system was driven at
$\xvec_0$ at time $0$. 
Numerically, $\Activity(\xvec, t; \xvec_0,L)$ is the estimated number
of topplings of site $\xvec$ at time $t$ after an initial charge at
$x_0$.

As explained above, this frequency is measured by recording the number
$n(\xvec,t;\xvec_0,L)$ of topplings that occur at each lattice site
$\xvec$ during the $t$th sweep across all active sites, which is the
$t$th round of parallel updates. The zeroth sweep
is the initial drive, and so at the first sweep
$\ave{n(\xvec,1;\xvec_0,L)}=\Ds(\xvec_0;\xvec_0,L)\delta_{\xvec,\xvec_0}$,
where $\Ds(\xvec_0;\xvec_0,L)$ is the (expected) density of particles at
the driven site. In order to derive time-resolved estimates, for each
time $t$ these records have to be summed over and divided by the total
number of drivings. If $n_i(\xvec,t;\xvec_0,L)$ is the record of
the activity after $i$ driving attempts (macroscopic time), we use the
estimator
\begin{equation}\elabel{def_activity_estimator}
\Activity(\xvec, t; \xvec_0,L)=\frac{1}{M} \sum_{i=1}^M
n_i(\xvec,t;\xvec_0,L)
\end{equation}
from a sample of $M$ (consecutively) attempted avalanches by driving the system at site
$\xvec_0$. To make the estimator well-defined for $t>T_i$, we define
$n_i(\xvec,t;\xvec_0,L)=0$ whenever $t$ exceeds $T_i$, the duration of
the $i$th avalanche ($T_i=0$ if no avalanche has occurred).

The time-dependence makes the numerics more difficult to handle compared
to the statistics in the quiescent state discussed above. Indeed, the response
function $\Activity(\xvec, t; \xvec_0,L)$ contains more information in
its space and time-dependence than, say, $\Ds(\xvec;\xvec_0,L)$ and even at fixed $\xvec_0$ the analysis
is numerically and analytically more difficult due to the additional
time-dependence.
To facilitate further analysis, 
we will focus mostly on various
integrals of the response function $\Activity(\xvec, t; \xvec_0,L)$.

By the definition of the local dynamics (toppling), the trajectories of
active particles are those of random walkers. On the other hand,
$\Activity(\xvec, t; \xvec_0,L)$ itself does \emph{not} 
obey the diffusion equation,\footnote{This does not contradict the time-integrated
activity, \Eref{TimeIntegratedActivity}, to obey the Poisson
equation 
$D\nabla^2 \TimeIntegratedActivity(\xvec, t; \xvec_0,L) = -(1/2)\delta(x,x_0)$
\cite{Dhar:1990a}, \Eref{time_integrated_diffusion_eqn}.}
 first
of all because active particles may
become trapped for a certain microscopic time, only to be re-activated
some time later. Were those resting times discounted, each individual active
particle would perform a random walk from the time it enters the system
by external drive to the time when it leaves the system through an
open boundary. However, regardless of how resting-times are discounted,
the density $\Activity(\xvec, t; \xvec_0,L)$ is never that of pure
diffusion, as there are fluctuations and correlations in the number of
active particles at different times. 

\Fref{response_sliced} shows time slices of the activity, which is,
according to \Fref{response_t} almost a slowly broadening Gaussian.
Below we discuss briefly in what sense a plain diffusion process is
recovered, but from \Fref{response_t} it is clear that the spatial
structure is not exactly but very close to a Gaussian, as demonstrated by the
slight mismatch of the data (full line) and an approximated Gaussian
with the same height  and roughly the same width (dashed line). One may argue
that the slight
deviation is due to
lattice effects or due to the parity conservation in the activity, as
the parity of the coordinate $x$ of sites active at a given time $t$ is
identical to that of $x_0+t$.  The slight difference is certainly not
due to the avalanche having reached the boundaries, as times are chosen
short enough.

The activity also shows a mild dependence on the system size, as shown
in \Fref{response_L}, but seems to converge.  One may think that this
is due to the probability of activity being triggered at all, which
is the occupation probability at the driving site, $\Ds(x_0;x_0,L)$,
because the (average) activity is reduced on the whole across \emph{all}
sites if the site driven is not occupied (and thus fails to topple).
However, this is not the case, as the data in \Fref{response_L} has been
rescaled accordingly. It is also not due to the avalanche having reached
the boundary, as times are chosen short enough again. As the time in
these figures is chosen to confine activity to the bulk, it seems most
likely that the reduction of activity is caused by \emph{all} sites in a
smaller system having a smaller occupation probability (\Fref{substrate_density_1D_bulk}), thus hindering
spreading of activity somewhat. At the same time, however, stationarity
is maintained, \ie the hindrance in the activity spreading does not
result in an accumulation of particles. That the time integral over all
activity is nevertheless identical to that of a random walk regardless of the system size, does not mean that the
activity in smaller systems, which is reduced at earlier times, must exceed that of bigger systems at later
times or last longer, because the path density of random walkers
increases with system size, as discussed below.

We will discuss the \emph{temporal} features of the response function in
further detail below. They show a very clear departure from a diffusion
process, rendering the present behaviour superdiffusive.

The random walker nature of individual particles can be captured
by summing over all times
\begin{equation}
\TimeIntegratedActivity(\xvec;\xvec_0,L) = \sum_{t=1}^\infty \Activity(\xvec, t; \xvec_0,L)
\elabel{TimeIntegratedActivity}
\end{equation}
which is the average number of topplings caused at site $\xvec$ by
driving at site $\xvec_0$. In directed sandpiles, this quantity can be
used to solve the system exactly, as higher order responses can be written in terms of products of such two-point functions
\cite{DharRamaswamy:1989,DharETAL:2015}. In the stationary state, every particle added
eventually leaves the system, following a random walker trajectory until
it reaches the dissipative boundary. One may thus think of each particle added as producing a
random walker trajectory from the source $\xvec_0$ to a point at the
boundary (even when the activity trace is more akin to a branching
process, following individual branches of the trajectories and allowing
occasional rests, produces random walker
paths). 

To ease notation, we adopt mostly a continuum perspective in the following.
Because each toppling results in \emph{two} particles being
moved, the total number of topplings at site $\xvec$ (per avalanche) is
thus half the density of Brownian paths (per particle),
$\TimeIntegratedActivity(\xvec;\xvec_0,L)=\half
\phihat_0(\xvec;\xvec_0,L)$ where
$\phihat_0(\xvec;\xvec_0,L)=\int_0^\infty \dint{t}
\phi_0(\xvec,t;\xvec_0,L)$ and $\phi_0(\xvec,t;\xvec_0,L)$ is the 
solution of the diffusion equation,
$\partial_t \phi_0 = D \nabla^2 \phi_0$, with diffusion constant
$D=1/(2d)$ and initial condition 
$\lim_{t\to0}\phi_0(\xvec,t)=\delta(\xvec-\xvec_0)$. 
On a lattice $\nabla^2$ is the lattice Laplacian and
$\delta(\xvec-\xvec_0)$ is to be replaced by $\delta_{\xvec,\xvec_0}$.
Carrying on in the continuum, it follows that 
$\TimeIntegratedActivity(\xvec;\xvec_0,L)$ solves 
the Poisson equation
\begin{equation}\elabel{time_integrated_diffusion_eqn}
D \nabla^2 \TimeIntegratedActivity(\xvec;\xvec_0,L) =
-(1/2)\delta(\xvec-\xvec_0)
\end{equation}
with $\xvec,\xvec_0\in(0,L+1)$ and
Dirichlet boundary conditions identical to those on the lattice, 
$\TimeIntegratedActivity(0;\xvec_0,L)=\TimeIntegratedActivity(L+1;\xvec_0,L)=0$,
which in one dimension produces 
\begin{subnumcases}{\elabel{triangular_profile}\!\!\!\!\!\!\!\!\!\TimeIntegratedActivity
(x;x_0,L)=\!\!}
\!\frac{x(L+1-x_0)}{2D(L+1)} &\!\!\!\!\!\!\!\!for $0\le x<x_0$ \\
\!\frac{x_0(L+1-x)}{2D(L+1)} &\!\!\!\!\!\!\!\!for $x_0\le x\le L+1$,
\end{subnumcases}
a solution that holds identically on the lattice. 

We note in passing that at $t\to0$ half the density of walker paths is
$\half \delta_{\xvec,\xvec_0}$ and thus clearly smaller than the
response $\Activity(\xvec, 0;
\xvec_0,L)=\Ds(\xvec_0;\xvec_0,L)\delta_{\xvec,\xvec_0}$ because
$1/2<\Ds(\xvec_0;\xvec_0,L)$, which means in turn 
that the return probability of activity is (at some later times) less than half of that of a random walker,
as otherwise the time integral of activity $\Activity(\xvec, t;
\xvec_0,L)$ cannot be exactly equal to half the integral over
$\phi_0(\xvec,t;\xvec_0,L)$.

The Poisson equation \Eref{time_integrated_diffusion_eqn} 
and, on the lattice, the corresponding difference equation
\begin{equation}\elabel{Poisson_lattice}
D \nabla^2 \TimeIntegratedActivity(\xvec;\xvec_0,L) =
-(1/2)\delta_{\xvec,\xvec_0}
\end{equation} 
with a lattice Laplacian on the left and a Kronecker-$\delta$
on the right
equally apply in higher dimensions. It may be interpreted and derived as a
continuity equation of particles being transported from one site to
another, two at each toppling. In the continuum it is easily solved in
higher dimensions by
\begin{multline}\elabel{triangular_profile_35}
\TimeIntegratedActivity(\xvec;\xvec_0,L) = \half
\left(\frac{1}{L'}\right)^{d-1}
\frac{2}{L+1}
\sum_{n=1}^\infty\\
\sum_{\mvec=(-\infty,\ldots,-\infty)}^{(\infty,\ldots,\infty)}
\frac{\sin(q_n x) \sin(q_n x_0)
\exp{\imag \kvec_{\mvec} \cdot (\yvec-\yvec_0)} }{D(q_n^2 + \kvec_{\mvec}^2)}
\end{multline}
where 
$\yvec\in[0,L']^{d-1}$ 
are the $d-1$ components of $\xvec$ in the periodic
directions and 
$x\in(0,L')$ 
is its component in the open direction,
correspondingly for $\xvec_0$. The $d-1$ dimensional vector
$\kvec_{\mvec}$ has components $k_{m_i}=2 \pi m_i/L'$ with
$m_i\in\Zset$ for $i=2,3,\ldots,d$, 
whereas $q_n=\pi
n/(L+1)$ is a scalar with 
$0<n\in\Nset^+$. 
The solution of \Eref{Poisson_lattice} is correspondingly
\begin{multline}\elabel{lattice_triangular_profile_35}
\TimeIntegratedActivity(\xvec;\xvec_0,L) = \half
\left(\frac{1}{L'}\right)^{d-1}
\frac{2}{L+1}
\sum_{n=1}^L\\
\sum_{\mvec=(0,\ldots,0)}^{(L'-1,\ldots,L'-1)}
\frac{\sin(q_n x) \sin(q_n x_0)
\exp{\imag \kvec_{\mvec} \cdot (\yvec-\yvec_0)} }{2D\big(
(1-\cos(q_n))
+
\sum_{i=2}^d (1-\cos(k_{m_i}))
\big)
}
\end{multline}
with 
$\yvec\in\{1,2,\ldots,L'\}^{d-1}$,
$x\in\{1,2,\ldots,L\}$,
$k_{m_i}=2 \pi m_i/L'$ with
$m_i\in\{0,1,\ldots,L'-1\}$ for $i=2,\ldots,d-1$, 
and 
$q_n=\pi n/(L+1)$ with
$n\in\{1,2,\ldots,L\}$.  

The continuum solution \Eref{triangular_profile_35}
still carries the signature of the lattice: there are $L'$ sites in
the periodic direction with site $y=0$ being identical to site $y=L'$
but only $L$ sites in the open direction, with activity on both sites $x=0$ and
$x=L+1=L'$ vanishing. The only quantities with the dimension of a length
on the right hand side of \Eref{triangular_profile_35} are therefore
$L'^{-d}$ from the pre-factors and $L'^{2}$ from $\kvec_{\mvec}^2 +
q_n^2$ in the denominator.  Anticipating the (simplified) scaling form
\Eref{time_integral_to_argue_vanishing_eta} we therefore notice that
\Eref{triangular_profile_35} can be written as $a b^{-z}
|\xvec-\xvec_0|^{-(d-2)} \FCtilde\left(\frac{\xvec-\xvec_0}{L'}\right)$
with dimensionless $a$, $b$ and $\FCtilde$. In case of the
time-integrated activity,
$\TimeIntegratedActivity(\xvec;\xvec_0,L)$, which is a two-point
response function, one
can therefore identify $L'$ as the correlation length analytically.

Integrating \Eref{triangular_profile_35} over the periodic directions
(sheets of constant $x$), as used later in \Eref{def_ActivityTimeAveSpave},
gives $L'^{d-1}$ times the integrand at $\kvec_{\mvec}=0$,
\begin{multline}\elabel{triangular_profile_35_integrated}
\int_0^{L'} \dXint{d-1}{y} \TimeIntegratedActivity(\xvec;\xvec_0,L) = 
L'^{d-1} \ActivityTimeAveSpave(x;\xvec_0,L) \\
=
\half
\frac{2}{L+1}
\sum_{n=1}^\infty
\frac{\sin(q_n x) \sin(q_n x_0)}{Dq_n^2}
\end{multline}
and on the lattice
\begin{multline}\elabel{lattice_triangular_profile_35_integrated}
\sum_{y'_2,\ldots,y'_d=(1,\ldots,1)}^{(L',\ldots,L')} \TimeIntegratedActivity(\xvec;\xvec_0,L) = 
L'^{d-1} \ActivityTimeAveSpave(x;\xvec_0,L) \\
=
\half
\frac{2}{L+1}
\sum_{n=1}^\infty
\frac{\sin(q_n x) \sin(q_n x_0)}{2D(1-\cos(q_n))}
\end{multline}
from \Eref{lattice_triangular_profile_35},
which recovers exactly \Eref{triangular_profile} with $D=1/(2d)$
dependent on the dimension. This is not surprising as integrating over
sheets of constant $x$ corresponds to considering hopping of particles
only as far as their $x$-coordinate is concerned. Given the periodicity
of $\TimeIntegratedActivity(\xvec;\xvec_0,L)$, the integral $L'^{d-1}
\ActivityTimeAveSpave(x;\xvec_0,L)$ obeys $D\partial_x^2 L'^{d-1}
\ActivityTimeAveSpave(x;\xvec_0,L)=-\half\delta(x-x_0)$, the
differential equation in one dimension with the reduced diffusion
constant of $D=1/(2d)$, as hops in only one of $d$ directions results in a
change of sheets.

\Fref{TimeIntegratedActivity} confirms the triangular shape of the
time-integrated activity \Eref{triangular_profile}. As discussed above,
the origin of the profile is somewhat trivial, but it has two important
implications: Firstly, the activity is \emph{shaped} by the boundaries.
As opposed to the nearly featureless density profile of the inactive
particles, \Fref{substrate_density_1D}, the activity is very strongly
affected by the presence of the boundary, as all activity ceases there.
Every particle added is eventually transported to the boundary
\cite{nonotePaczuskiBassler:2000}. Secondly, because time integrals of
the response are \emph{exactly} random walker profiles, the
$\kvec$-dependence of a (suitable) propagator in a field theory will not
renormalise. A non-trivial dynamical exponent $z$, which is often
obtained through the renormalisation of the diffusion constant, will
have to be obtained through the renormalisation of the time-dependence.
At frequency $\omega=0$,
the full propagator in a field theory reads exactly $1/(D\kvec^2)$, 
whereas the frequency-dependence may deviate from
the tree-level $-\imag \omega$ in almost
arbitrary form, provided only that it vanishes at
$\omega=0$. Anticipating some of the discussion below, we note that the propagator
being $1/(D\kvec^2)$ implies $\eta=0$.

\begin{figure}
\includegraphics*[width=0.95\linewidth]{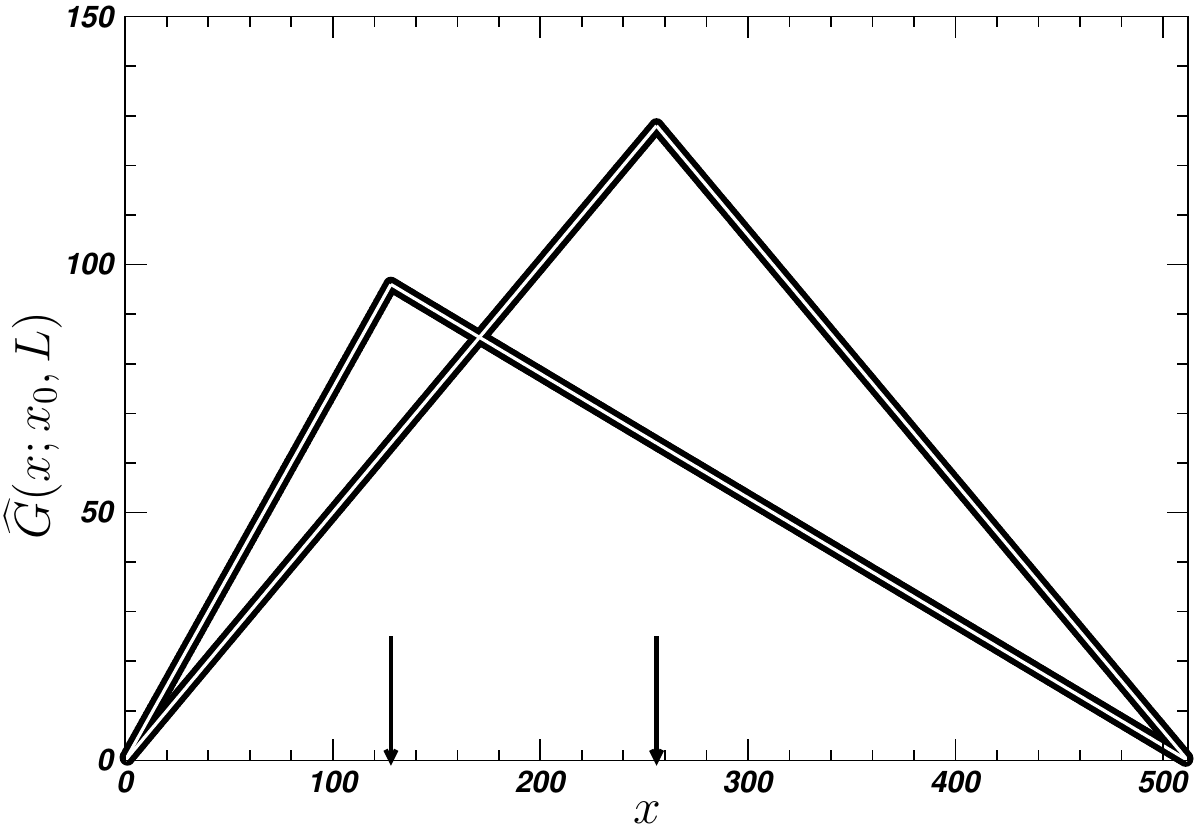}
\caption{\flabel{TimeIntegratedActivity}
The time integrated activity
$\TimeIntegratedActivity(\xvec;\xvec_0,L)$ according to
\Eref{TimeIntegratedActivity} is in fact half the number of times a
random walker starting at $x_0$ passes through a particular site $x$
before leaving at the boundary. The activity (the response
$\Activity(x, t; x_0,L)$ to driving at $x_0$) has a complicated
dependence on time, because of its frequent stops and restarts, but once
that is integrated out, the resulting densities are those of random
walkers.
The exact solution, \Eref{triangular_profile}, shown as a white
line, is virtually indistinguishable from the numerical results ($L=511$
driven at $x_0=256$ or $x_0=128$ as indicated by the arrows).
}
\end{figure}

\subsubsection{Spatially integrated activity}
\slabel{shape_of_ava_1D}
\begin{figure*}
\subfigure[Total activity as a function of time.]{\includegraphics*[width=0.45\linewidth]{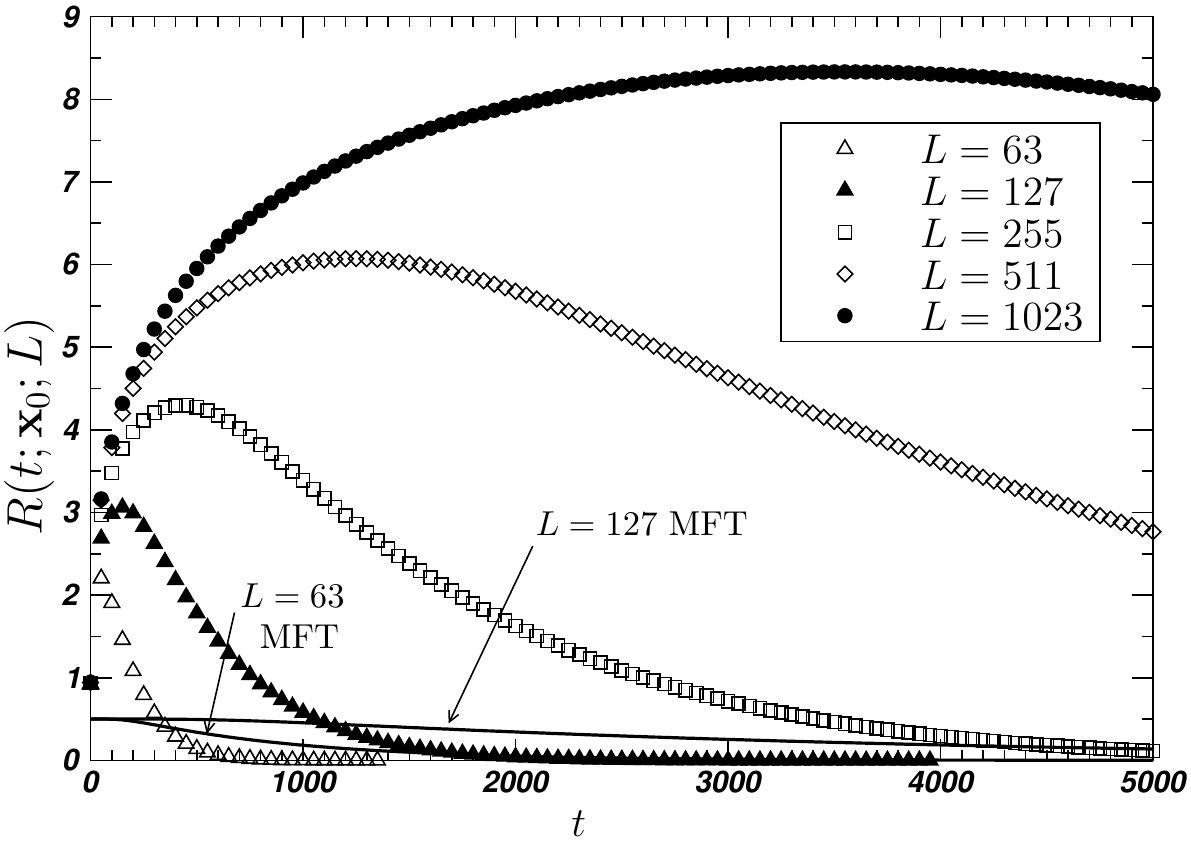}\flabel{ava_abs}}
\subfigure[Collapse of the total activity as a function of time.]{\includegraphics*[width=0.45\linewidth]{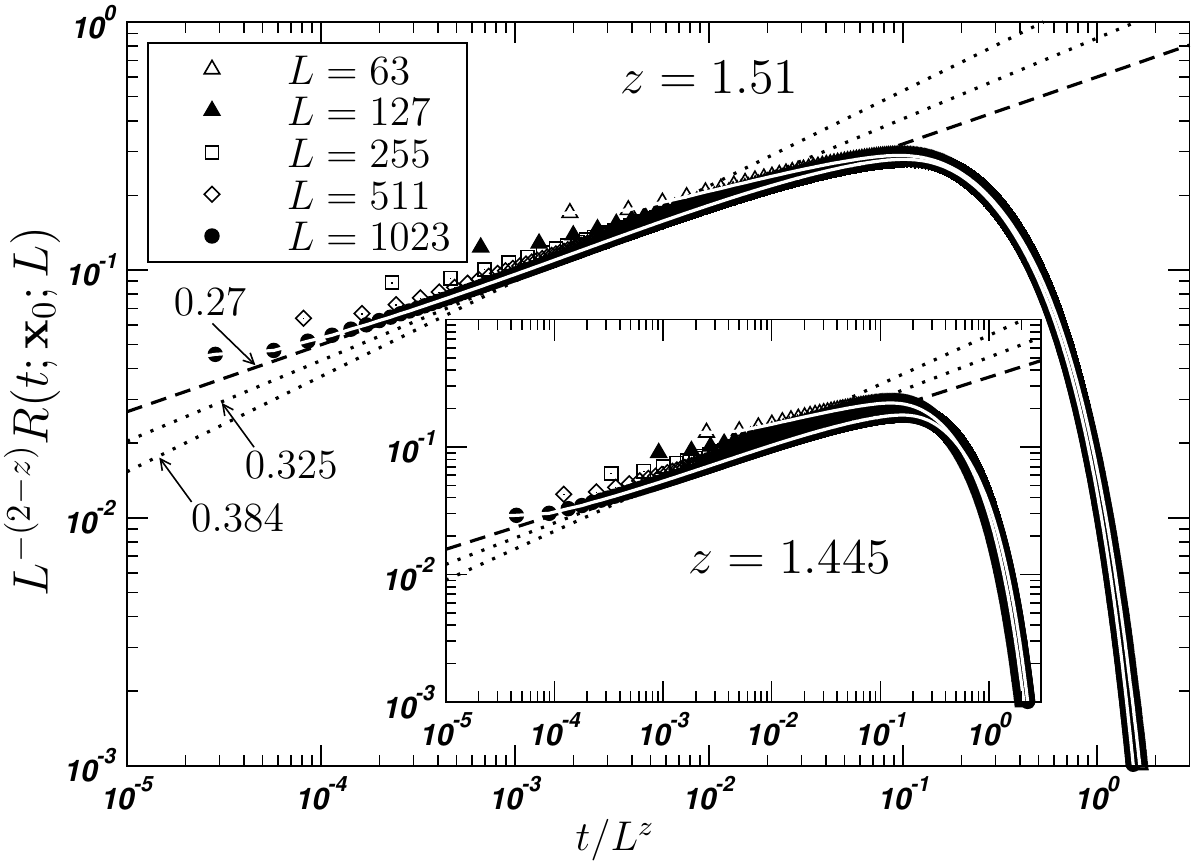}\flabel{ava_rel}}
\caption{\flabel{ShapeAvaAbs}
The total (spatially integrated) activity 
$\ShapeAvaAbs(t;\xvec_0,L)$ as a function of time $t$ for various system
sizes $L$ in one dimension, driven at $x_0=(L+1)/2$.  \subfref{ava_abs}
Activity as a function of absolute (microscopic) time (symbols, pruned)
and comparison to the (badly matching) mean field theory
\Eref{shape_ava_abs_MFT} (full lines). Error bars (not shown) are
much smaller than symbols.  \subfref{ava_rel} Collapse of
$\ShapeAvaAbs(t;\xvec_0,L)$ for a range of systems sizes $L$ (as
indicated) according to \Eref{ShapeAvaAbs_collapse}.
The dashed line in the main panel shows a powerlaw with exponent
$0.27$, the apparent behaviour of $\ShapeAvaAbs(t;\xvec_0,L)$ in small
$t$. The dotted lines show the expected behaviour $(2-z)/z=0.384(10)$
with $z=1.445(10)$ from literature \cite{HuynhPruessnerChew:2011} and
$(2-z)/z=0.325$ from $z=1.51$ used in the
collapse.  The white lines on top of the symbols show the data for
$L=63$ and $L=1023$ to allow for an easier assessment of the quality of
the collapse.  The inset shows the same plot for the
literature value $z=1.445$.
}
\end{figure*}

Instead of integrating the response function over time, to reduce the
number of independent variables, one may just as well
integrate in space. The
simplest version of this quantity is the spatial integral of the
activity
\begin{equation}\elabel{def_ShapeAvaAbs}
\ShapeAvaAbs(t;\xvec_0,L) = \sum_{\xvec} \Activity(\xvec, t; \xvec_0,L) \ ,
\end{equation}
\ie the total (spatially integrated) activity at time $t$, which in a
numerical implementation corresponds to the height of the stack of
active sites  with each entry being weighted by $\lfloor z_{\xvec}(t)/2 \rfloor$,
the activity at that site. 
The 
spatial activity integral is
closely related to the order parameter of many absorbing state phase
transitions \cite{Luebeck:2004}, the spatially averaged activity density
$\rho_a=\ShapeAvaAbs(t;\xvec_0,L)/L^d$. In fact
$\ShapeAvaAbs(t;\xvec_0,L)$ is the area under the activity ``slices''
shown in \Fref{response_sliced}. Numerical results, shown in
\Fref{ava_abs}, clearly differ across system sizes 
on the large time scale. Even on the very short time scale
($t=1,\ldots,10$, not shown separately)
$\ShapeAvaAbs(t;\xvec_0,L)$ appears to differ systematically
for all system sizes considered (although displaying some convergence). If this
is solely due to the slightly decreased occupation density by inactive
particles, $\Ds(\xvec;\xvec_0,L)$, then the effect is cumulative and
highly non-linear, as $\ShapeAvaAbs(t=0;\xvec_0,L)$ barely varies
between different system sizes. For large $t$ the activity eventually
reaches the boundary of the system. As is clear from the collapse in \Fref{ava_rel} and
further discussed below, the position of the maximum total activity,
$\ShapeAvaAbs(t_{\max};\xvec_0,L)$, scales with the dynamical exponent,
$t_{\max}\propto L^z$.

Employing again a continuum approximation (where we identify $L=L'$ to
ease notation), in an MFT, the spatially integrated
activity $\ShapeAvaAbs(t;\xvec_0,L)$ is given by
the propagation of the activity profile of a single
active particle, which undergoes a Poissonian branching or extinction
with equal rates, subject to Dirichlet boundary conditions. In one dimension the resulting profile is
the spatial integral of the density $\half\frac{2}{L+1}\sum_{n=1}^L
\sin(x_0 q_n) \sin(x q_n) \exp{-D q_n^2 t}$, see \Erefs{triangular_profile_35_integrated}
and \eref{lattice_triangular_profile_35_integrated}, with momenta
$q_n=n\pi/(L+1)$
\citep{Pruessner_aves:2013}, as discussed above, which gives\footnote{The
integral $\int_0^{L+1} \dint{x} \sin(q_n x)=2/q_n$ for odd $n$ needs to be
replaced
by $\sum_{x=1}^{L} \sin(q_n x) = \sin(q_n)/(1-\cos(q_n))$ on the lattice.}
\begin{equation}\elabel{shape_ava_abs_MFT}
\ShapeAvaAbs_{\MFT}(t;\xvec_0,L) = 
\frac{2}{\pi} \sum_{n=1,\text{odd}}^L \frac{\sin(x_0 q_n)}{n} \exp{-D q_n^2 t} \ .
\end{equation}
\Fref{ava_abs} shows this profile as well. The mean field theory differs
very clearly from the one-dimensional Manna Model in a number of points: 
Firstly, the tail of the mean field activity drags out for very long times, even for moderately
large systems, not least as to make up for
\Eref{ave_s_as_time_integral} below, the ``sum rule'' relating the mean
avalanche size and the time integral over the (total) activity. In
comparison, avalanches in the Manna Model are ``short and sharp''.
Secondly, by
construction, the activity in the mean field theory never exceeds $1/2$,
whereas the maximum activity in the Manna Model seems to increase with
system size, clearly exceeding unity even for the smallest system sizes
studied.
This is particularly clear at $t=0$ where the simple mean field assumes
an activity of $1/2$, whereas in
the Manna Model activity is triggered with the occupation probability at
the driven site.

The scaling of $\ShapeAvaAbs(t;\xvec_0,L)$ can be determined by making
the
usual (Ornstein-Zernike-like) scaling ansatz of the
response \cite{HansenMcDonald:2006,Taeuber:2014},
\begin{multline}
\Activity(\xvec, t; \xvec_0,L) = a |\xvec-\xvec_0|^{-(d-2+\eta+z)} \\
\times \FC\left(\frac{\xvec-\xvec_0}{L},\frac{\xvec-\xvec_0}{b t^{1/z}}\right)
\ ,
\elabel{def_eta_z}
\end{multline}
where we assume, for simplicity, translational invariance, even when our
systems are not translationally invariant in the $x$-direction.
\Eref{def_eta_z} may be regarded of the definition of the anomalous
dimension $\eta$ and the dynamical exponent $z$. The dimensionless
scaling function $\FC(u,v)$ turns off correlations beyond the system
size and confines them to a region of linear extent proportional to 
$t^{1/z}$ at
the short time scale. Dimensional consistency is restored by metric
factors $a$ and $b$.  Taking the limit $L\to\infty$ in \Eref{def_eta_z},
the spatial integral of $\Activity$ gives
$\ShapeAvaAbs(t;\xvec_0,L)\propto t^{(2-\eta-z)/z}$.
We expect
this scaling behaviour to hold for $t\ll L^z$; in fact, re-writing
\Eref{def_eta_z} as 
\begin{multline}\elabel{def_eta_z_again}
\Activity(\xvec, t; \xvec_0,L) = a |\xvec-\xvec_0|^{-(d-2+\eta+z)}\\
\times \FCtilde\left(\frac{t}{b' L^z},\frac{\xvec-\xvec_0}{b t^{1/z}}\right)
\end{multline}
and integrating over $\xvec$ at finite $L$ gives
\begin{equation}\elabel{ShapeAvaAbs_scaling}
\ShapeAvaAbs(t;\xvec_0,L) = a \left(\frac{t}{b}\right)^{(2-\eta-z)/z}
\FCtilde_0\left( \frac{t}{b' L^z} \right)
\end{equation}
even for finite $L$, or alternatively 
\begin{equation}\elabel{ShapeAvaAbs_collapse}
\ShapeAvaAbs(t;\xvec_0,L) = \atilde L^{2-\eta-z}
\FCtilde_0'\left(\frac{t}{\btilde L^{z}}\right)
\end{equation}
with suitable metric factors and scaling function, as used in
\Fref{ava_rel}.

It turns out that $\eta$ in fact vanishes,  as suggested
in the discussion after \Eref{triangular_profile_35_integrated}. Firstly, this is implied by
the sum rule arising from the temporal integral over
$\ShapeAvaAbs(t;\xvec_0,L)$, which is the average
avalanche size, 
\begin{equation}
\elabel{ave_s_as_time_integral}
\ave{s}(\xvec_0,L) 
= \sum_{t=0}^{\infty} \ShapeAvaAbs(t;\xvec_0,L)
\simeq \int_0^\infty \dint{t} \ShapeAvaAbs(t;\xvec_0,L)
\end{equation}
written as an integral for convenience. 
In the present case of centre driving,
the average avalanche size
scales in $L$ like  $\ave{s}\propto
L^2$ and $\eta=0$ follows from using the scaling form
\Eref{ShapeAvaAbs_collapse} in the integrand of
\Eref{ave_s_as_time_integral}.

A more subtle demonstration that $\eta=0$ follows from the time integral
of the activity, which reduces the activity to random walker
trajectories.
Taking the time integral of \Eref{def_eta_z}
gives 
\begin{equation}\elabel{time_integral_to_argue_vanishing_eta}
\TimeIntegratedActivity(\xvec;\xvec_0,L)=
a b^{-z} |\xvec-\xvec_0|^{-(d-2+\eta)} \FChat\left(\frac{\xvec-\xvec_0}{L}\right)
\end{equation}
with a new, suitably defined scaling function $\FChat$. This
observable, $\TimeIntegratedActivity$, is the average number of
topplings at site $\xvec$ per particle added at site $\xvec_0$, as
discussed at the beginning of \Sref{active_state_1D}, see
\Eref{TimeIntegratedActivity}. An Ornstein-Zernike correlation function
\cite{HansenMcDonald:2006}, as the one generated by the density of
Brownian paths has $\eta=0$ and must coincide with
\Eref{time_integral_to_argue_vanishing_eta}, so $\eta=0$ follows.
Alternatively, one may consult $\TimeIntegratedActivity$ in \Eref{triangular_profile} and
\Eref{triangular_profile_35}, which indeed behave like $x^{2-d}
\FChat(x/L)$ for fixed $x_0$.

Taking $\eta=0$ henceforth,
\Fref{ava_rel} shows a good collapse on the basis of
\Eref{ShapeAvaAbs_collapse} with 
$z=1.51$, in poor agreement with the literature value of $z=1.445(10)$
based on the scaling
of avalanche durations
\cite{HuynhPruessnerChew:2011}.
The collapse based on $z=1.445$ is shown in
the inset of \Fref{ava_rel}. White lines for the data of $L=63$ and
$L=1023$ have been added on top of the symbols to assess the quality of
the collapse, which, away from the tail, is clearly worse for
$z=1.445$ than for $z=1.51$.

Apart from the collapse in $L$, according to 
\Eref{ShapeAvaAbs_scaling}, there is also scaling in $t$ at large
enough but early times. Clearly, for fixed $t$ the total activity
$\ShapeAvaAbs(t;\xvec_0,L)$ converges in large $L$, as for sufficiently
large $L$ the activity no longer reaches the boundaries, which therefore
become irrelevant. Apart from small (but possibly cumulative) effects
due to the density of inactive particles,
for fixed $t$ a regime exists where $\ShapeAvaAbs(t;\xvec_0,L)\propto
t^{(2-z)/z}$ independent of (large) $L$, so that $\FCtilde_0$ in \Eref{ShapeAvaAbs_scaling},
which
carries all $L$-dependence, is
constant, or equivalently, that $\FCtilde'_0$ in
\Eref{ShapeAvaAbs_collapse} behaves like a power law itself,
$\FCtilde_0'(u)\propto u^{(2-z)/z}$. 
Therefore
in \Eref{ShapeAvaAbs_scaling}
$t^{(2-z)/z}$  shapes $\ShapeAvaAbs(t;\xvec_0,L)$
on the short time scale,  
and the
scaling function $\FCtilde_0$ on the long time scale.

The initial power law regime is clearly visible in 
\Fref{ava_rel}, which
suggests $(2-z)/z\approx 0.27$ and thus $z\approx 1.574\ldots$, again
quite off the expected value of $(2-z)/z=0.384(10)$ from
$z=1.445(10)$ \citep{HuynhPruessnerChew:2011}, also shown in the plot.
A third slope shown in \Fref{ava_rel}, $(2-z)/z\approx0.325$ is
determined by $z$ giving the best collapse, $z\approx1.51$.  However,
measuring exponents by fitting a section of the data against a straight
line in a double logarithmic plot ignores the r{\^o}le of the scaling
function and is generally prone to errors
\citep{ClausetShaliziNewman:2009,DelucaCorral:2014,ChristensenETAL:2008}.
The significance of the slopes shown in \Fref{ava_rel} is therefore that
both slopes of $0.27$ and $0.325$ (corresponding to $z=1.574$ and
$z=1.51$ respectively) seem to be consistent with the data,
whereas the literature value of $z=1.445(10)$ fails, producing a slope
of $0.384$, which is clearly off. 

\subsubsection{Temporal shape of the avalanche}
\slabel{temp_shape_of_ava}
The Manna Model differs from the MFT above in that particles and thus
activity display some complicated ``resting'', but otherwise
trajectories are random walks.  Time integrals over the activity
therefore remove any non-trivial behaviour, whereas space integrals
retain it. It is thus worthwhile to look for ways of extracting
universal features from $\ShapeAvaAbs(t;\xvec_0,L)$ or similar
quantities.

\begin{figure}[t]
\includegraphics*[width=0.95\linewidth]{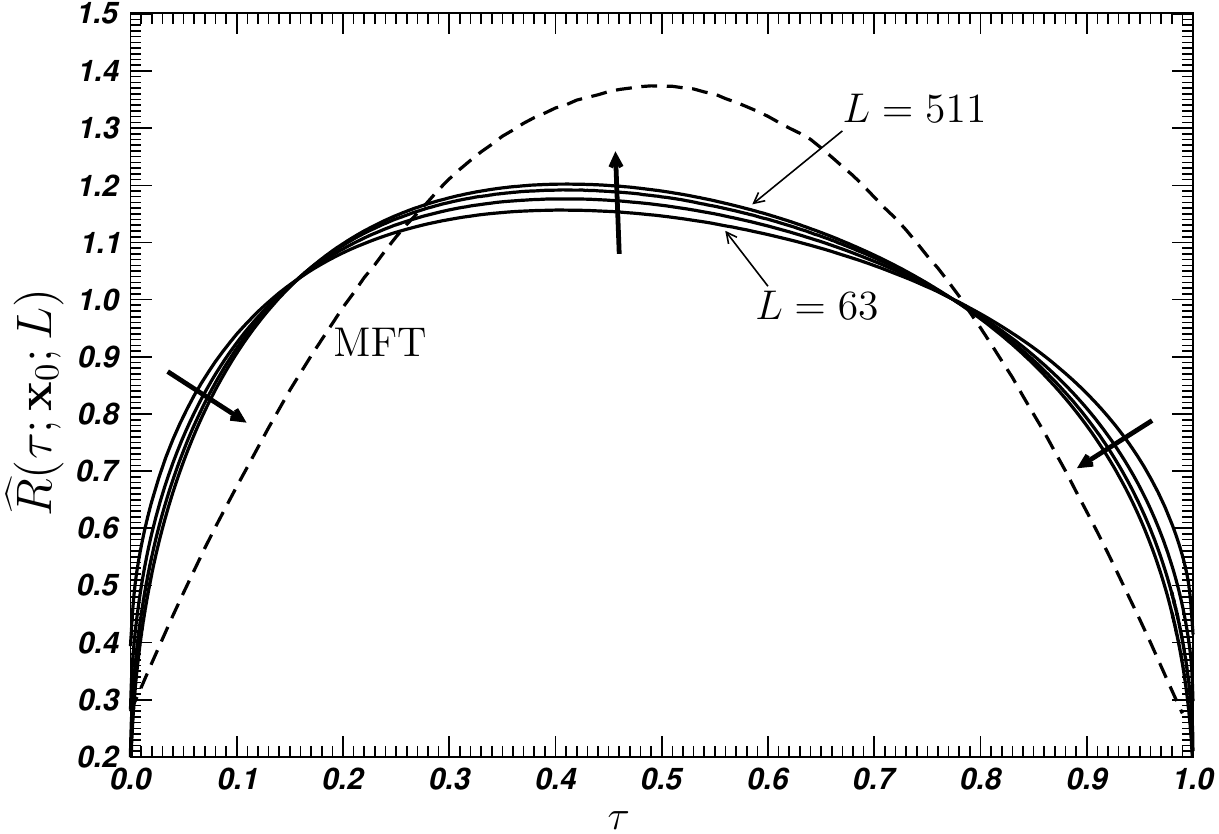}
\caption{\flabel{ava_rescaled} The rescaled activity $\ShapeAvaRel(\tau;
\xvec_0,L)$
(``temporal shape of the avalanche''), \Eref{def_ShapeAvaRel}, in one-dimensional systems of size
$L=63,127,255,511$ driven at the centre, $x_0=(L+1)/2$.  The rescaling
(see main text) maps the times of each (spatially integrated) activity
time series to the interval $\tau\in[0,1]$ and the data is normalised so
that the integral under the curve is unity.  The numerical data are
shown as full lines, as the error bars are exceedingly small. The thick
arrows point in the direction of increasing system size.  The data seems
to suggest that there is slow convergence to a shape one might suspect
to be universal.  It is noticeably lopsided compared to the mean-field
theory (MFT) shown as a dashed line.  The mean-field theory is based on
a branching random walk on a (finite) lattice ($L=255$). Numerical data
for larger systems are computationally prohibitively expensive, see
\Sref{mft}.
}
\end{figure}

\begin{figure*}
\subfigure[Width $\msd(t;\xvec_0,L)$ of the propagator as a function of time $t$, for different system sizes $L$.]{\includegraphics*[width=0.45\linewidth]{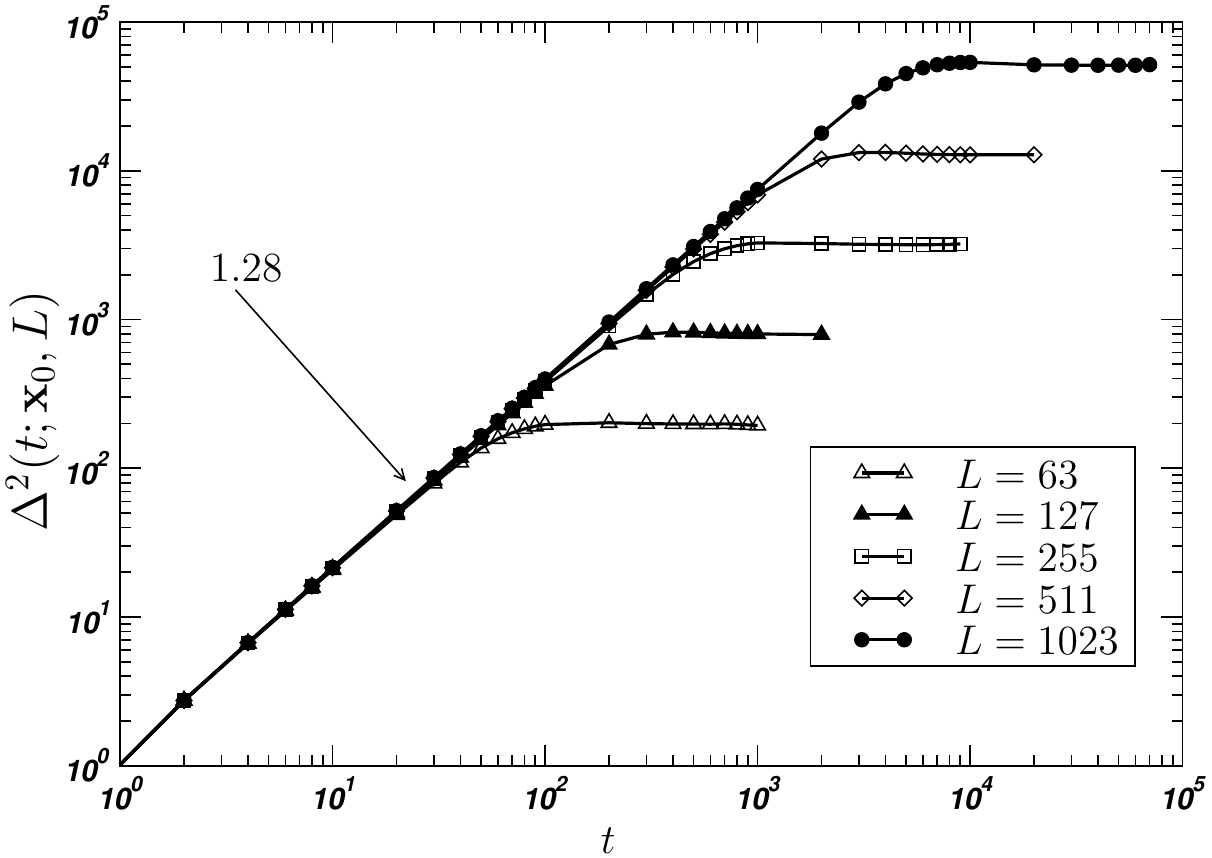}\flabel{msd}}
\subfigure[Collapse of $\msd(t;\xvec_0,L)$ for different system sizes $L$.]{\includegraphics*[width=0.45\linewidth]{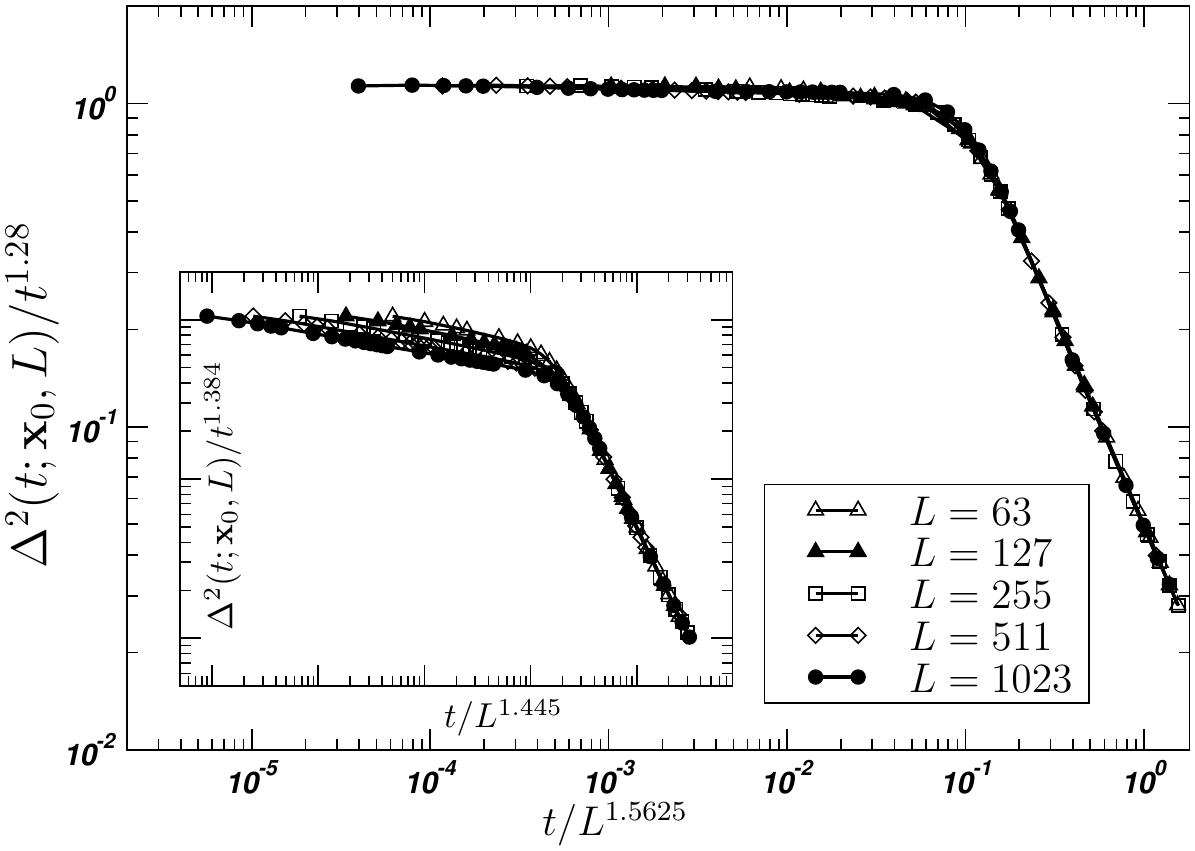}\flabel{msd_collapse}}
\caption{\flabel{msd_plots} Width $\msd(t;\xvec_0,L)$ of the response
propagator $\Activity(x, t; x_0,L)$, as defined in \Eref{def_msd}, in
one dimension for centre drive, $x_0=(L+1)/2$, and for different system sizes
$L$.  \subfref{msd} The apparent scaling in time $t$ 
suggests $2/z\approx1.28$ and thus 
a
dynamical exponent 
$z=1.5625$ very different
from 
$z=1.445(10)$ determined
from the moments of the avalanche duration
\cite{HuynhPruessnerChew:2011}. The
scaling is confirmed by the collapse shown in \subfref{msd_collapse}.
The inset of that figure shows the same collapse with the latter literature
value
of $z=1.445(10)$.
}
\end{figure*}

Because $\ShapeAvaAbs(t;\xvec_0,L)$ is bound to scale in order to
accommodate the average avalanche size, no convergence of
$\ShapeAvaAbs(t;\xvec_0,L)$ can be expected in large $L$ at large
$t \propto L^z$. To observe
convergence for all $t$, the profile has to be rescaled by the  duration of the avalanche and the 
mean activity.  Theory has access (at least at MFT level) to an approximation of the spatial
integral of the activity conditional to a certain time of termination,
$T$, the duration of the avalanche, but it is difficult to
condition in addition to a certain avalanche size. Numerically, on the
other hand, this would be trivial: If
$\ShapeAvaAbs_i(t;\xvec_0,L)=\sum_{\xvec} n_i(\xvec,t;\xvec_0,L)$ is an individual measurement of the
space-integrated activity at time $t$ of an avalanche that has size $s_i$
and duration $T_i$, then $\ShapeAvaAbs_i(t;\xvec_0,L)/(s_i/T_i)$ would
be the relevant quantity to consider.
Instead we note that
\begin{equation}\elabel{instantaneous_ava_size}
s_i=\sum_{t=0}^{T_i-1} \ShapeAvaAbs_i(t;\xvec_0,L)
\end{equation}
is the avalanche size if $T_i$ is the duration and averaging over the rescaled measurements
$\ShapeAvaAbs_i(\tau T_i;\xvec_0,L)$,
\[
\overline{f}(\tau) = \frac{1}{M} \sum_{i=1}^M
\ShapeAvaAbs_i(\left\lfloor \tau T_i \right\rfloor;\xvec_0,L)
\]
gives $\int_0^1\dint{\tau} \overline{f}(\tau) = \frac{1}{M} \sum_{i=1}^M s_i/T_i$, so that normalising by
(the estimate of) $\ave{s/T}$ produces
\begin{equation}\elabel{def_ShapeAvaRel}
\ShapeAvaRel(\tau; \xvec_0,L) =
\frac{1}{\ave{s/T}} \frac{1}{M} \sum_{i=1}^M \ShapeAvaAbs_i(\left\lfloor
\tau T_i \right\rfloor;\xvec_0,L) \ ,
\end{equation}
a quantity that has a unit-integral and may therefore be expected to
converge.
Closely related to this is a quantity sometimes referred to as the ``temporal shape of the
avalanche'' \citep{LeDoussalWiese:2013,ThieryLeDoussalWiese:2015}, first studied
by Kuntz and Sethna \cite{KuntzSethna:2000} (see also \cite{BaldassarriColaioriCastellano:2003}),
and closely connected to the ($1/f$) power spectrum \cite{LaursonAlavaZapperi:2005}.
To make $\ave{s/T}$ in \Eref{def_ShapeAvaRel} well-defined in case of $T_i=0$
(and thus $s_i=0$), one may consider the above derivation with $T_i$
replaced by $T_i+\epsilon>T_i$ and take the limit $\epsilon\to0$.

\Fref{ava_rescaled} shows
$\ShapeAvaRel(\tau; \xvec_0,L)$ for different system sizes $L$,
demonstrating the expected convergence. Notably, the graph displays a
slight, unexpected asymmetry that is absent in the mean-field theory of
a branching process. 

Because the time averaged total activity $\ave{s/T}$ diverges (slowly)
for  $L\to\infty$ and $\ShapeAvaAbs_i(t;\xvec_0,L)$ at $t=0$ is exactly
the finite occupation density at $\xvec_0$ and expected to be small at
$t=T_i$, in the thermodynamic limit,
$\ShapeAvaRel(\tau; \xvec_0,L)$ is expected to vanish at $\tau=0$ and
$\tau=1$ (and one may speculate that the MFT gives asymptotically
$\ShapeAvaRel(\tau)=6\tau(1-\tau)$).
In a suitable theory, one may redefine time and 
observables such that $\ShapeAvaAbs_i(t;\xvec_0,L)$ is exactly unity at
these points.

\begin{figure*}
\subfigure[Collapse of the activity at fixed relative distance from the
driving site using the literature value $z=1.445$.]{\includegraphics*[width=0.45\linewidth]{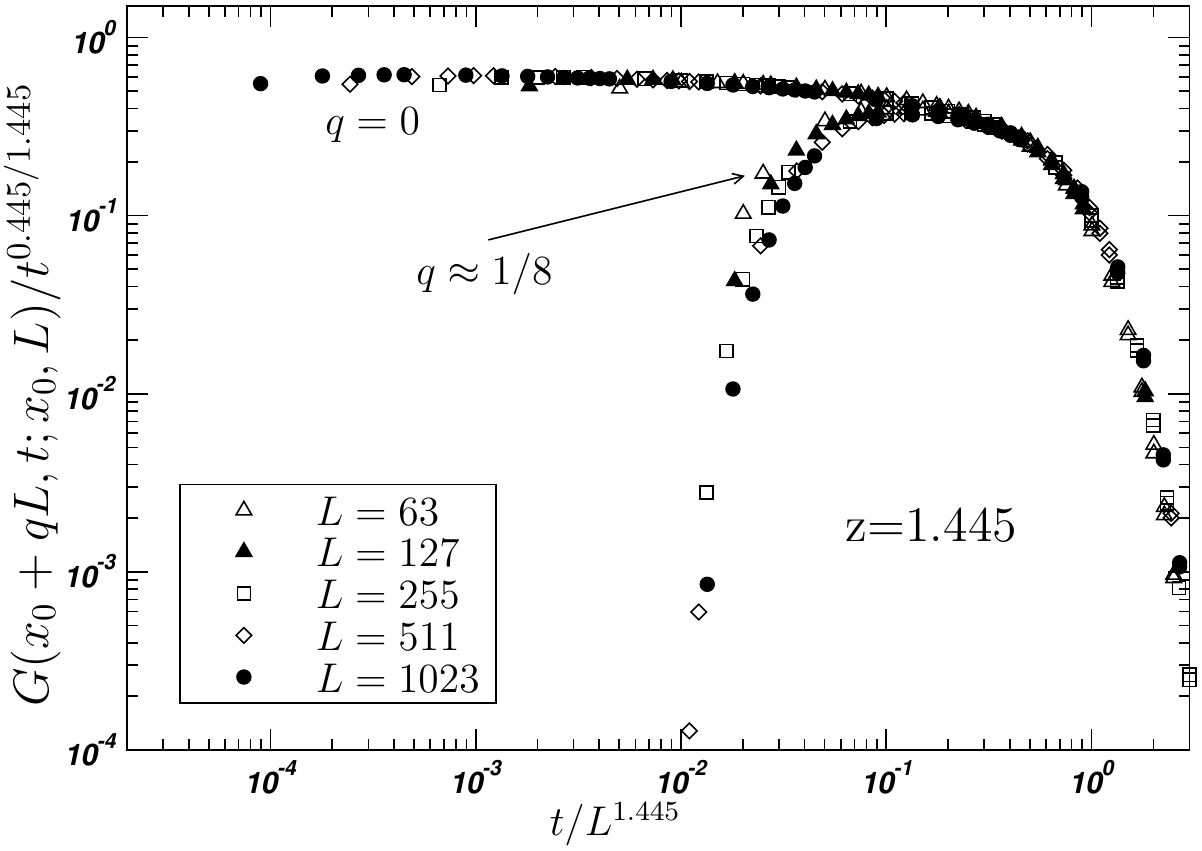}\flabel{slice}}
\subfigure[Collapse of the activity at fixed relative distance from the
driving site using $z=1.59$, see also \Fref{full_collapse}.]{\includegraphics*[width=0.45\linewidth]{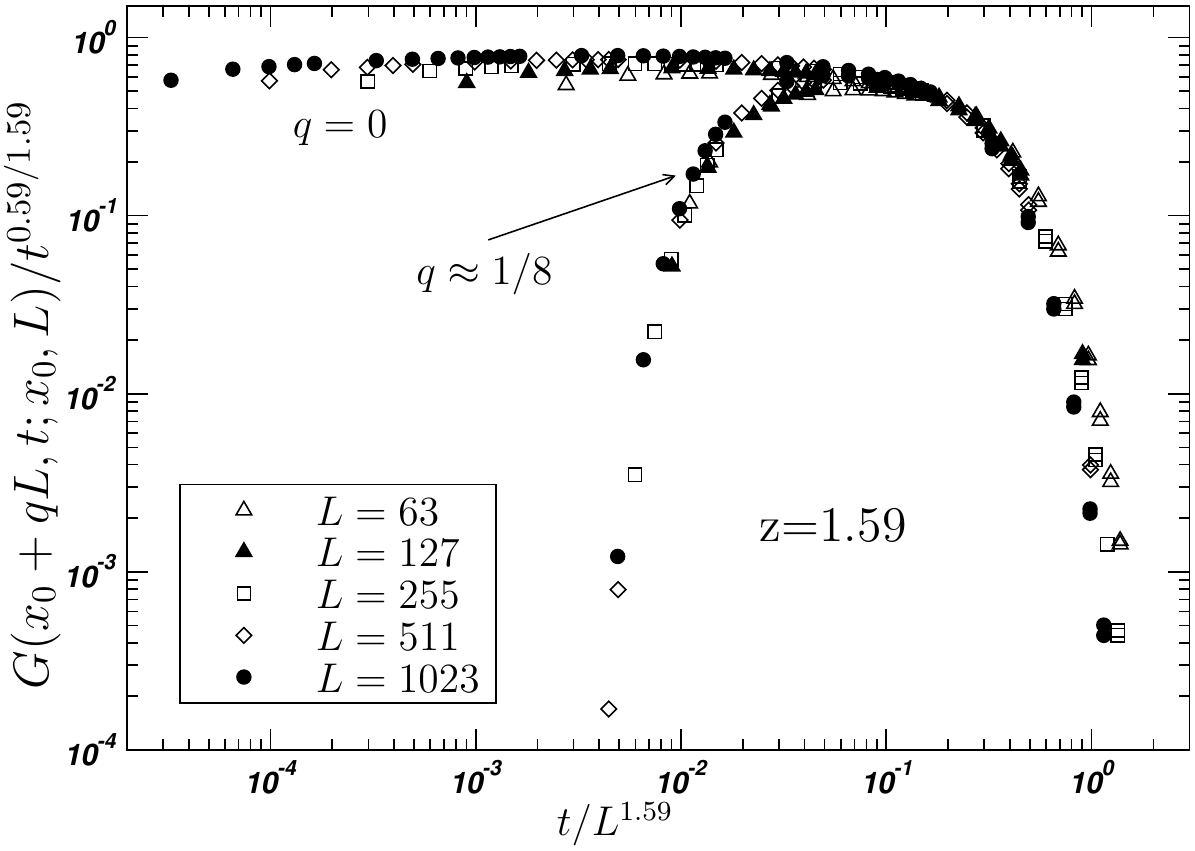}\flabel{slice_1.59}}
\caption{\flabel{slices} The activity $\Activity(x_0+qL, t; x_0,L)$ in
one dimension for fixed $q=0$ and $q\approx 1/8$ collapsed by plotting
$\Activity(x_0+qL, t; x_0,L) t^{(z-1)/z}$ against $t/L^z$, using two
different exponents, \subfref{slice} $z=1.445$ (literature
\cite{HuynhPruessnerChew:2011}) and \subfref{slice_1.59} $z=1.59$
(larger than any estimate above, but see \Fref{full_collapse}).}
\end{figure*}

\subsubsection{Width of the response}
\slabel{msd_1D}
In the following we want to characterise further the deviation of the
activity density from plain diffusion.  In \Fref{response_sliced} we
have demonstrated that the spatial distribution of activity
$\Activity(x, t; x_0,L)$ at $(x,t)$ in response to driving at $x_0$ at
$t=0$ is very close to Gaussian. We may proceed by determining the
kurtosis \etc, but as mentioned above, it is difficult to attribute any
deviation correctly, because there are several sources for corrections:
finiteness of the lattice, boundaries, discretisation in space,
discretisation in time and separation into even and odd sublattices.
Only the time integral of the activity can be mapped exactly to a random
walk, which is \emph{approximated} by a Gaussian in the continuum. 
It turns out that the \emph{temporal evolution} of the spatial distribution
of activity is strongly superdiffusive.
To see this more clearly, we may
calculate the spatial variance of the \emph{normalised}
$\Activity(\xvec, t; \xvec_0,L)$, using the spatial integral $\ShapeAvaAbs(t;\xvec_0,L)$,
\Eref{def_ShapeAvaAbs}. The width of the response $\Activity(x, t;
x_0,L)$ may be defined as
\begin{equation}
\msd(t;\xvec_0,L) = \ShapeAvaAbs^{-1}(t;\xvec_0,L) \sum_{\xvec}
(\xvec-\xvec_0)^2 \Activity(x, t; x_0,L) \ .
\elabel{def_msd}
\end{equation}
In the present definition, $\msd(t;\xvec_0,L)$ looks very much like a
mean squared displacement, except that many particles contribute to
$\Activity(x, t; x_0,L)$ simultaneously and only very few have actually
been displaced starting from $\xvec_0$, the origin. In fact, many may
have been moved \emph{towards} $\xvec_0$ and most may have moved only a
couple of sites.  

\Fref{msd} shows a very clear power law dependence of
$\msd(t;\xvec_0,L)$ on (small) time $t$, 
scaling faster than linear in $t$ for intermediate
times (below a cutoff set by the system size), thus rendering the process
superdiffusive. 
To relate this 
to the results
above, we integrate
$|\xvec-\xvec_0|^2\Activity$ using 
\Eref{def_eta_z_again} over $\xvec$, which gives
\begin{equation}\elabel{msd_scaling}
\msd(t;\xvec_0,L) = \left(\frac{t}{b}\right)^{2/z} 
\frac{\FCtilde_2\left(\frac{t}{b' L^z}\right)}
{\FCtilde_0\left(\frac{t}{b' L^z}\right)}
\ ,
\end{equation}
using \Eref{ShapeAvaAbs_scaling}, regardless of
$\eta=0$.

If $\msd(t;\xvec_0,L)\propto t^{2/z}$ (expected for $t\ll L^z$ but not
guaranteed as the scaling function may contribute, \Eref{msd_scaling}),
\Fref{msd} suggests $2/z\approx 1.28$ and thus $z\approx1.5625$ rather
than $z=1.445(10)$ of \cite{HuynhPruessnerChew:2011}. This is
confirmed by a collapse, \Fref{msd_collapse}.  
The value of $z\approx1.5625$ from the $t$-dependence of
$\msd(t;\xvec_0,L)$ is reasonably consistent with the
value of $z\approx1.574$ from the
$t$-dependence of $\ShapeAvaAbs(t;\xvec_0,L) \propto t^{(2-z)/z}$ in 
\Fref{ava_rel} ($(2-1.574)/1.574\approx0.27$). The literature value of 
$z=1.445(10)$ produces a rather dissatisfying collapse shown in the inset
of \Fref{msd_collapse}, that improves in the tail (large $t$) only because
$\msd$ converges to $L^2$ and so $\msd/t^{2/z'}$ plotted versus $(L^2/t^{2/z'})^
{-z'/2}=t/L^{z'}$ produces a straight line in a double logarithmic plot 
with slope $-z'/2$
for any $z'$.
We will discuss the range of results for $z$
further in \Sref{conclusion}.

\subsubsection{Further spatial scaling of the response}
A \emph{quantitative} test for scaling is to extract ``moments'' by
integrating out all but one independent variable. This procedure leads
to the measurements normally taken in SOC
\cite{DeMenechStellaTebaldi:1998}, such as moments of the avalanche size
(for example, the temporal integral of $\ShapeAvaAbs(t;\xvec_0,L)$ gives
$\ave{s}$). The usual caveats apply, in particular finite size
corrections, which we have largely ignored in the present analysis. A
\emph{qualitative} test of scaling is to attempt a collapse of the data,
such as the one for $\ShapeAvaAbs(t;\xvec_0,L)$ in \Fref{ava_rel}.
However, for $\Activity(\xvec, t; \xvec_0,L)$ this is difficult to attain
in the form \eref{def_eta_z} as there are at least \emph{three} independent
parameters, $|\xvec-\xvec_0|$ (assuming translational invariance), $t$ and
$L$, listed in order of
increasing sparseness. In principle such a collapse can be done in
three-dimensional plots, but the small range of $L$ compared
to the high density of points in $|\xvec-\xvec_0|$ and the still fairly
large range and number of measurements in $t$, makes it difficult to
assess the quality of such a collapse, which becomes nearly useless when projected into two
dimensions.

To investigate further the spatial dependence of the activity $\Activity(x, t;
x_0,L)$, we consider fixed
$(x-x_0)/L=q$, so that according to \Eref{def_eta_z} with $\eta=0$,
\begin{equation}\elabel{activity_fixed_q}
\Activity(x_0+qL, t; x_0,L) = a |qL|^{-(d-2+z)} \FC\left(|q|,\frac{t}{b |qL|^z}\right) \ ,
\end{equation}
which for fixed $q$ ought to collapse when plotting $\Activity(x_0+qL,
t; x_0,L) L^{d-2+z}$ against $t/L^z$ (for the earliest to the latest
times $t$).  \Fref{slices} shows a collapse
for $x=x_0$ ($q=0$) as well as $x=(L+1)/8$ ($q\approx1/8$) with centre driving,
$x_0=(L+1)/2$.
While the literature value of $z=1.445$ works fairly well, \Fref{slice}, the
collapse is
relatively insensitive against different choices of the dynamical
exponent ($z=1.59$ is shown in \Fref{slice_1.59}, exceeding even
$z=1.574$ above, see \Fref{ava_rel}, but identical to the $z$ used in
\Fref{full_collapse}).  

Assuming instead $L^z\gg t$ produces a
collapse of $\Activity$ on the basis of \Eref{def_eta_z} when plotting
$\Activity(\xvec, t; \xvec_0,L) |\xvec-\xvec_0|^{-(d-2+z)}$ against
$|\xvec-\xvec_0|^z/t$ for different $t$ and $L$, \Fref{full_collapse}.
The main panel shows a fairly neat collapse in the tail, while the 
inset shows a somewhat broader tail which, however, seems to cover a 
wider range of data, incorporating even the data marked by ``off''.
Picking, however, the data as to exclude those that do not produce a
collapse, presumably on the basis that $L^z\gg t$ is violated, is a form
of biased selection.

\begin{figure}
\includegraphics[width=0.95\linewidth]{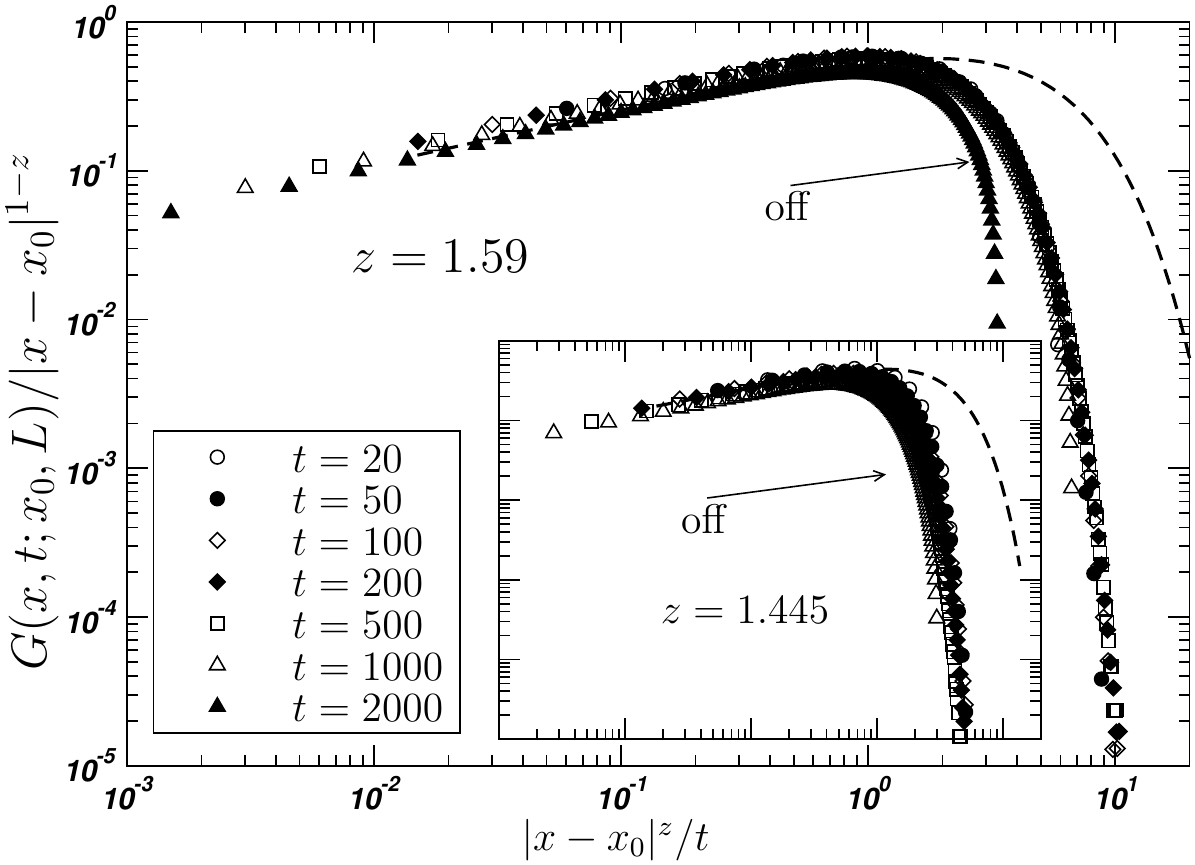}
\caption{\flabel{full_collapse}
Attempt of a collapse of the activity (response) $\Activity(\xvec, t;
\xvec_0,L)$ with $L=511$, according to \Eref{def_eta_z} using $z=1.59$ (see
\Fref{slice_1.59}) in the main panel and $z=1.445$ in the inset.
The data were selected so that \emph{apparently} $L^z\gg t$, except for
$t=2000$ marked by ``off''. The dashed lines shows the Gaussian
of \Fref{response_L}.
}
\end{figure}

In summary, the scaling of the response $\Activity(\xvec, t; \xvec_0,L)$
is exactly as expected in a critical finite system: on the short time
scale the characteristic length is set by the time $t^{1/z}$,
\Erefs{def_eta_z} and \eref{ShapeAvaAbs_scaling} (demonstrated in
\Fref{ava_rel}), and on the long time scale by the system size $L$ in the form $L^z$,
as discussed in \Sref{shape_of_ava_1D} (\Eref{ShapeAvaAbs_scaling} and discussion towards the end).
Indeed, with the
time-dependence integrated out, activity has all characteristics of a
random walk (\Fref{TimeIntegratedActivity}), so that all the non-trivial features are to be found in the
time-dependence. While collapses such as \Fref{ava_rel} and
\Fref{msd_plots} confirm the presence of scaling, the exponent $z$ we found
in one dimension varied:
In \Fref{ava_rel}
$z\approx1.51$ from the collapse but
$z\approx1.574$ from the scaling in $t$ of $\ShapeAvaAbs(t;\xvec_0,L)$,
in \Fref{msd}
$z\approx1.5625$ from the collapse and the scaling in $t$ of the width $\msd(t;\xvec_0,L)$
and in \Fref{slices}
no clear outcome ($z=1.445$ but also $z=1.59$) for the scaling of
the activity $\Activity(x, t; x_0,L)$. On the other
hand, the dynamical exponent determined in the literature from the
scaling of the cutoff of the avalanche duration is $z=1.445(10)$
\cite{HuynhPruessnerChew:2011}. We will discuss this discrepancy further
in \Sref{conclusion}.

\subsubsection{Activity-activity correlations}
\slabel{actact_corrs_1D}
There are very little spatial correlations in the density of the inactive
particles (\Fref{substrate_correlations_1D}). This is very different for
the activity.
The correlation function to be
considered next is, strictly speaking, a three-point function, as it
measures the correlations of activity at different sites $\xvec_1$ and
$\xvec_2$ in a system driven at $\xvec_0$. Although we have also
considered data where all three sites are distinct, generally statistics
is better for $\xvec_0=\xvec_1$ (or $\xvec_0=\xvec_2$, which amounts to
the same), so we have focused on that case.

\begin{figure*}
\subfigure[Spatial activity-activity correlation function
$\ActAct_u(x_2,x_0,t;x_0,L)$ collapsing for different times $t$,
according to \eref{actact_initial} and \eref{actact_saturation}.]{\includegraphics*[width=0.45\linewidth]{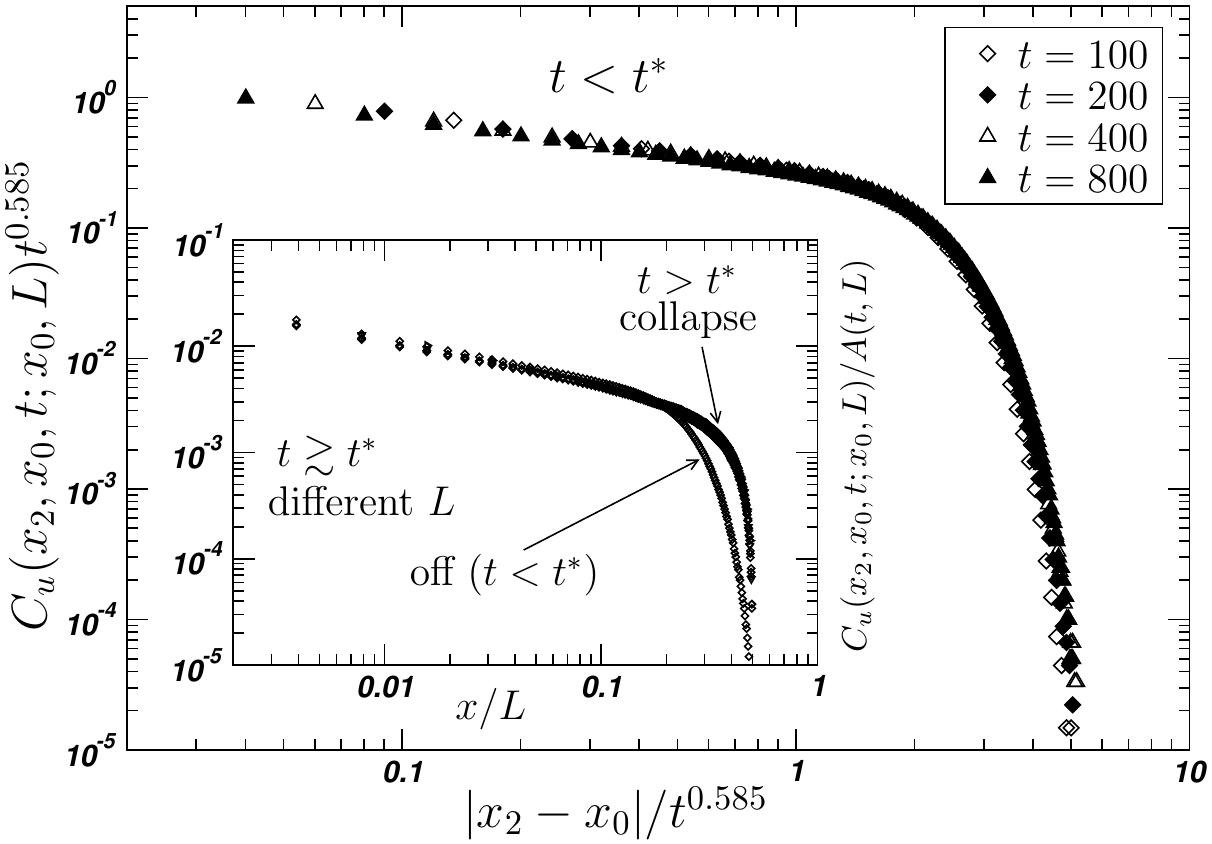}\flabel{actact_time_sliced}}
\subfigure[Time averaged unconnected spatial activity-activity
correlation function $\ActActTimeAve_u(x_2,x_0;x_0,L)$ collapsing for
different system sizes $L$ according to \eref{actact_scaling} with
$\mu=0.58$.]{\includegraphics*[width=0.45\linewidth]{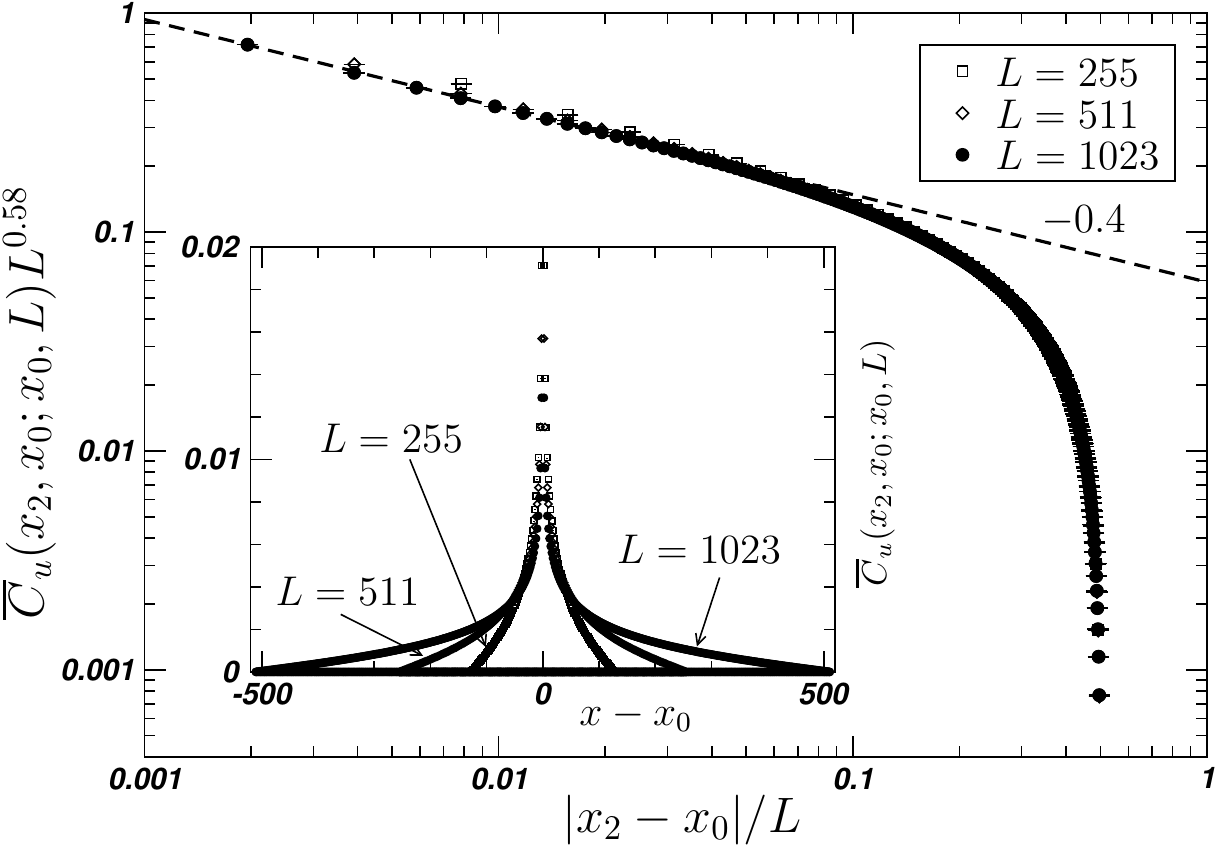}\flabel{actact_collapse}}
\caption{\flabel{actact}Spatial correlations of the activity.
\subfref{actact_time_sliced} Activity-activity correlation function
$\ActAct_u(x_2,x_0,t;x_0,L)$, \Eref{def_ActAct_u}, at different (equal) times and for
various system sizes $L$ and with centre drive, $x_0=(L+1)/2$. The
main panel shows a collapse for $L=511$ by rescaling both axes by a
suitable power of the time $t$, here $\lambda=0.585$. The collapse
\Eref{actact_initial}, however,
works only up until some saturation time $t^*$ from when on
$\ActAct_u(x_2,x_0,t;x_0,L)$ displays essentially the same shape with a
time-dependent pre-factor, \Eref{actact_saturation}. This is shown in the inset, where data for
different times and system sizes ($t=3000,5000,7000$ for $L=511$ as well
as $t=2000$ for $L=255$ and $t=500$ for $L=127$) is made to collapse by
plotting $\ActAct_u(x_2,x_0,t;x_0,L)$ divided by a suitable amplitude $A(t,L)$
against $x/L$. That amplitude drops roughly exponentially in time (shown
are $A(t,511)=0.59^{(t-1000)/2000}$, but also $A(2000,255)=0.513$ and $A(500,127)=1.35$);
it is, for fixed system
size $L$, to
large part given by the fraction of avalanches that last until $t$.  The
set marked ``off'' shows data for $L=511$ and $t=1000$, before the
saturation.  \subfref{actact_collapse} Collapse \Eref{actact_scaling} of the time-averaged
spatial correlation function
$\ActActTimeAve_u(\xvec_2,\xvec_0;\xvec_0,L)$,
\Eref{def_ActActTimeAve_u}, on a double logarithmic scale using $\mu=0.58$.
That the distance has to be
rescaled by the system size, is a reminder of the latter coinciding with
the correlation length. The straight, dashed line shows a power law with
exponent $-0.4$.  Inset: Correlation function on a linear scale. Because
of the parallel update,
$n_i(\xvec_2,t;\xvec_0,L)n_i(\xvec_1,t;\xvec_0,L)$ vanishes for all $t$ if
$\xvec_1$
and $\xvec_2$ are on sublattices of different parity resulting in
vanishing
$\ActActTimeAve_u(\xvec_2,\xvec_0;\xvec_0,L)$, thus not
shown in the logarithmic main panel.
}
\end{figure*}

This correlation function is also a function of two times relative to the
time of driving. We decided to consider only equal time correlations,
with the aim to extract interesting
\emph{spatial} behaviour. The estimators for the unconnected
activity-activity correlation function
\begin{equation}\elabel{def_ActAct_u}
\ActAct_u(\xvec_2,\xvec_1,t;\xvec_0,L) = \frac{1}{M} \sum_{i=1}^M
n_i(\xvec_2,t;\xvec_0,L)n_i(\xvec_1,t;\xvec_0,L) 
\end{equation} 
and the connected correlation function 
\begin{multline}\elabel{def_ActAct_c}
\ActAct_c(\xvec_2,\xvec_1,t;\xvec_0,L) = \frac{1}{M} \sum_{i=1}^M
n_i(\xvec_2,t;\xvec_0,L)n_i(\xvec_1,t;\xvec_0,L)\\ - \frac{1}{M}
\sum_{i=1}^M n_i(\xvec_2,t;\xvec_0,L) \frac{1}{M} \sum_{i=1}^M
n_i(\xvec_1,t;\xvec_0,L) 
\end{multline} 
are based on the same
measurements of local activity as \Eref{def_activity_estimator}.
Choosing $\xvec_1=\xvec_0$ still leaves us with four independent
variables. To capture the temporal evolution, we chose to take samples
only at $t=1,2,3\ldots,9,10,20,\ldots,90,100,200,\ldots9000$  for a
range of different system sizes, all driven at the centre. 
However, as mentioned in \Sref{observables}, activity vanishes on sites whose distance to the
driven site $\xvec_0$ has a parity different from that of $t$, and in
particular $n_i(\xvec_0,t;\xvec_0,L)=0$ strictly for all $\xvec_0$ at odd $t$,
so that both 
$\ActAct_u(\xvec_2,\xvec_0,t;\xvec_0,L)$
and
$\ActAct_c(\xvec_2,\xvec_0,t;\xvec_0,L)$
vanish at odd $t$.
From
the data shown in \Fref{actact_time_sliced}, it is clear that the
unconnected correlation function
$\ActAct_u(\xvec_2,\xvec_1,t;\xvec_0,L)$ collapses in one dimension for
different, early $t\ll t^*(L)$ according to
\begin{equation}\elabel{actact_initial} 
\ActAct_u(x_2,x_0,t;\xvec_0,L)=
A_0 t^{-\lambda} \CC\left( \frac{t}{b |x_2-x_0|^{1/\lambda}} \right)
\end{equation} 
with some amplitude $A_0$, exponent $\lambda$ (here
$\lambda=0.585$ similar to the exponent $\mu$ in \Eref{actact_scaling}
below, whereas $1/z=0.692(5)$ \citep{HuynhPruessnerChew:2011}) and scaling
function $\CC$, but ``saturates'' for $t\gg t^*(L)$ like (inset of
\Fref{actact_time_sliced})
\begin{equation}\elabel{actact_saturation}
\ActAct_u(x_2,x_0,t;\xvec_0,L) = A(t,L) \CC'(|x_2-x_0|/L) 
\end{equation}
where $A(t,L)$ is no longer a power-law in $t$ (as effectively in
\eref{actact_initial}), but is instead dominated by an exponential in
$t$. For fixed $L$, that amplitude is essentially the fraction of ``survivors'', \ie
the probability of an avalanche lasting at least until $t$.  Data for $t
\gtrsim t^*(L)$ is shown in the inset of \Fref{actact_time_sliced},
together with data of the correlation function not quite at saturation
(labelled ``off'' as opposed to ``collapse'').  The evolution of the
activity-activity correlation function is therefore compatible with the
classic narrative of correlations spreading throughout the system as the
avalanche unfolds
\cite{Grinstein:1995,DickmanVespignaniZapperi:1998,Lise:2002}, until the
effective correlation length (the cutoff length $t^\lambda$ in
\Eref{actact_initial}) reaches the boundaries, suggesting $\lambda=1/z$
(see below).  For essentially all the time thereafter and thus most of
the time, the correlation function has the form
\eref{actact_saturation}, although numerical data for very long times
becomes very noisy.

We are unable to offer an explanation for the scaling form
\Eref{actact_initial}, because of the many different variables and
scales involved. It depends on at least two different points in space
and on a time that needs to be small enough so that
\Eref{actact_initial} applies, determining $t^*(L)$ implicitly. The most
striking feature of the scaling form is that the exponent in the
pre-factor $t^{-\lambda}$ is identical to the exponent in the argument of
the scaling function, which we expect to be $1/\lambda=z$. However,
$1/\lambda\approx1.71$ (based on $\lambda=0.585$ in
\Fref{actact_time_sliced}) is greater than any $z$ measured above.
Below we will derive the scaling of the time-averaged correlation
function $\ActActTimeAve_u$, but it is difficult to relate it to the
scaling of \eref{actact_initial}, as the latter requires $t\ll t^*(L)$.

Repeating the analysis for the connected correlation function
$\ActAct_c$, \Eref{def_ActAct_c}, shows rather poor collapses.  It
remains somewhat unclear why that happens. The second term in
\eref{def_ActAct_c} has a fairly small contribution, yet big enough to
spoil most collapses. A collapse like \Fref{actact_time_sliced} for
$\ActAct_c$ is rather noisy and dissatisfying, producing no reasonable
estimate of $\lambda$ according to \Eref{actact_initial}.

\begin{figure*}
\subfigure[Connected 
correlation function $K_c(t,t+\tau;\xvec_0,L)$]{\includegraphics[width=0.45\linewidth]{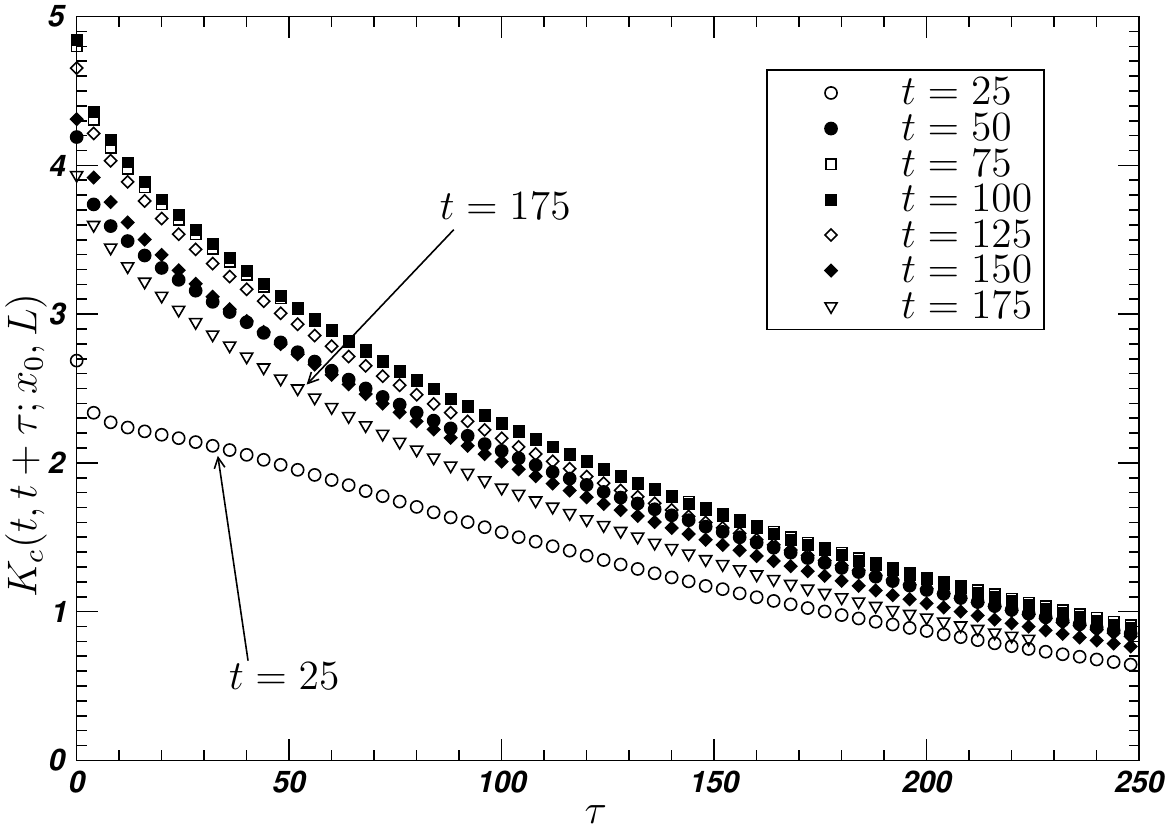}\flabel{K_c_nearly_same}}
\subfigure[Unconnected 
correlation function $K_u(t,t+\tau;\xvec_0,L)$]{\includegraphics[width=0.45\linewidth]{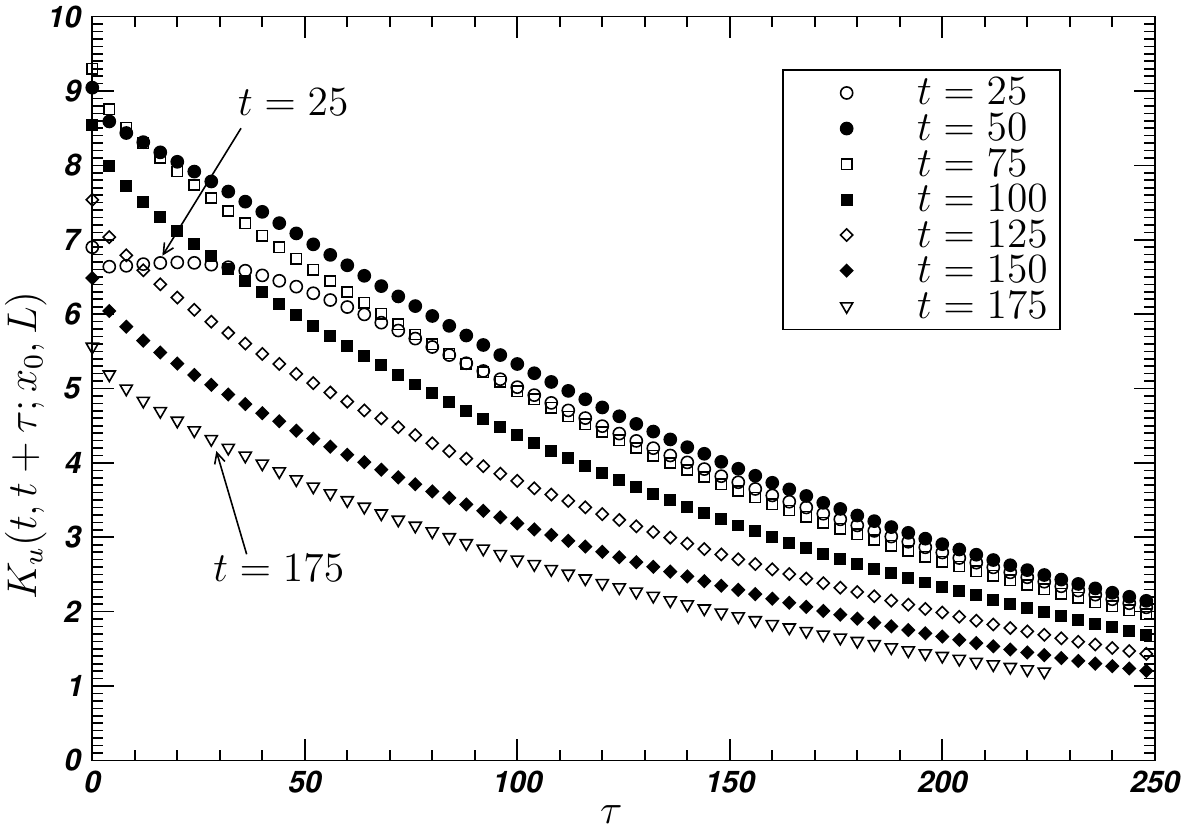}\flabel{K_u_nearly_same}}
\caption{\flabel{K_nearly_same}The two-time correlation
functions $K_{c,u}(t,t+\tau;\xvec_0,L)$, \Erefs{def_K_u} and \eref{def_K_c},
for a range of times
$t=25,50,\ldots,175$ against the time lag $\tau$ (pruned) for centre
driving of a one-dimensional system of size $L=63$. \subfref{K_c_nearly_same}
For
$t\ge50$ and
$t\le150$ the connected correlation functions are very similar. About $52\%$
of
avalanches have a duration of $T\ge50$.
\subfref{K_u_nearly_same} For $t\ge50$ the shape of correlation functions
differ
very little, while their amplitude clearly does.}
\end{figure*}

If the activity correlations can be regarded as ``almost stationary''
(or quasi-stationary as in fixed energy sandpiles
\cite{DickmanVespignaniZapperi:1998,VespignaniETAL:1998,Luebeck:2004}),
it is justified to take their time average. If $T_i$ is the duration of
the $i$th avalanche, then
\begin{equation}\elabel{def_ActivityTimeAve}
\ActivityTimeAve(\xvec;\xvec_0,L) = \frac{1}{\sum_{i=1}^M T_i}
\sum_{i=1}^M\sum_{t=1}^{T_i} n_i(\xvec,t;\xvec_0,L)
\end{equation}
estimates the time-averaged activity at site $\xvec$ \emph{conditional to
activity (somewhere)} \cite{Pruessner:2007b}, as
otherwise time stops passing. Because $\ave{\sum_{i=1}^M T_i}=M\ave{T}$, the
time-average $\ActivityTimeAve(\xvec;\xvec_0,L)\ave{T}$ is the average number of
topplings of site $\xvec$ per avalanche, \ie 
\begin{equation}
\ActivityTimeAve(\xvec;\xvec_0,L) \ave{T} = \TimeIntegratedActivity(\xvec;\xvec_0,L) 
\end{equation}
see \Erefs{def_activity_estimator} and \eref{TimeIntegratedActivity}.
Correspondingly, the unconnected time-averaged correlation function may be written as
\begin{multline}\elabel{def_ActActTimeAve_u}
\ActActTimeAve_u(\xvec_2,\xvec_1;\xvec_0,L) \\
= 
\frac{1}{\sum_{i=1}^M T_i} \sum_{i=1}^M\sum_{t=1}^{T_i} n_i(\xvec_2,t;\xvec_0,L)n_i(\xvec_1,t;\xvec_0,L)
\end{multline}
and the connected one as
\begin{multline}\elabel{def_ActActTimeAve_c}
\ActActTimeAve_c(\xvec_2,\xvec_1;\xvec_0,L) \\
= 
\frac{1}{\sum_{i=1}^M T_i} \sum_{i=1}^M\sum_{t=1}^{T_i} n_i(\xvec_2,t;\xvec_0,L)n_i(\xvec_1,t;\xvec_0,L)\\
- \ActivityTimeAve(\xvec_2;\xvec_0,L) \ActivityTimeAve(\xvec_1;\xvec_0,L)  \ .
\end{multline}
The unconnected time-averaged correlation function for $x_1=x_0$ (in one dimension)
is
shown in \Fref{actact_collapse} and displays a remarkably good collapse 
\begin{equation}\elabel{actact_scaling}
\ActActTimeAve_u(x_2,x_0;x_0,L) = L^{-\mu}
\GC\left(\frac{|x_2-x_0|}{L}\right)
\end{equation}
for different system sizes with $\mu\approx0.58$ (for an earlier
estimate
of $\mu=0.658$ based on uniform driving see
\cite{McAteerETAL:2016}). The denominator in the
argument of the scaling function $\GC(|x_2-x_0|/L)$ should be regarded
as the correlation length, which is in fact proportional to the system
size, exactly as expected for a finite system at the critical point
\cite{BakTangWiesenfeld:1987,BakTangWiesenfeld:1988,Luebeck:2004,Pruessner:2012:Book,WatkinsETAL:2016}.
The exponent $\mu$ that scales the amplitude in \Eref{actact_scaling}
can be related via a sum-rule to known scaling exponents, such as the
scaling of the activity variance $\Delta \rho_a/L^d$
\citep{Luebeck:2004,McAteerETAL:2016} (see Eq.~(2.36) in
\cite{Luebeck:2004}, $\Delta \rho_a L^{-d}$ is the variance of the
density $\rho_a$ and $\Delta \rho_a$ is the spatial integral of the
variance), 
\begin{equation}
\elabel{def_gamma_dash}
\frac{\Delta \rho_a}{L^d} = 
\frac{1}{L^{2d}} \int\ddint{x_1}\ddint{x_2}
\ActActTimeAve_u(\xvec_2,\xvec_1;\xvec_0,L) 
\propto L^{\gamma'/\nu_\perp-d}
\end{equation}
where we have used the notation of \cite{Luebeck:2004}, except that we
use $d$ for the 
spatial dimension (denoted by $D$ in \cite{Luebeck:2004}). If the scaling
of $\ActActTimeAve_u(\xvec_2,\xvec_1;\xvec_0,L)$ is essentially translationally
invariant,
as suggested in \Eref{actact_scaling},
then $\Delta \rho_a L^{-d} \propto L^{-\mu}$ from \Eref{def_gamma_dash} and thus
$\mu=d-\gamma'/\nu_\perp=2\beta/\nu_\perp$. The estimate
$\mu\approx0.58$ above (\Fref{actact_collapse}) is perfectly in line with
literature values of $1-\gamma'/\nu_\perp=0.59(4)$ in one dimension
\cite{Luebeck:2004}.

Alternatively, one may relate $\ActActTimeAve_u$ to the second moment of
the avalanche size.  Given that the instantaneous avalanche size is
\Eref{instantaneous_ava_size}, the estimator for the second moment 
$\ave{s^2}=M^{-1}\sum_{i=1}^M s_i^2$
is
\emph{exactly} equivalent to
\begin{equation}\elabel{exact_s2}
\ave{s^2} = 
\frac{1}{M} \sum_{i=1}^M\sum_{t_1,t_2=1}^{T_i} 
\sum_{\xvec_1,\xvec_2}
n_i(\xvec_2,t_2;\xvec_0,L)n_i(\xvec_1,t_1;\xvec_0,L)
\end{equation}
resulting in a corresponding sum rule on the unconnected correlation
function. It requires, however, the
\emph{two (three) point}, \emph{two time} correlation function. Assuming that it follows
essentially \Eref{actact_scaling} with $x_1/x_2$ and $t_1/t_2$ as
additional arguments in the scaling function, gives
\begin{equation}\elabel{alternative_scaling_relation}
\ave{s^2} \propto \ave{T} L^{2d+z} L^{-\mu}
\end{equation}
where $\ave{T}$ undoes the normalisation in \Eref{def_ActActTimeAve_u}
compared to \Eref{exact_s2}, $L^{2d}$ is due to the double space
integral (or sum, \Eref{exact_s2}), $L^z$ is due to the additional time
integral in \Eref{exact_s2} compared to \Eref{def_ActActTimeAve_u} and,
finally, $L^{-\mu}$ is due to \Eref{actact_scaling}. Given that
$\ave{s^2}\propto L^{D(3-\tau)}$, $\ave{T}\propto L^{z(2-\alpha)}$ and
$D(1-\tau)=z(1-\alpha)$ (all exponents explained in detail in
\cite{Pruessner:2012:Book}), one arrives at $\mu=2(d+z-D)$ which is in
line with $\mu=d-\gamma'/\nu_\perp=2\beta/\nu_\perp$ because what is
denoted $D$ here is $D_f+z$ in \cite{Luebeck:2004} and so $d+z-D$ here is in
fact $\beta/\nu_\perp$ there. However, $\mu=2(d+z-D)$ gives
$\mu=0.38(5)$ using the values of $z=1.445(10)$ and $D=2.253(14)$ of
\cite{HuynhPruessnerChew:2011}. The mismatch with $\mu=0.59(4)$
mentioned above cannot be explained by the dynamical exponent of
$z=1.393(37)$ measured in \cite{Luebeck:2004}, which gives an even
smaller value of $2(d+z-D)$. Rather, it is the poor match of
$D_f+z=d-\beta/\nu_\perp$ in \cite{Luebeck:2004}, which gives
$D=2.11(4)$ based on measurements in the fixed energy sandpile (FES)
version of the Manna Model, and $D=2.253(14)$ of
\cite{HuynhPruessnerChew:2011} taken in the SOC mode. 

The slope of $\GC(q)$, \Eref{actact_scaling}, that is so clearly visible
in \Fref{actact_collapse} (dashed line) may be captured by fitting against a power law
with a cutoff, so that $\GC(q)\propto q^{-0.4}$ for small $z$.  For
sufficiently small $x_2-x_0$ and sufficiently large $L$, the correlation
function $\ActActTimeAve_u(x_2,x_0;x_0,L)$ thus behaves like
$L^{-\mu+0.4} |x_2-x_0|^{-0.4}$. It is difficult to see how to relate
this behaviour to known exponents through scaling relations.

The connected time-averaged correlation function
$\ActActTimeAve_c(\xvec_2,\xvec_1;\xvec_0,L)$ collapses as well, but is
much less sensitive to a change of the exponent $\mu$ in
\Eref{actact_scaling}. While it is compatible to $\mu=0.58$, with some
small deviations, its tail ($|\xvec_2-\xvec_1|/L$ close to unity) still
collapses even for $\mu=0.3$.  That the tail collapses so easily is of
course unsurprising, as $\ActActTimeAve_c(\xvec_2,\xvec_1;\xvec_0,L)$ 
vanishes as sites become uncorrelated, \ie when
$|\xvec_2-\xvec_1|$ approaches $L$, and a sharp drop is very
insensitive to rescaling.

\subsubsection{Two-time activity correlations and $1/f$-noise}
Historically, SOC was introduced as an ``explanation for $1/f$ noise''
\cite{BakTangWiesenfeld:1987}, even when that notion quickly took a less
prominent place \cite{Pruessner:2012:Book,WatkinsETAL:2016}. The initial
measurement \cite{BakTangWiesenfeld:1987} of the power-law
characteristics of the power spectrum of the BTW Model was
soon revised by Jensen \etal \cite{JensenChristensenFogedby:1989}, who
had to made a number of drastic assumptions in order to link the power
spectrum to the avalanche duration distribution. 
A more recent, detailed numerical
analysis by Laurson \etal \cite{LaursonAlavaZapperi:2005} found non-trivial
$1/f^{\alpha}$ noise in both the BTW and the Manna Model.

In the following, 
we analyse the correlations in the spatially integrated activity signal 
\begin{equation}
\nhat_i(t;\xvec_0,L) = \sum_{\xvec} n_i(\xvec,t;\xvec_0,L)
\end{equation}
of avalanche $i$ with duration $T_i$, which, of course, varies among
avalanches. That makes the calculation of correlations somewhat
ambiguous,\footnote{Apart from allowing for vanishing $n_i$ (see below),
one may, for example, rescale time to the unit interval or weight avalanches of
different durations differently.} in particular when taking averages over ensembles. 
We intend to carry out the following analysis in the spirit
of
the original link of SOC and $1/f$ spectra. To arrive at a single time series,
we effectively concatenate consecutive $\nhat_i(t;\xvec_0,L)$ by inserting
an infinite trail of zeros. 

\begin{figure}
\includegraphics[width=0.95\linewidth]{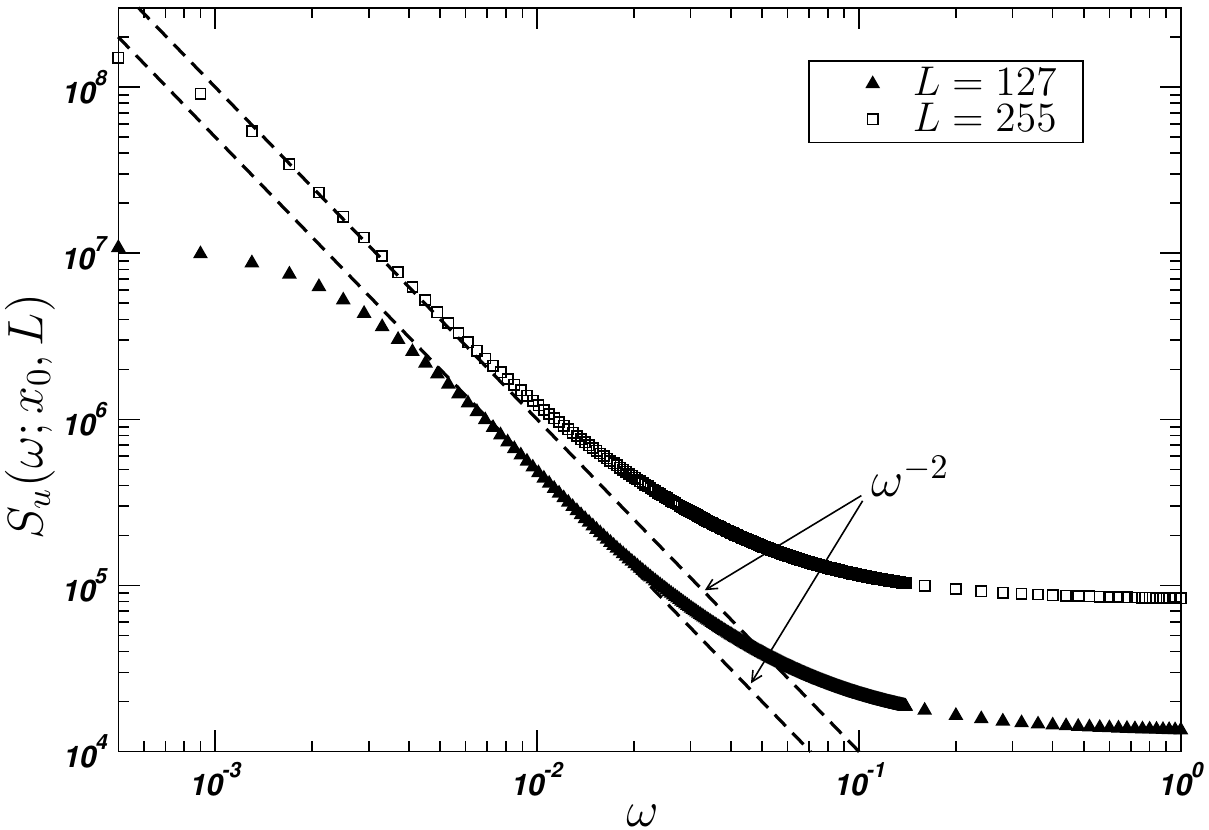}
\caption{\flabel{power_spectrum}Double logarithmic plot of the (pruned)
power spectrum $S_u(\omega;x_0,L)$, \Eref{def_S_u}, of the unconnected
correlation function $K_u$, \Eref{def_K_u}, in a the centre-driven
one-dimensional Manna Model of size $L$ as indicated. There is a
possibly increasing (with $L$) intermediate regime that displays some
power law scaling $\propto \omega^{-2}$ (dashed line). }
\end{figure}

We will make use of the estimators (\cf \Erefs{def_ActAct_u} and
\eref{def_ActAct_c})
\begin{equation}
K_u(t_2,t_1;\xvec_0,L) = \frac{1}{M} \sum_{i=1}^M \nhat_i(t_2;\xvec_0,L)
\nhat_i(t_1;\xvec_0,L) \ ,
\elabel{def_K_u}
\end{equation}
and
\begin{multline}
K_c(t_2,t_1;\xvec_0,L) \\
= K_u(t_2,t_1;\xvec_0,L) 
- \ShapeAvaAbs(t_2;\xvec_0,L) \ShapeAvaAbs(t_1;\xvec_0,L) \ ,
\elabel{def_K_c}
\end{multline}
where the second term makes use of $\ShapeAvaAbs(t;\xvec_0,L)$, which is
the expectation of $\nhat_i(t;\xvec_0,L)$,
\Erefs{def_activity_estimator} and \eref{def_ShapeAvaAbs}.  To
make $\nhat_i(t;\xvec_0,L)$ well-defined for all $t$, we take
$\nhat_i(t;\xvec_0,L)=0$ for $t>T_i$. This together with \Eref{def_K_u} that
never mixes different avalanches, amounts to analysing correlations
in concatenated activity
histories as if each was succeeded by an infinite trail of zeros, thereby
implementing separation of time scales. Other choices have been made in the
literature, notably by Laurson \etal who suggest the effect of trailing zeros
is negligible \cite{LaursonAlavaZapperi:2005}.

As the process is not time-homogeneous, $K_{c,u}(t_2,t_1;\xvec_0,L)$ are
functions of two times $t_1$ and $t_2$. 
In the past, this feature was to large extent ignored. In
\cite{JensenChristensenFogedby:1989} the authors assume that $\nhat_i$ is a
top-hat function with duration $T_i$ and height $s_i/T_i$, approximating
$K_u$ by a suitably weighted integral of convolutions of that top-hat.

From \Fref{ShapeAvaAbs} it is clear that in the Manna Model, $\nhat_i$
is not overly well approximated by a top-hat. In the following, we want
to explore the shape of the unconnected and connected correlation
functions $K_u$ and $K_c$. The first question that arises is whether it
is justified to assume that $K_u$ and $K_c$ are essentially functions of
the time lag $|t_2-t_1|$, but otherwise independent of $t_1$.
\Fref{K_c_nearly_same} shows that this is indeed the case for a range of
$t_1$ at least as far as the connected correlation function is concerned. This feature is
clearly less pronounced for the unconnected correlation function, \Fref{K_u_nearly_same},
which,
however, is closer to the one studied in the past with regard to its $1/f$
characteristics in the BTW Model \cite{JensenChristensenFogedby:1989,LaursonAlavaZapperi:2005}.

Carrying on along the lines of Jensen \etal
\cite{JensenChristensenFogedby:1989} but also Laurson \etal \cite{LaursonAlavaZapperi:2005},
the
unconnected two-time correlation function is time-integrated (not
time-averaged, as usual, \eg \cite{Gardiner:1997}) over $t$,
\begin{equation}\elabel{def_Kbar_u}
\Kbar_u(\tau;\xvec_0,L) = \sum_{t=0}^\infty K_u(t,\tau;\xvec_0,L) 
\end{equation}
and Fourier-transformed,
\begin{equation}\elabel{def_S_u}
S_u(\omega;\xvec_0,L) = 2 \sum_{\tau=0}^{\infty} \cos(\omega\tau) \Kbar_u(\tau;\xvec_0,L)
\end{equation}
which resembles a sigmoidal shape, as shown in \Fref{power_spectrum}, quite
different to what Laurson \etal \cite{LaursonAlavaZapperi:2005} obtain using
a different concatenation scheme.
The intermediate drop may be approximated by a power law, $\omega^{-2}$,
consistent with similar findings in the BTW Model
\cite{JensenChristensenFogedby:1989}, but it is difficult to trace that
behaviour back to the decay of temporal correlations.

In fact, there is little non-trivial, asymptotic long-time behaviour in $K_u$ or $K_c$ at all.
What displays scaling in these quantities is their cutoff (because of
the scaling of durations), but not the
correlation functions themselves, \Fref{K_c_nearly_same}. The approach by
Laurson \etal \cite{LaursonAlavaZapperi:2005} has clearly been more successful
in that respect.

One might be tempted to repeat the above analysis for the connected
correlation function $K_c$ as defined in \Eref{def_K_c}. 
However, taking an average like \Eref{def_Kbar_u} over the unconnected
part $\ShapeAvaAbs(t_2;\xvec_0,L) \ShapeAvaAbs(t_1;\xvec_0,L)$ remains
ambiguous. It is arguably to be replaced by the square of the
average activity $\ave{s}/\ave{T}$, in which case, however, the time
averaged $K_c$ no longer drops to $0$ for $t\to\infty$ as normally expected
for a connected correlation function. Given that our main interest in
the present correlation functions was to make contact with the historic research
focus, namely $1/f$-noise as studied in
\cite{JensenChristensenFogedby:1989}, we did not pursue this approach
any further.

\begin{figure}
\includegraphics*[width=0.95\linewidth]{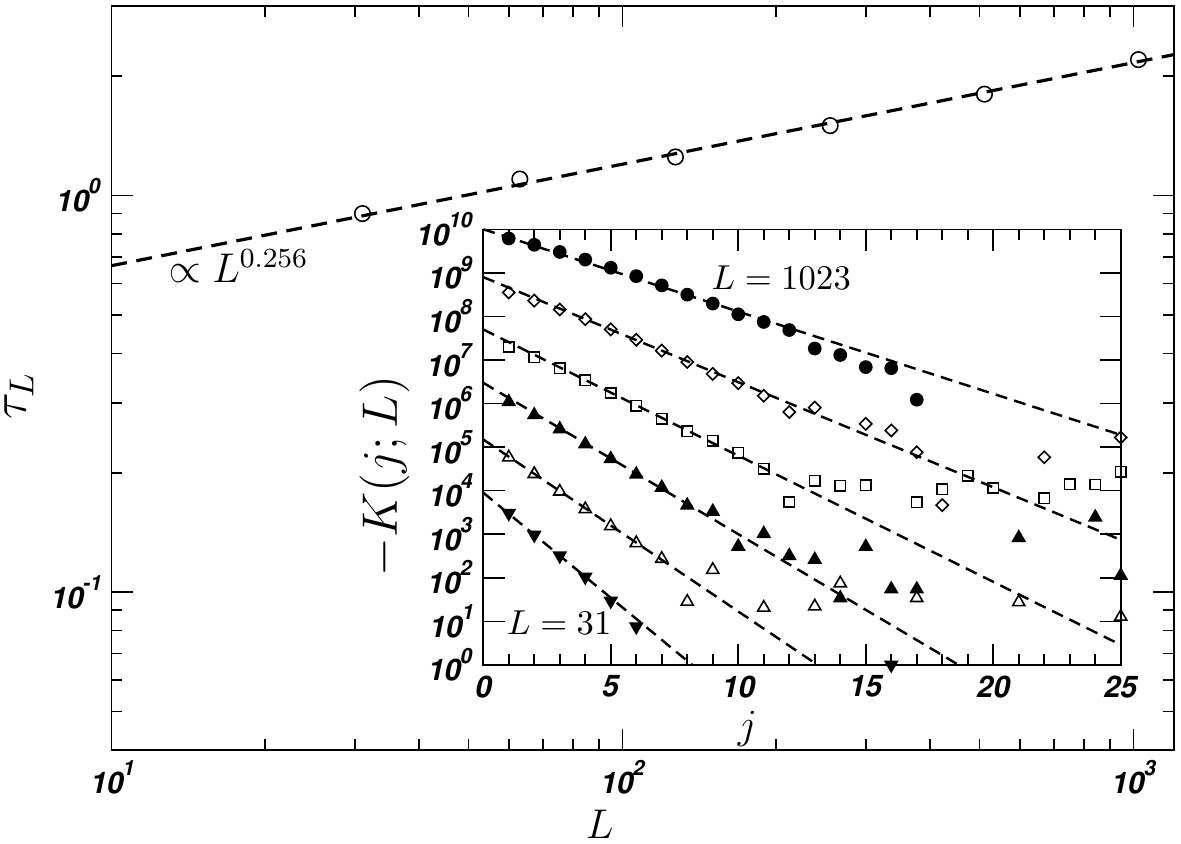}
\caption{\flabel{ava_ava_corrs_1D}
Scaling of the macroscopic avalanche-avalanche correlation time
$\tau_L$ with the system size $L$. The power law fitted (dashed line),
$\tau_L\propto L^{0.256}$, is compatible with the prediction
$\tau_L\propto L^{D-2}$ with $D=2.253(14)$
\cite{HuynhPruessnerChew:2011}.
In one dimension, correlation times are rather short, well under $10$
for the system sizes considered.  Inset: The negative of the actual
correlation function $\AvaAvaCorr(j;L)$ for $L=31,63,\ldots,1023$ with
the exponential fits, $A_L \exp{-j/\tau_L}$ shown as dashed lines.
}
\end{figure}

\subsubsection{Macroscopic time correlations}
\slabel{ava_ava_corr}
We finally consider the correlations in the size $s_i$ of consecutive
avalanches, $\ave{s_i s_{i+j}}-\ave{s}^2$
as a function of $j$, estimated via \cite{Anderson:1971}
\begin{multline}\elabel{def_AvaAvaCorr}
\AvaAvaCorr(j;L) \\
= \frac{1}{M-j} \sum_{i=1}^{M-j} s_i s_{i+j} - 
 \frac{1}{M-j} \sum_{i=1}^{M-j} s_i 
 \frac{1}{M-j} \sum_{i=1}^{M-j} s_{i+j} \ .
\end{multline}
In the Oslo Model \cite{ChristensenETAL:1996} the avalanche size can be
interpreted as the displacement of an interface pulled on one end over a
rough surface
\cite{PaczuskiBoettcher:1996,Pruessner:2003,Pruessner:2012:Book} and the
macroscopic correlation time is, in this model, therefore related to the
interface's roughness, which scales like $L^{\roughness}$ with
$\roughness=D-d$. In the present case, neither the mapping exists nor
can the driving be interpreted easily as a pulling force. Following
nevertheless the same argument as in \cite{PaczuskiBoettcher:1996}, the
typical total number of topplings needed to avoid an overlap between
initial and final interface configuration scales like
$L^{\roughness+d}$. The number of avalanches to be triggered to reach
essentially independence is therefore of order
$L^{\roughness+d}/\ave{s}$, which in the present case scales like
$L^{\roughness+d-2}$ as $\ave{s}\propto L^2$, \Eref{ave_ava} with
$x_0=(L+1)/2$. In other words, the macroscopic correlation time should
scale like $D-2=0.253(14)$ \cite{HuynhPruessnerChew:2011}.

It is difficult to extract good estimates of the macroscopic correlation time
from the data. As observed in directed models
\cite{WelinderPruessnerChristensen:2007}, correlations for $j>0$ are
anti-correlations, $\AvaAvaCorr(j;L)<0$, as large avalanches are
normally followed by smaller ones and vice versa.
\Fref{ava_ava_corrs_1D} shows $-\AvaAvaCorr(j;L)$ in a semi-logarithmic
plot in the inset, suggesting an exponential decay of correlations,
which are quite short-lived in one dimension (and obviously
orders of magnitude shorter than the transient). Fitting them (by eye),
$\AvaAvaCorr(j;L) = A_L \exp{-j/\tau_L}$ with amplitude $A_L$, and
plotting the resulting estimates for the correlation times $\tau_L$
yields a scaling of approximately $\tau_L\propto L^{0.256}$, as shown in
\Fref{ava_ava_corrs_1D}, compatible with $D-2$ quoted above.

\section{Higher dimensions}
\slabel{higher_dimensions}
\begin{figure}
\includegraphics*[width=0.95\linewidth]{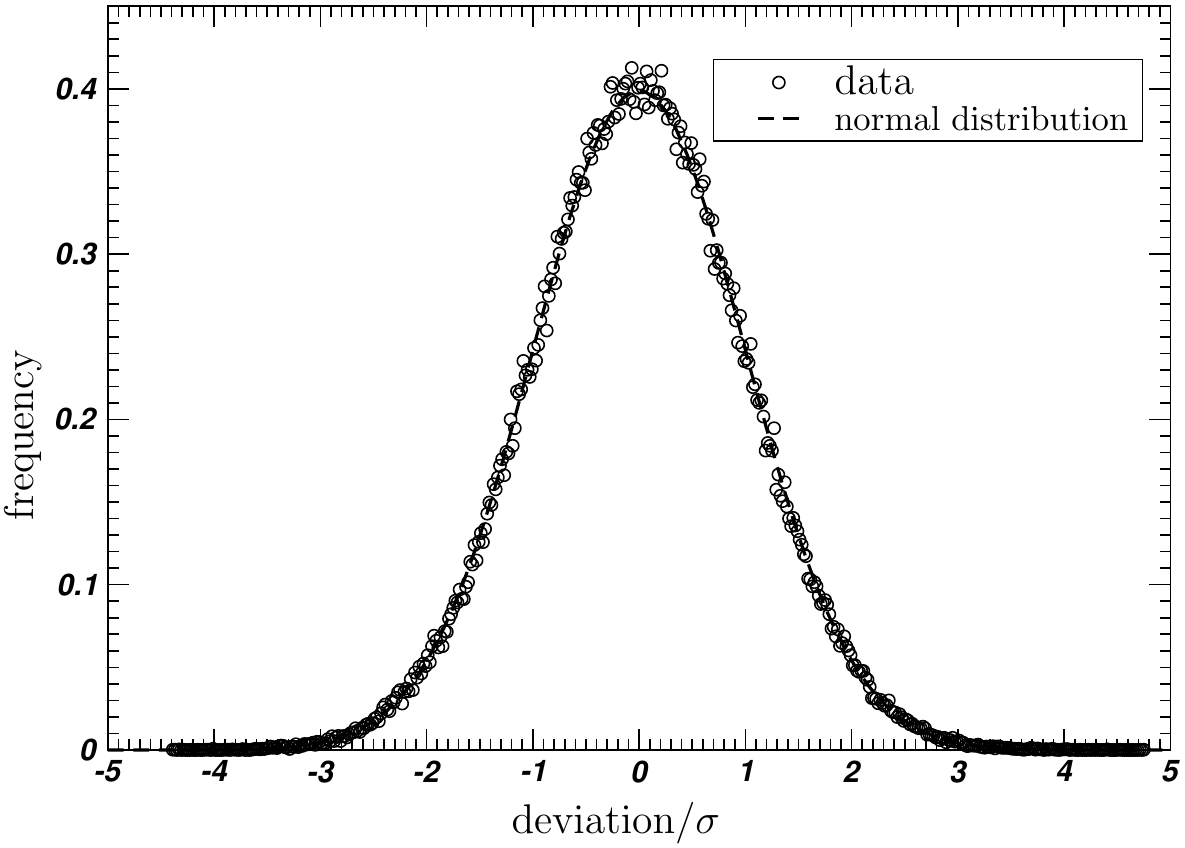}
\caption{\flabel{3D_deviation_distribution}A histogram of the deviation
of the local density $\Ds(\xvec;\xvec_0,L)$ from the average
$\Dsbar(x;\xvec_0,L)$ in a sheet of constant $x$, in units of standard
deviations of the density within a sheet, for a three dimensional system of size $63\times64\times64$.
Within numerical error, the data follows a normal
distribution shown as a dashed line.
}
\end{figure}

We have repeated many of the measurements discussed above in three and
five dimensions. Because the upper critical dimension of the Manna Model
is $d_c=4$ \cite{LuebeckHucht:2002,LuebeckHeger:2003b,Luebeck:2004}, $d=3$ is expected to be much
closer to mean-field results (and in that sense better behaved), whereas
$d=5$ is expected to reproduce them at least as far as universal
quantities are concerned.

As the most interesting observables to consider in higher dimensions, we
have selected the density of inactive particles in the quiescent state
(\cf \Sref{quiescent_state_1D}), 
the spatially integrated activity (\cf \Sref{shape_of_ava_1D}), 
the temporal shape of the avalanche (\cf \Sref{temp_shape_of_ava}),
the width of the response (\cf
\Sref{msd_1D}), the spatial activity-activity correlation function
(\cf \Sref{actact_corrs_1D}), and the macroscopic time
avalanche-avalanche correlations (\cf \Sref{ava_ava_corr}).

Numerically, dimensions greater than one pose the disadvantage of high
memory requirements for comparatively ``small'' system sizes as far as
linear extent is concerned. In particular in five dimensions, lattices
of linear extent beyond $L=63$ are difficult to realise. The largest
lattice we used in five dimensions therefore was $L=95$. 
We have avoided scanning lattices as much as possible (see discussion before
\Eref{summation_by_parts}), so that CPU-time requirements are mostly determined by the average
avalanche size $\propto L^2$.
The largest
lattice in three dimensions was $L=511$. As discussed above, periodic
boundary conditions were applied in all but one (open) direction.

The periodic boundary conditions may suggest that observables do not
depend on the coordinates $y_2,\ldots,y_d$ orthogonal to the open direction
(parameterised by $x$). 
We will indeed consider in the following certain observables with respect
to their sheet-average (spatial
average at constant $x$) and their deviation from it.
Yet, given centre driving at one single site
$x_0=(L+1)/2$, $y_{2,\ldots,d}=0$, translational invariance is broken in
every direction.\footnote{If one is not interested in the particular
symmetries of the system, this can be cured by driving (randomly) across
entire sheets of constant $x$.} 
This is expected to be reflected in the observables, for example the
response function: Periodic boundary conditions or not, activity that
begins in a point will not spread instantaneously across entire sheets
of constant $x$ (which are of size $L'^{d-1}$). 

\subsection{Three dimensions}
\begin{figure*}
\subfigure[Sheet-averaged density profile in relative coordinates.]{\includegraphics*[width=0.45\linewidth]{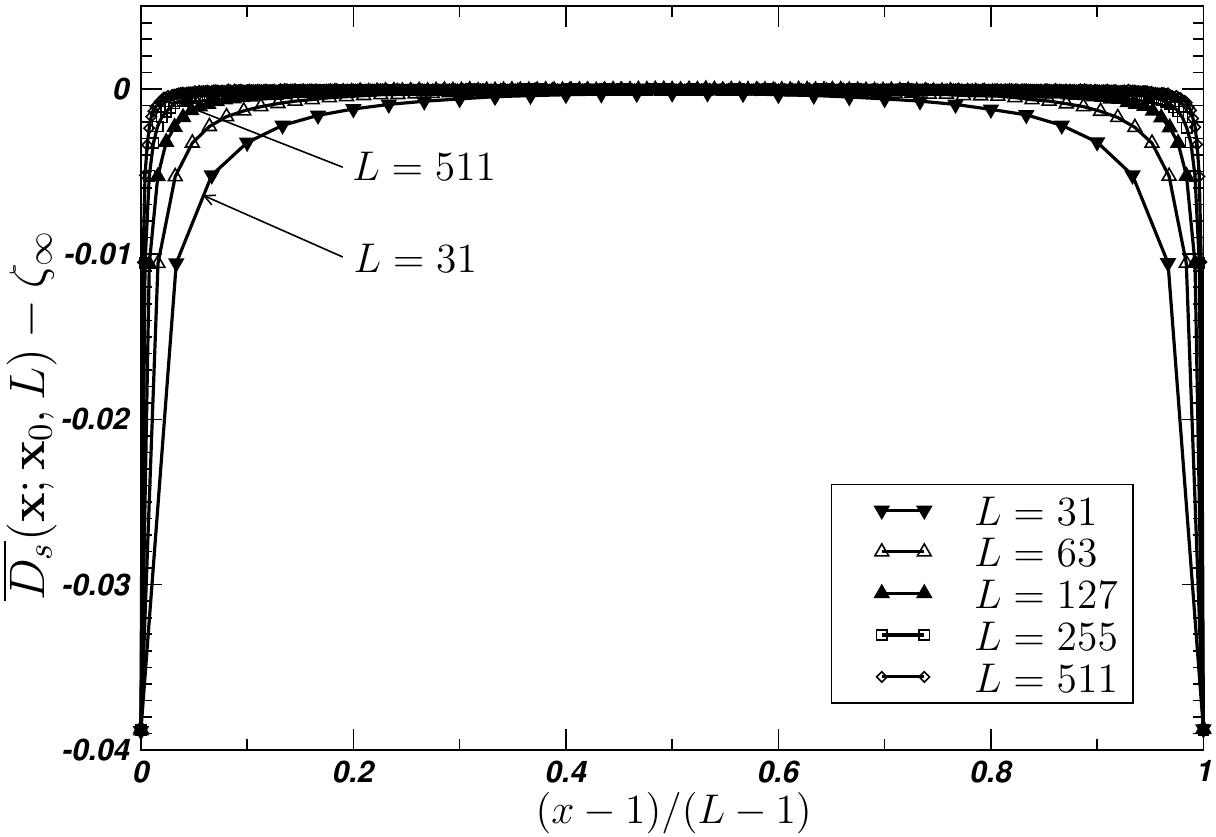}\flabel{substrate_density_3D_relpos}}
\subfigure[Sheet-averaged density profile close to the boundary.]{\includegraphics*[width=0.45\linewidth]{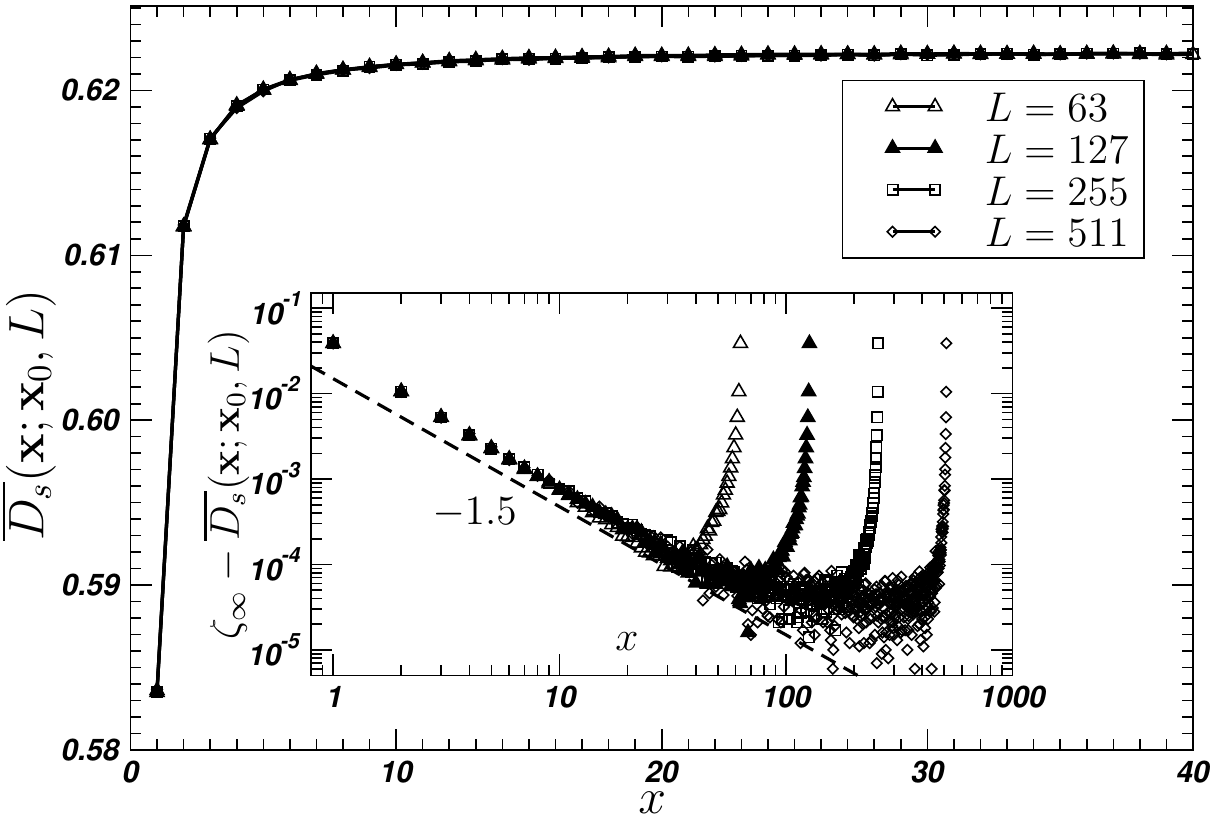}\flabel{substrate_density_3D_boundary}}
\caption{\flabel{substrate_density_3D}The sheet-averaged density of
inactive particles in three dimensions,
$\Dsbar(x;\xvec_0,L)$,
as deviation from the bulk value ($\zeta_{\infty}=0.622325$
\cite{HuynhPruessner:2012b}) in the quiescence state at stationarity in
three dimensions, \cf \fref{substrate_density_1D}. 
Lines to guide the eye. 
\subfref{substrate_density_3D_relpos} Plotting $\Dsbar(x;\xvec_0,L)-\zeta_
{\infty}$ versus the relative position $(x-1)/(L-1)$ shows that the deviation
of the density from the bulk value decays quickly away from the boundaries,
producing an increasingly sharp shoulder with increasing system size.
\subfref{substrate_density_3D_boundary} Close to the boundary, on an absolute
scale $x$ (distance from the boundary), the density profile shows very quick
convergence with increasing $L$. There is no noticeable difference between
different $L$. However, as shown in the inset, the deviation from the bulk
density $\zeta_{\infty}$
shows a (noisy) power law decay away from the boundary, roughly with exponent
$-1.5$
(dashed line) and with a cutoff linear in $L$. 
}
\end{figure*}

It turns out, however, that even with (non-translational invariant) centre driving the density of inactive particles
displays translational invariance within sheets of constant $x$. It is rather futile
to attempt to visualise that by plotting the density profile for
constant $y_2,y_3$ (coordinates within a sheet).
\Fref{3D_deviation_distribution} shows instead a histogram of the number
of standard deviations by which the (estimated) density at each and every point
within the sheets deviates from the average within a sheet, 
\begin{equation}
\Dsbar(x;\xvec_0,L)=\sum_{y_2,y_3} \Ds(\xvec;\xvec_0,L) \ ,
\end{equation}
where $\xvec=(x,y_2,y_3)$.  The result is a distribution very close to a
normal distribution (also shown), suggesting that the small deviations
of local densities from the sheet average $\Dsbar(x;\xvec_0,L)$ are
possibly random and independent rather than systematic. 
However, confirming that by direct measurements of correlations is difficult,
firstly because of the statistical noise (as seen in \Fref{substrate_correlations_1D_zoom})
and secondly because of the 
wide range of correlations to consider. For example the two
point function $C_s(\xvec_2,\xvec_1;\xvec_0,L)$, \Eref{def_Cs},
depends on the positions relative to the open ends, $x_1$ and $x_2$, as well
as the $d-1$ displacements in the periodic direction,
$\yvec_1-\yvec_2$. Again, integrated observables, such
as the window-averaged density \Eref{sigma2_from_C_s}, might be better suited to reveal
(anti-) correlations.

Given the (observed) translational invariance in the periodic
directions, \Fref{substrate_density_3D} shows the sheet-averaged density
profile $\Dsbar(x;\xvec_0,L)$ across the open direction in
three-dimensional systems as the deviation from the bulk value
$\zeta_{\infty}=0.622325$ \cite{HuynhPruessner:2012b}.  This data shows
little qualitative difference compared to \Fref{substrate_density_1D},
except that the density remains much closer to the bulk value
throughout, even close to the boundaries and even for comparatively small
systems. The range of the ordinate of \Fref{substrate_density_3D_relpos} is about an order of magnitude
smaller than in one dimension.  
\Fref{substrate_density_3D_boundary}
shows that the profile close to the boundary displays no discernible
difference among the system sizes considered; there is certainly no
finite size scaling of the amplitude of the profile. 
However, similar to one dimension,
\Fref{substrate_density_1D}, boundary effects decay like a power law. 
We estimate the
exponent to be around $-1.5$ (as shown in the inset of \Fref{substrate_density_3D_boundary}), 
but a reliable estimate is hampered by the high level of noise in the data and does
not improve much when the data is binned. For comparison (see discussion in
\Sref{quiescent_state_1D}) $1/\nu_\perp=1.7(4)$ \cite{Luebeck:2004}, compatible
with the exponent estimated here.

Integrating the local activity over the entire lattice produces the
total activity, $\ShapeAvaAbs(t;\xvec_0,L)$, as defined in
\Eref{def_ShapeAvaAbs}. As discussed in \Sref{shape_of_ava_1D}, we
expect plotting $L^{2-z} \ShapeAvaAbs(t;\xvec_0,L)$ against
$t/L^{z}$ to produce a collapse.  In $d=3$ dimensions this
is indeed the case and works very well using $2-z=0.223$,
based on the dynamical exponent $z=1.777(4)$ as
obtained in \cite{HuynhPruessner:2012b}; \Fref{ShapeAvaAbs_3D} shows a
corresponding collapse like the one obtained in \Fref{ava_rel} in $d=1$
dimensions.

Also shown in this figure is the slope of the rescaled total activity in
rescaled time, which we expected to display an exponent of $(2-z)/z$
(see the discussion in \Sref{shape_of_ava_1D}). While this was not fully confirmed
in one dimension, it is perfectly in line with the findings in three
dimensions.

Normalising the activity profile in the form introduced in
\Eref{def_ShapeAvaRel} produces the ``temporal shape of the avalanche'' shown
in
\Fref{ava_rescaled_3D} together with the mean field theory introduced in
\Sref{shape_of_ava_1D} and \Sref{temp_shape_of_ava}. The curves are remarkably close given
that
the
mean field theory should apply only in dimensions of $d\ge d_c=4$. 
The data still shows a slight skew which is difficult to see by naked
eye. The mean field data displays a similar slant (possibly a finite size
effect), but in the opposite
direction. The collapse thus improves when plotting
$\ShapeAvaRel(1-\tau; \xvec_0,L)$ against the MFT.

\begin{figure}
\includegraphics*[width=0.95\linewidth]{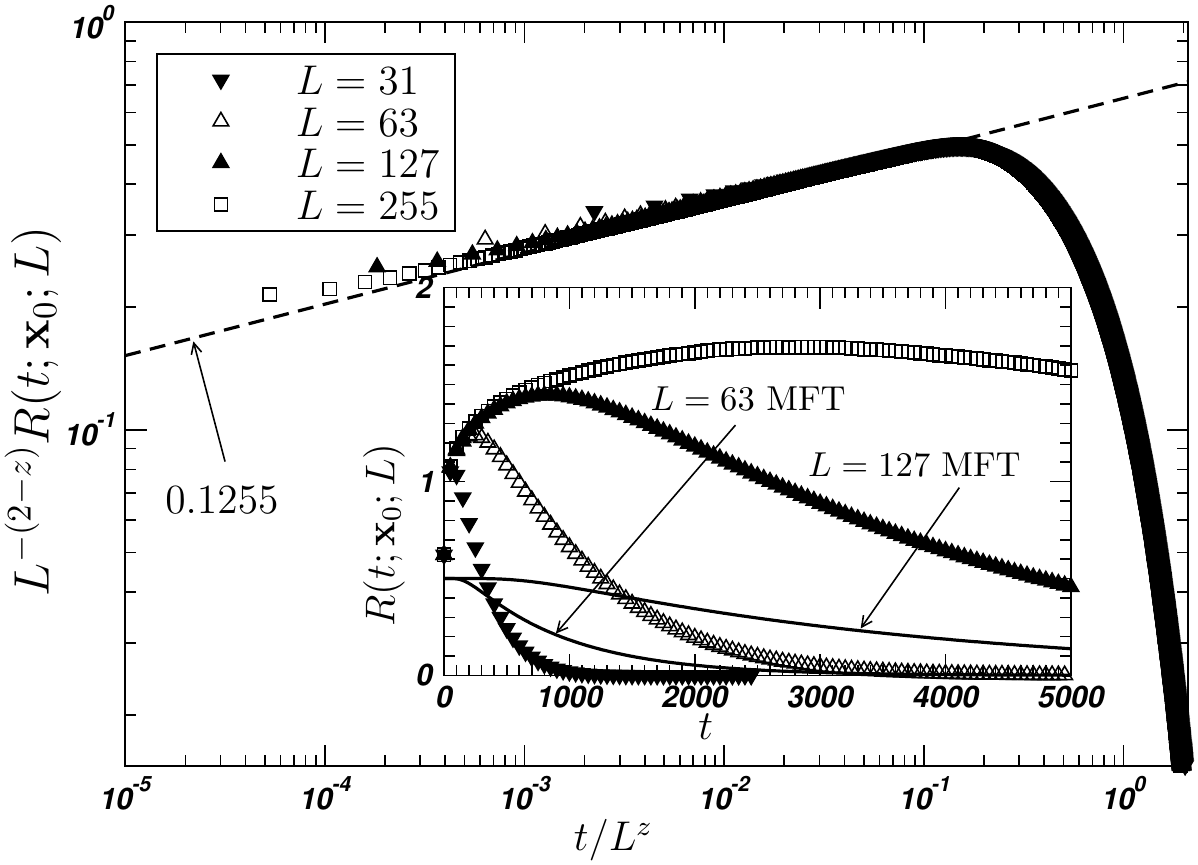}
\caption{\flabel{ShapeAvaAbs_3D}Collapse of the spatially integrated
activity, $\ShapeAvaAbs(t;\xvec_0,L)$, \Eref{def_ShapeAvaAbs}, for the
centre driven Manna Model in $d=3$ dimensions (see \Fref{ava_abs} for
the result in one dimension), plotting
$\ShapeAvaAbs(t;\xvec_0,L)L^{-(2-z)}$ against
$t/L^{z}$ using the literature value of $z=1.777(4)$
\cite{HuynhPruessner:2012b}. The dashed line
shows the expected slope $(2-z)/z=0.1255(25)$, in line with the
discussion in \Sref{shape_of_ava_1D}. The inset shows the (pruned) data on a
linear scale together with the MFT as in \Fref{ava_abs}.}
\end{figure}
\begin{figure}
\includegraphics*[width=0.95\linewidth]{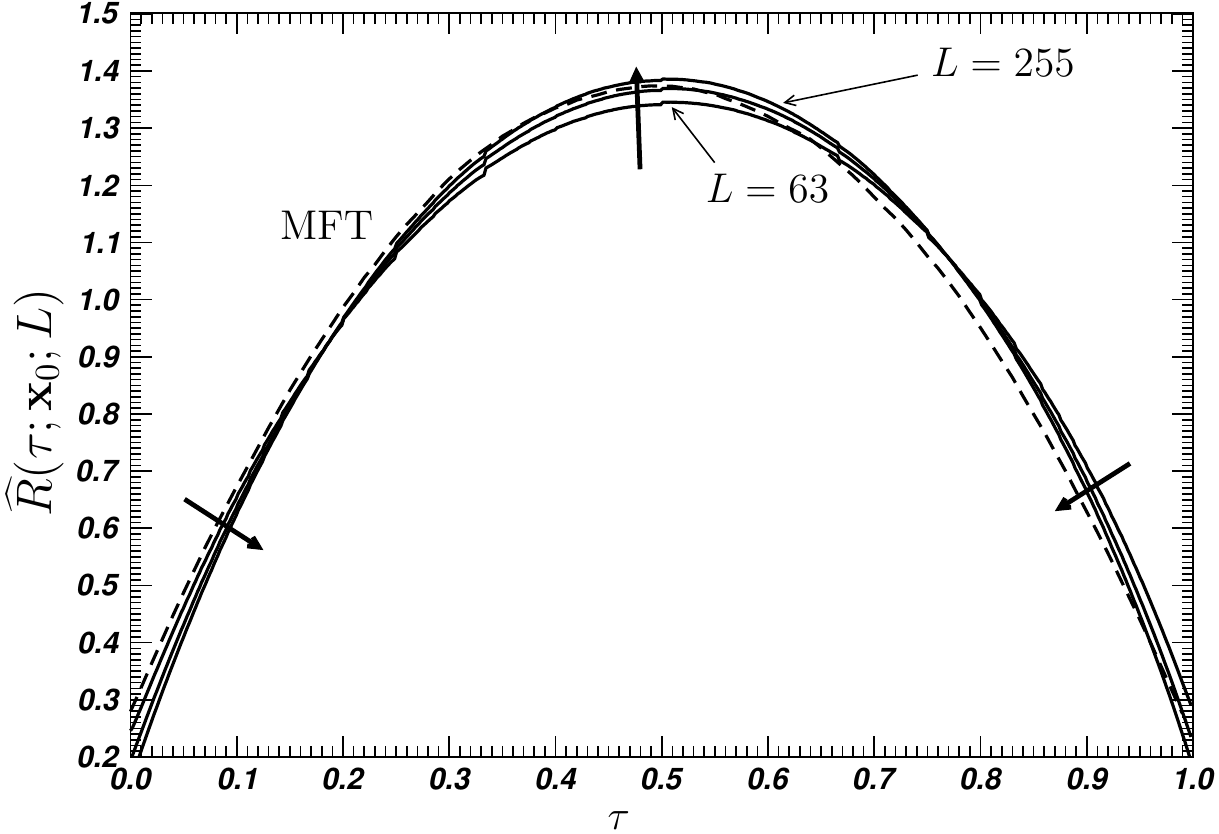}
\caption{\flabel{ava_rescaled_3D} The rescaled activity
$\ShapeAvaRel(\tau; \xvec_0,L)$ in three-dimensional systems of size
$L=63,127,255$ driven at the centre, $x_0=(L+1)/2$, to be compared to
\Fref{ava_rescaled}.  The thick arrows point in the direction of
increasing system size.  The rescaling maps the spatially integrated
activity to the interval $\tau\in[0,1]$. Normalisation is applied so
that the integral under the curve is unity.  The data is remarkably
close to that of the mean-field theory, shown as a dashed line,
($L=255$, see \Fref{ava_rescaled}).  }
\end{figure}
\begin{figure}
\includegraphics[width=0.95\linewidth]{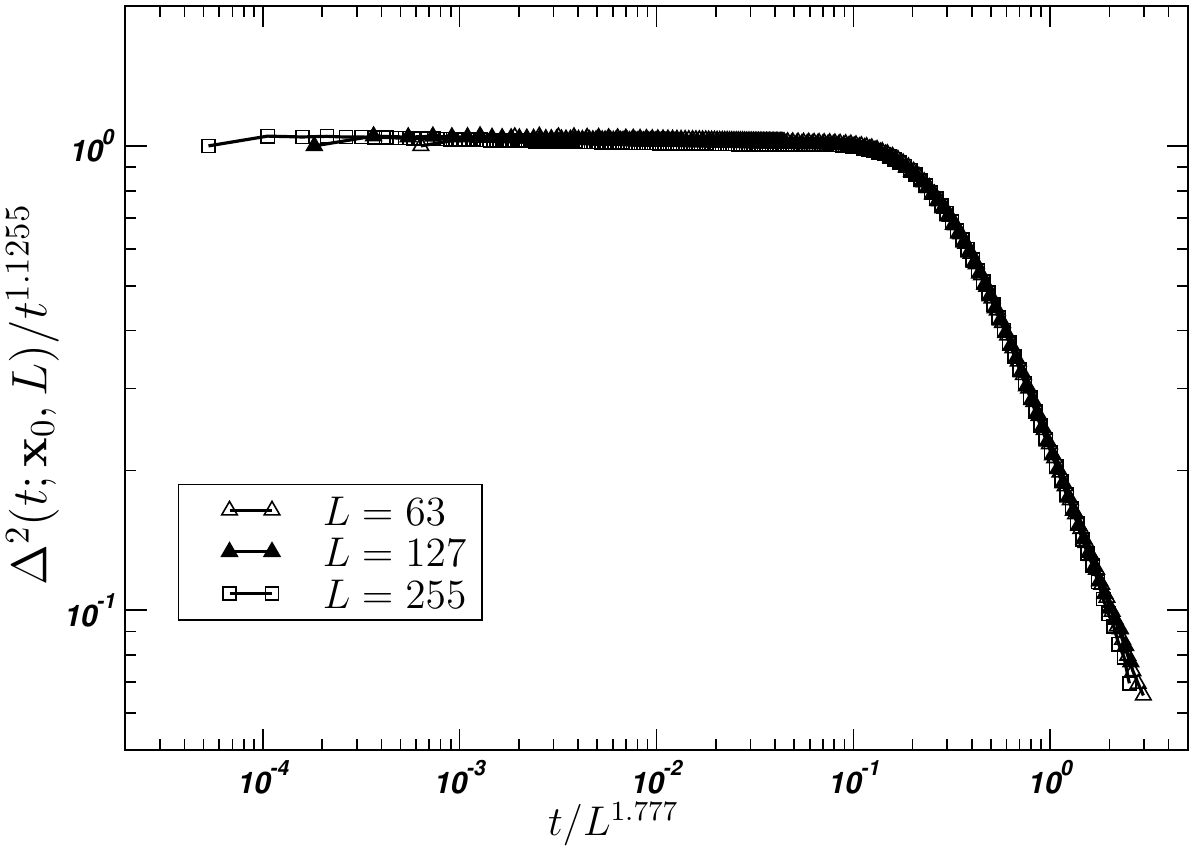}
\caption{\flabel{msd_3D}
Collapse of the width $\msd(t;\xvec_0,L)$ (\Eref{def_msd}) in the
three-dimensional, centre driven Manna Model. Data is binned. Both axes
have been rescaled using the dynamical exponent $z=1.777(4)$ found in
the literature \cite{HuynhPruessner:2012b}, with $2/z\approx1.1255$.
}
\end{figure}

The width of the response propagator, $\msd(t;\xvec_0,L)$, as defined in
\Eref{def_msd} produces a collapse if rescaled suitably, as shown for
$d=1$ in \Fref{msd}. In \Sref{msd_1D} it was demonstrated that in one
dimension the collapse requires a dynamical exponent, $z\approx1.5625$,
that deviates quite clearly from the expected value of $z=1.445(10)$
from the literature.  In \Fref{msd_3D} we show that the data in three
dimensions is compatible with the expected value, $z=1.777(4)$
\cite{HuynhPruessner:2012b}, apparently validating the scaling arguments
in \Sref{msd_1D}. The possible causes of the mismatch in one dimension is
discussed further in \Sref{conclusion}.

\begin{figure*}[t]
\subfigure[Time and sheet-averaged unconnected spatial activity-activity correlation function $\ActActTimeAveSpave_u(x_2,x_0;x_0,L)$, \Eref{def_ActActTimeAveSpave_u}, collapsing for different system sizes $L$ according to \eref{actact_scaling} with $\mu=2.92$.]{\includegraphics*[width=0.45\linewidth]{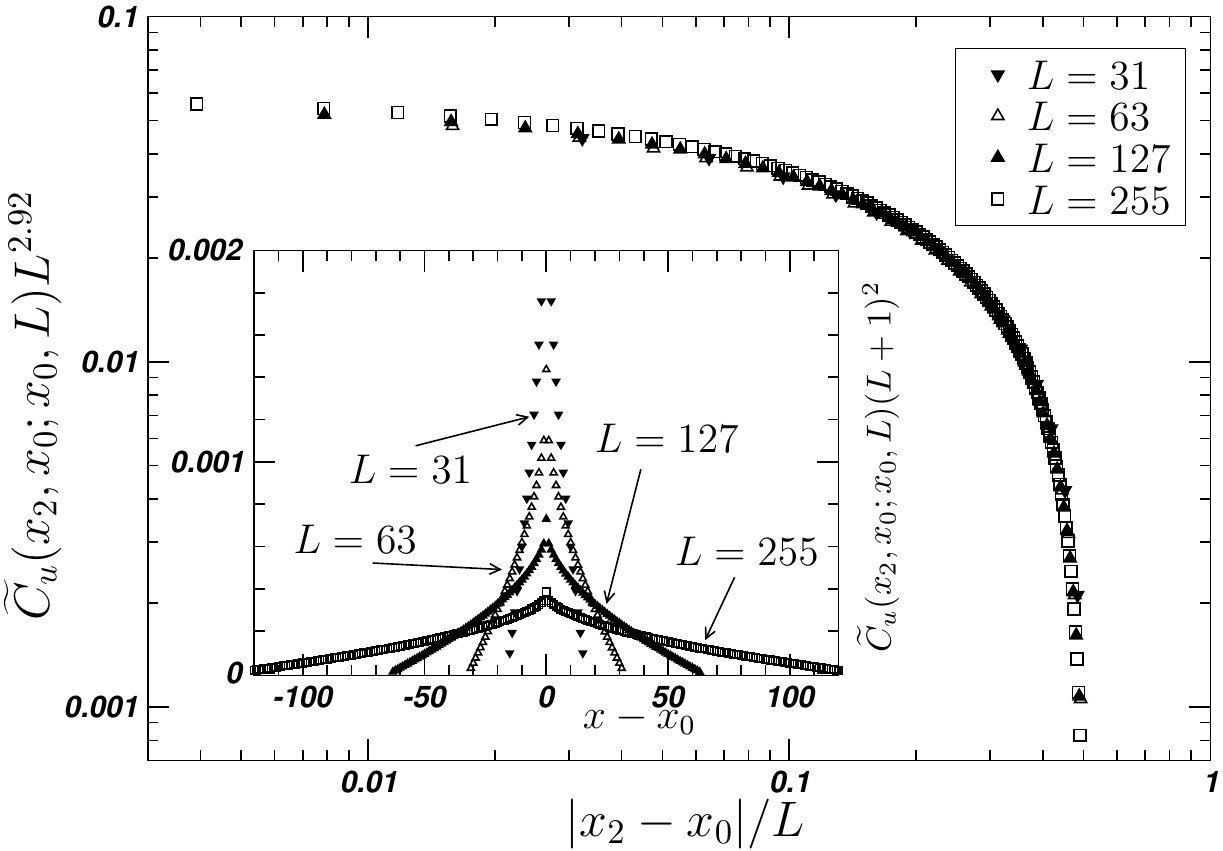}\flabel{actact_u_collapse_3D}}
\subfigure[Time and sheet-averaged connected spatial activity-activity correlation function $\ActActTimeAveSpave_c(x_2,x_0;x_0,L)$, \Eref{def_ActActTimeAveSpave_c}, collapsing for different system sizes $L$ according to \eref{actact_scaling} with $\mu=2.79$.]{\includegraphics*[width=0.45\linewidth]{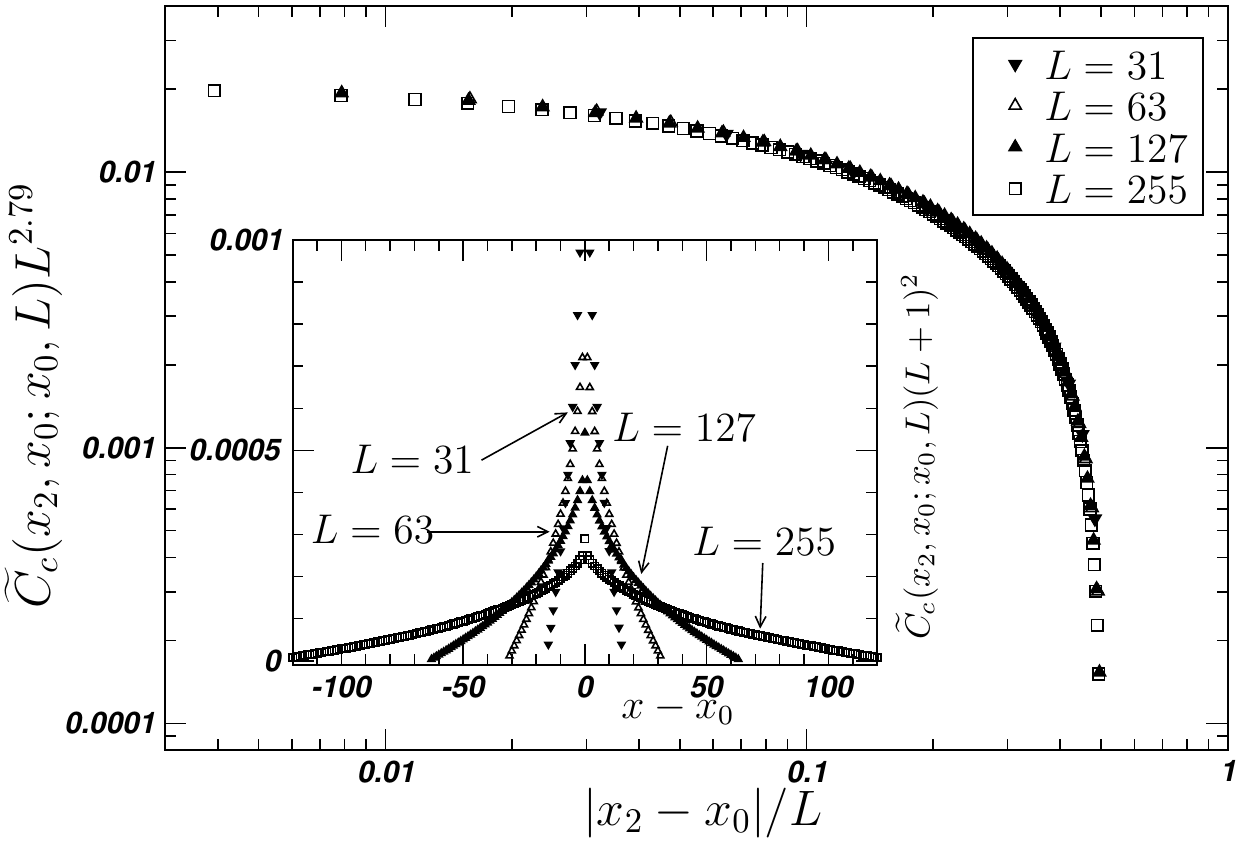}\flabel{actact_c_collapse_3D}}
\caption{\flabel{actact_collapses_3D} Collapses according to
\Eref{actact_scaling} of the two point spatial correlation function
$\ActActTimeAveSpave_u$ and $\ActActTimeAveSpave_c$ as defined in
\Erefs{def_ActActTimeAveSpave_u} and \eref{def_ActActTimeAveSpave_c}
respectively for different three-dimensional system sizes as indicated.
The exponent $\mu$ used is largely consistent with literature values. The
insets show the correlation function on linear scales, however
rescaled by the size of a sheet, $(L+1)^2$, so that different system
sizes $L$ can be shown simultaneously. See also \Fref{actact_collapse}.  }
\end{figure*}

As mentioned above, the propagation of activity (the response function)
is certainly not expected to be translationally invariant, as activity
surely is correlated to the position of the initial driving.
However, it is numerically very challenging to analyse spatial
data as a function of all $d$ components. We have therefore decided to
study activity and its correlations after spatially averaging in the
periodic direction. This way we obtain enough statistics and yet can
still test for the expected scaling behaviour. 

As indicated in \Erefs{triangular_profile_35_integrated} and \eref{lattice_triangular_profile_35_integrated},
averaging (or, for that purpose, summing) the time-integrated activity
over the periodic direction reproduces the one-dimensional profile
\Eref{triangular_profile}. The resulting profiles of the
time-integrated activity summed over the periodic direction look
therefore up to a pre-factor exactly like \Fref{TimeIntegratedActivity}.

Activity-activity correlations on the other hand show some clear
non-trivial scaling as a function of space.  Defining the
time-integrated, sheet-averaged activity in terms of the numerical
observable $n_i(\xvec,t;\xvec_0,L)$ as
\begin{equation}\elabel{def_ActivityTimeAveSpave}
\ActivityTimeAveSpave(x;\xvec_0,L) = 
\frac{1}{\sum_{i=1}^M T_i} \sum_{i=1}^M\sum_{t=1}^{T_i} 
\frac{1}{L'^{d-1}}\!\! \sum_{y'_2,\ldots,y'_d} 
n_i(\xvec,t;\xvec_0,L)
\end{equation}
produces a function of essentially only one variable, $x$, the position
in the open direction with the shape shown in
\Fref{TimeIntegratedActivity}, see \Eref{lattice_triangular_profile_35_integrated}. The activity-activity correlation
functions may be defined in the 
same vein,
\begin{multline}\elabel{def_ActActTimeAveSpave_u}
\ActActTimeAveSpave_u(x_2,x_1;\xvec_0,L) = 
\frac{1}{\sum_{i=1}^M T_i} \sum_{i=1}^M\sum_{t=1}^{T_i} 
\frac{1}{L'^{d-1}}\\
\times 
\sum_{y''_2,\ldots,y''_d} 
n_i(\xvec_2,t;\xvec_0,L)
\sum_{y'_2,\ldots,y'_d} 
n_i(\xvec_1,t;\xvec_0,L)
\end{multline}
and
\begin{multline}\elabel{def_ActActTimeAveSpave_c}
\ActActTimeAveSpave_c(x_2,x_1;\xvec_0,L) = 
\frac{1}{\sum_{i=1}^M T_i} \sum_{i=1}^M\sum_{t=1}^{T_i} 
\frac{1}{L'^{d-1}} \\
\times 
\sum_{y''_2,\ldots,y''_d} 
n_i(\xvec_2,t;\xvec_0,L)
\sum_{y'_2,\ldots,y'_d} 
n_i(\xvec_1,t;\xvec_0,L)\\
- \ActivityTimeAveSpave(x_2;\xvec_0,L) \ActivityTimeAveSpave(x_1;\xvec_0,L)  \ ,
\end{multline}
with 
$\xvec_1=(x_1,y'_2,y'_3,\ldots,y'_d)$ 
and
$\xvec_2=(x_2,y''_2,y''_3,\ldots,y''_d)$.  
Noticeably, $\xvec_1$ and
$\xvec_2$ have each $d-1$ dashed components which are summed over
independently, with the intention to render $\ActActTimeAveSpave_{c,u}$
correlation functions of spatial averages rather than spatially averaged
correlation functions. This is a matter of choice, motivated by the
independence of the two coordinates in \Eref{def_gamma_dash}. 

Both correlation functions collapse very nicely under suitable
rescaling, similar to \Fref{actact_collapse}. \Fref{actact_collapses_3D}
shows the collapses by plotting
$\ActActTimeAveSpave_{u,c}(x_2,x_1;\xvec_0,L) L^{\mu}$ against
$|x_2-x_1|/L$, \Eref{actact_scaling}, with $\mu=2.92$ for the
unconnected correlation function $\ActActTimeAveSpave_u$ and $\mu=2.79$
for $\ActActTimeAveSpave_c$. Just like in one dimension, we have chosen
$x_1=x_0$, \ie one sheet in the two-point correlation function is the
one where the driving takes place. The exponent $\mu$ should be compared
to $1-\gamma'/\nu_\perp=2.74(3)$ from \cite{Luebeck:2004} or
$2(d+z-D)=2.81(3)$ \cite{HuynhPruessner:2012b}, see the discussion after
\Eref{actact_scaling}.  These two values from scaling relations do not fully agree, but not as
significantly as in one dimension, where they suggest $0.59(4)$ and
$0.38(5)$ respectively. Moreover, the collapses with $\mu=2.92$ and
$\mu=2.79$ are relatively better compatible with these literature values in both
cases (in contrast to $\mu=0.58$ in one dimension, which is much less
compatible with $0.38(5)$ than with $0.59(4)$).  One may wonder that
the unexpectedly inconsistent scaling is a feature of one dimension (see
also \cite{LuebeckHeger:2003a}).

\begin{figure}
\includegraphics[width=0.95\linewidth]{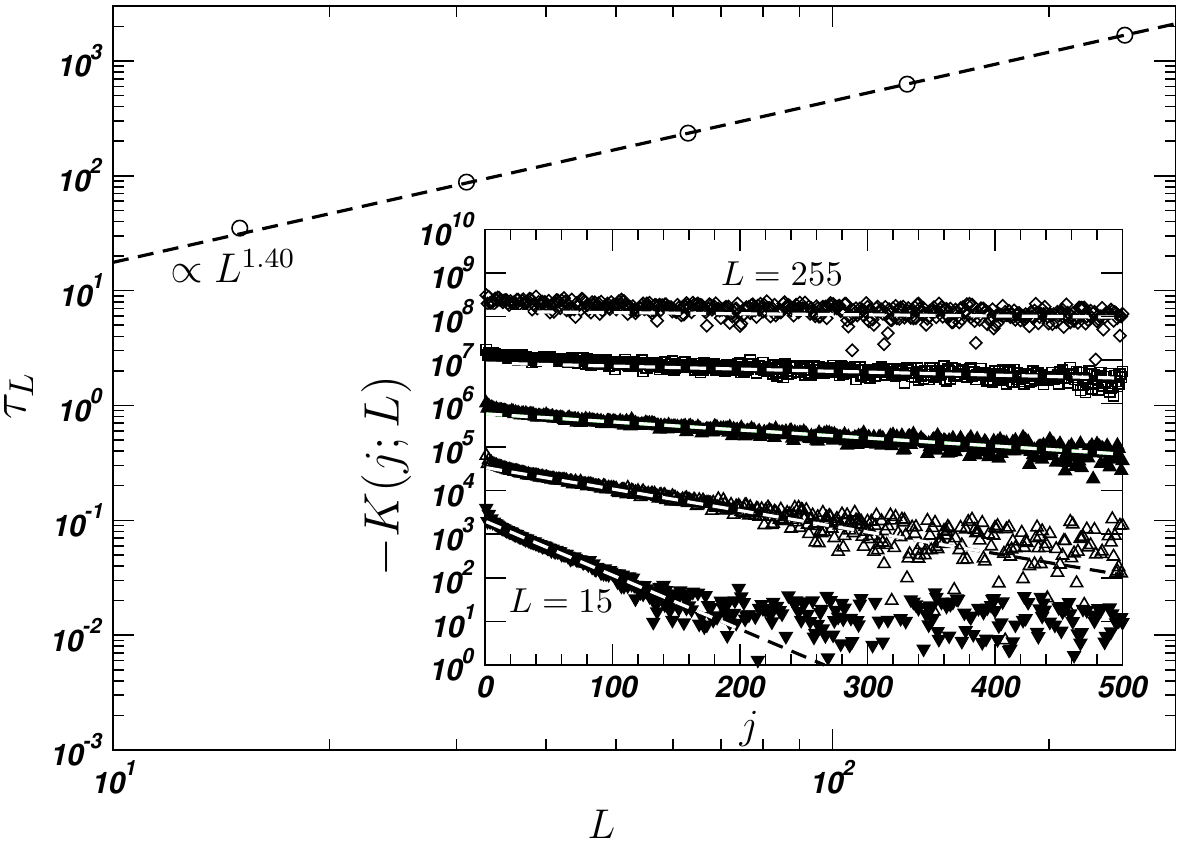}
\caption{\flabel{ava_ava_corrs_3D}
Scaling of the macroscopic avalanche-avalanche correlation time
$\tau_L$ with the system size $L$. The power law fitted (dashed line),
$\tau_L\propto L^{1.40}$ is compatible with the prediction
$\tau_L\propto L^{D-2}$ with $D=3.370(11)$ \cite{HuynhPruessner:2012b}. In three dimensions,
correlation times are very long and rise quickly compared to one dimension (\cf
\Fref{ava_ava_corrs_1D}). The inset shows the negative of the actual
correlation function
$\AvaAvaCorr(j;L)$ for $L=15,31,\ldots,255$ with the exponential fits
(by eye),
$A_L \exp{-j/\tau_L}$, shown as dashed lines.
}
\end{figure}

Finally we consider correlations of avalanche sizes on the macroscopic
time scale. Compared to one dimension, where correlation ``times''
(measured in number of avalanches attempted) are fairly short and do not
increase dramatically with system size, correlations in three dimensions
last much longer and rise significantly. \Fref{ava_ava_corrs_3D} shows
the correlation function (in the inset, \Eref{def_AvaAvaCorr}) and the
estimated correlation times, which change from around $35$ avalanches at
$L=15$ to about $1680$ avalanches at $L=255$. Fitting the correlation
time against the system size results in an exponent of $1.40$, which is
not too far off the expected value of $D-2=1.370(11)$
\cite{HuynhPruessner:2012b} (see discussion after
\Eref{def_AvaAvaCorr}).

\subsection{Five dimensions}
\begin{figure*}
\subfigure[Sheet-averaged density profile in relative coordinates.]{\includegraphics*[width=0.45\linewidth]{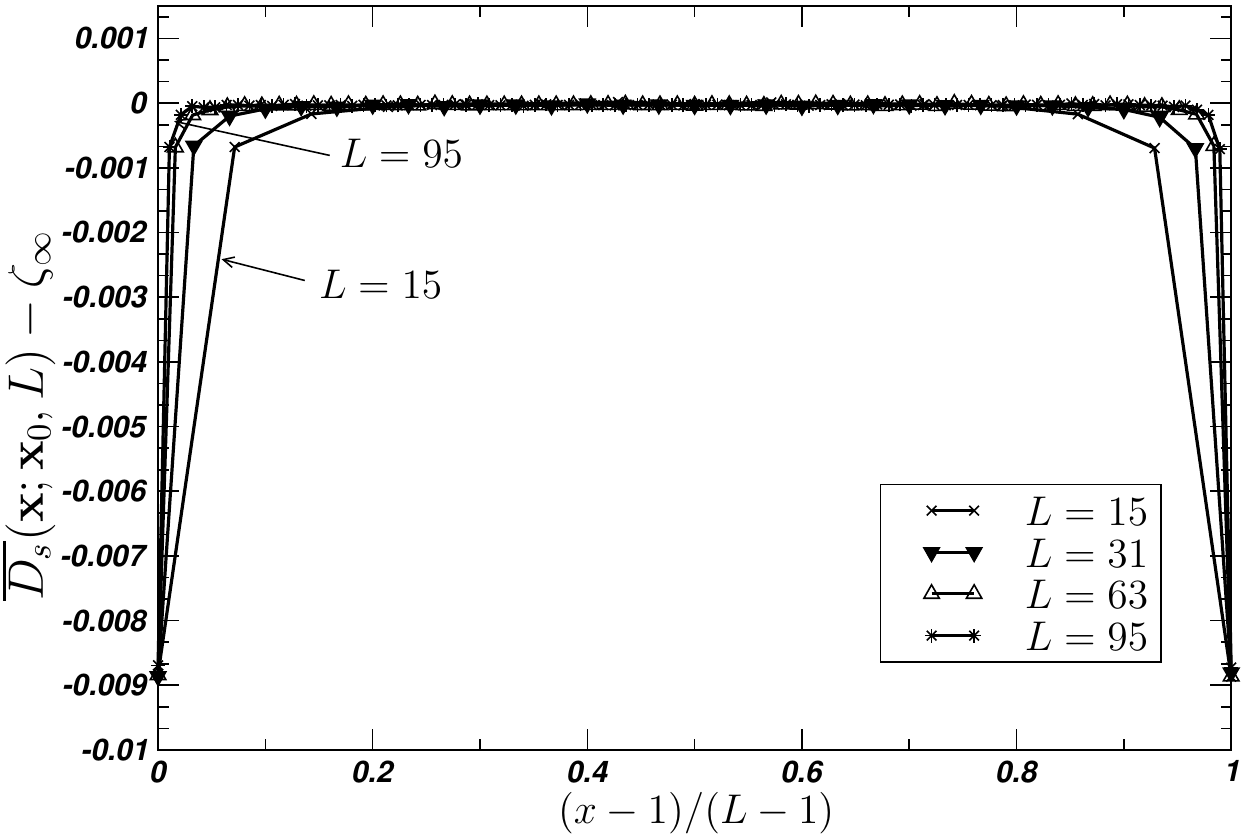}\flabel{substrate_density_5D_relpos}}
\subfigure[Sheet-averaged density profile close to the boundary.]{\includegraphics*[width=0.45\linewidth]{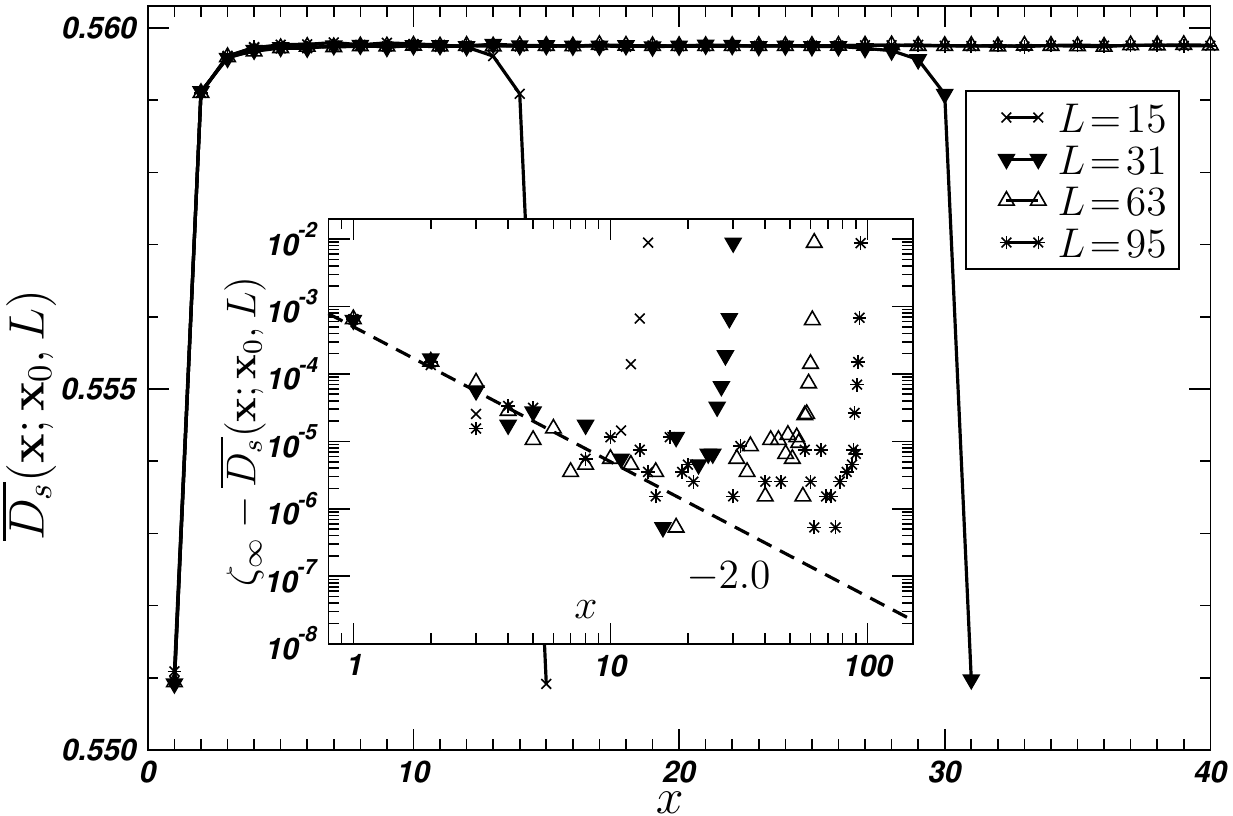}\flabel{substrate_density_5D_boundary}}
\caption{\flabel{substrate_density_5D}The density of inactive particles
in the quiescent state at stationarity in five dimensions as the
deviation from the bulk density $\zeta_\infty=0.559780(5)$, \cf
\Frefs{substrate_density_1D} and \ref{fig:substrate_density_3D}.
As in three dimensions, the observable has been taken as a spatial
average in the periodic direction (sheets of constant $x$). 
\subfref{substrate_density_5D_relpos} The shoulder of the density close to
the boundary is even sharper than in three dimensions, \Fref{substrate_density_3D_relpos}.
\subfref{substrate_density_5D_boundary} Different system sizes show essentially
identical shoulders (spatial scale identical to \Fref{substrate_density_1D_boundaries}
and \Fref{substrate_density_3D_boundary}).
As shown in the inset, we were unable to detect a clear power law decay away from
the boundary of the deviation
of the density from its bulk
value $\zeta_\infty$, because the deviations are minute, the data comparatively
noisy and any estimate for the exponent clearly dependent on the estimate
of $\zeta_\infty$, which we adjusted here to $0.55975$. 
The line for $-2$ is shown for comparison to \Fref{substrate_density_3D_boundary}.
The numerical data is shifted by $1$ on the $x$-axis to produce a cleaner power law.
}
\end{figure*}

Five dimensions are supposedly above the upper critical dimension and so
the Manna Model should display the same asymptotes and, in particular, the
same scaling, as its Mean Field Theory
\cite{LeDoussalWiesePruessner:2016:MFT_unpublished}, where $z=2$ and
$D=4$ \cite{Luebeck:2004}. Because of the high dimensionality, the
linear extent of the systems studied numerically is rather limited. The largest
lattice we used was thus $L=95$ with $95\cdot96^4\approx8\cdot10^9$
sites. With a particle density of $\zeta_{\infty}\approx0.55$ that means
one cannot reasonably expect the system to equilibrate within $10^7$
avalanches when starting from an empty lattice. For these very large
systems, we decided to determine the bulk density on a smaller lattice
(starting from $L=15$, having $983040<10^6$ sites) over the course of $10^8$
avalanches, and use that density uniformly as the initialisation for the next
bigger lattice. In each case, we dismissed $2\cdot10^7$ avalanches as
transient.

As seen already in the sharpening of the density profile from one
dimension, \Fref{substrate_density_1D}, to three dimensions,
\Fref{substrate_density_3D}, in five dimensions, \Fref{substrate_density_5D}, the density across
different four-dimensional sheets of constant $x$ is nearly the same
everywhere in the system. Only around $x=1,2,3$ and $x=L,L-1,L-2$ a
deviation from the bulk density $\zeta_{\infty}$ is actually noticeable
but still is minute.  To determine $\zeta_{\infty}$, we have fitted
$\zeta_{L}$ against $\zeta_{\infty}+a L^{-\epsilon}$ with fitting
parameters $\zeta_{\infty}$, $a$ and $\epsilon$. From $L=15,31,63,95$ we
found $\zeta_{\infty}=0.559780(5)$ (and $\epsilon=0.94(1)$) with a goodness of fit of $0.70$,
which is to some extent owed to the small number of data points and the
comparatively large number of fitting parameters. In the inset of \Fref{substrate_density_5D_boundary}
we used $\zeta_{\infty}=0.55975$ instead, as it produced a more
systematic dependence of the apparent cutoff on the system size. 
In that figure, we show that the small deviations of the substrate density 
from the bulk value close to the boundaries is reproduced for different 
system sizes and may display some power law dependence on the distance 
(similar to \Frefs{substrate_density_1D_boundaries} and \ref{fig:substrate_density_3D_boundary}).
Unfortunately the data is too noisy to extract a reliable estimate of the exponent.
\Fref{substrate_density_5D_boundary} shows exponent $-2$ for comparison.

While the particle density is non-universal and difficult to capture
theoretically, the total activity $\ShapeAvaAbs(t;\xvec_0,L)$,
\Eref{def_ShapeAvaAbs} (\Sref{shape_of_ava_1D}), as shown in \Fref{ShapeAvaAbs_5D} and its normalised form $\ShapeAvaRel(\tau;
\xvec_0,L)$, \Eref{def_ShapeAvaRel} (\Sref{temp_shape_of_ava}), 
as shown in \Fref{ava_rescaled_5D}, display universal scaling, as shown in \Frefs{ava_rel} and \ref{fig:ava_rescaled}, 
in one dimension and in \Frefs{ShapeAvaAbs_3D} and \ref{fig:ava_rescaled_3D} in three dimensions.
The former, $\ShapeAvaAbs(t;\xvec_0,L)$, is based on the collapse of the rescaled activity as a
function of suitably rescaled time, the latter, $\ShapeAvaRel(\tau;
\xvec_0,L)$ on the universal shape of
the activity profile averaged after scaling it to the interval $[0,1]$.

\Fref{ShapeAvaAbs_5D} shows the activity $\ShapeAvaAbs(t;\xvec_0,L)$ as a function of rescaled time
using the exponent $z=2$, \Eref{ShapeAvaAbs_collapse}.
Indeed, no initial slope $\propto t^{(2-z)/z}$ is visible in
$\ShapeAvaAbs(t;\xvec_0,L)$, as expected for $z=2$,
\Eref{ShapeAvaAbs_scaling} (see
\Sref{shape_of_ava_1D}). Accordingly, the collapse occurs
when time is rescaled by $L^2$. The inset shows a comparison to the MFT
as introduced in \Sref{mft}.  While the match is far from perfect, it
shows a noticeable improvement over the data for $d=3$,
\Fref{ShapeAvaAbs_3D}, and the data for $d=1$, \Fref{ShapeAvaAbs}. It could surely be improved further by rescaling
the activity so that it matches, initially, the density of immobile
particles, $\zeta_{\infty}$, as the activity at $t=0$ is exactly the
probability of the initial particle (supplied by the external drive) arriving at an occupied site.  A qualitative difference to the situation
in $d=3$ and $d=1$ (\Fref{ShapeAvaAbs_3D} and \Fref{ava_abs}
respectively) is the fact that the activity drops almost monotonically, apart
from a very slight initial increase. In $d=3$ and $d=1$
it shows a very clear maximum, that exceeds unity in both cases.

\begin{figure}
\includegraphics*[width=0.95\linewidth]{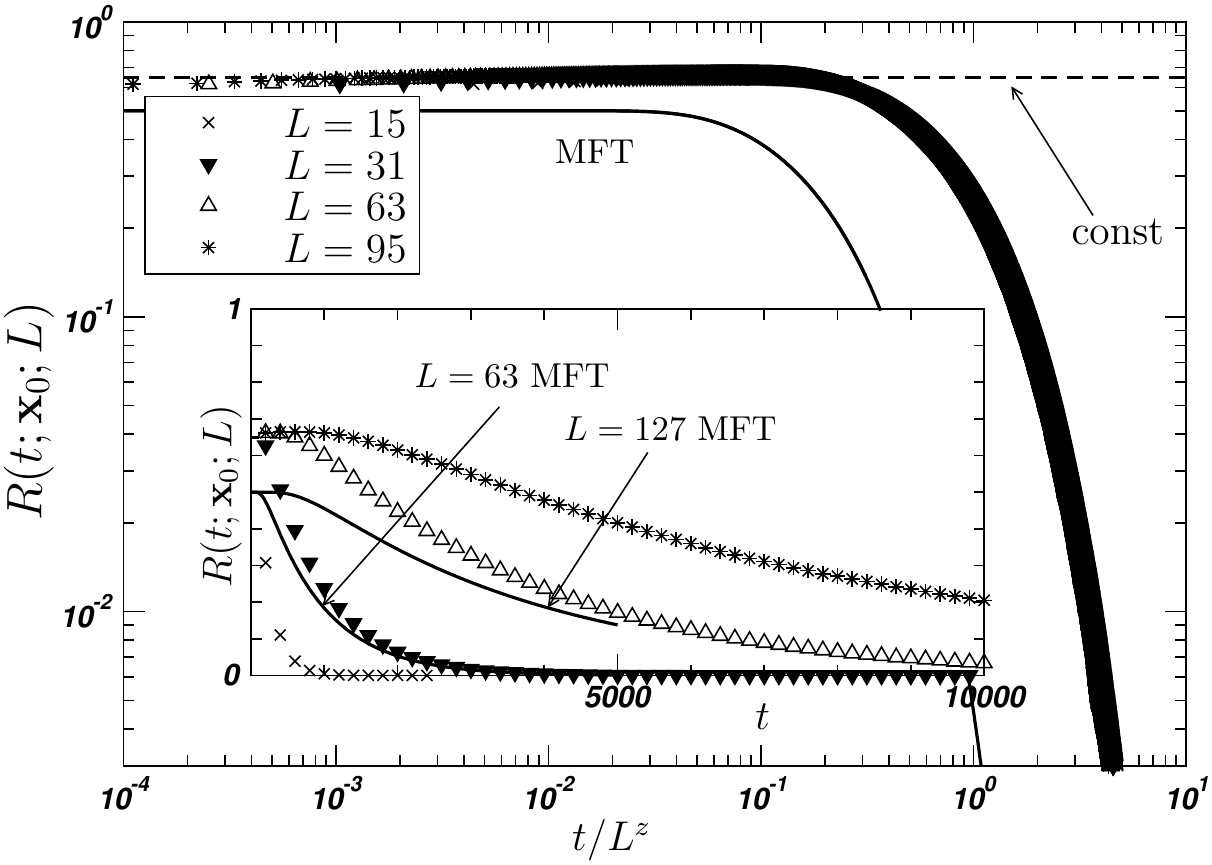}
\caption{\flabel{ShapeAvaAbs_5D}Collapse of the spatially integrated
activity, $\ShapeAvaAbs(t;\xvec_0,L)$, for the centre driven Manna Model
in five dimensions (see \Fref{ShapeAvaAbs} and \Fref{ShapeAvaAbs_3D} for the data in 
one and three dimensions respectively), plotting
$\ShapeAvaAbs(t;\xvec_0,L)L^{2-z}$
against $t/L^{z}$ using the literature value of $z=2$
\cite{Luebeck:2004}. The dashed line shows the
expected slope $(2-z)/z=0$, the full line the collapse ($L=63,127$) of the
MFT (labelled).
The inset shows the (pruned) data on a linear scale together with the MFT as in \Fref{ava_abs}.
}
\end{figure}

\Fref{ava_rescaled_5D} shows a comparison of the (specially normalised)
activity profile $\ShapeAvaRel(\tau;
\xvec_0,L)$ according to \Eref{def_ShapeAvaRel} and the profile
expected from mean field theory, \ie the branching random walk
(\Sref{shape_of_ava_1D}), which are expected to share the same universal shape.
The data for the five-dimensional lattice
is in good agreement with the MFT. Because of the large avalanche duration exponent
$\alpha=2$ and avalanche size exponent $\tau=3/2$ \cite{Pruessner:2012:Book} in five dimensions,
relatively many avalanches are short in time and small in size, so that
the data used for $\ShapeAvaRel(\tau; \xvec_0,L)$ in
\Eref{def_ShapeAvaRel} is mostly rather stepped. That results in visible
artefacts as seen in \Fref{ava_rescaled_5D}, made worse by the fact that
the artefacts align for different system sizes. The situation can be
improved by using Poissonian updating which smears out the steps or by
introducing a cutoff to suppress avalanches below a certain size. 

\begin{figure}
\includegraphics*[width=0.95\linewidth]{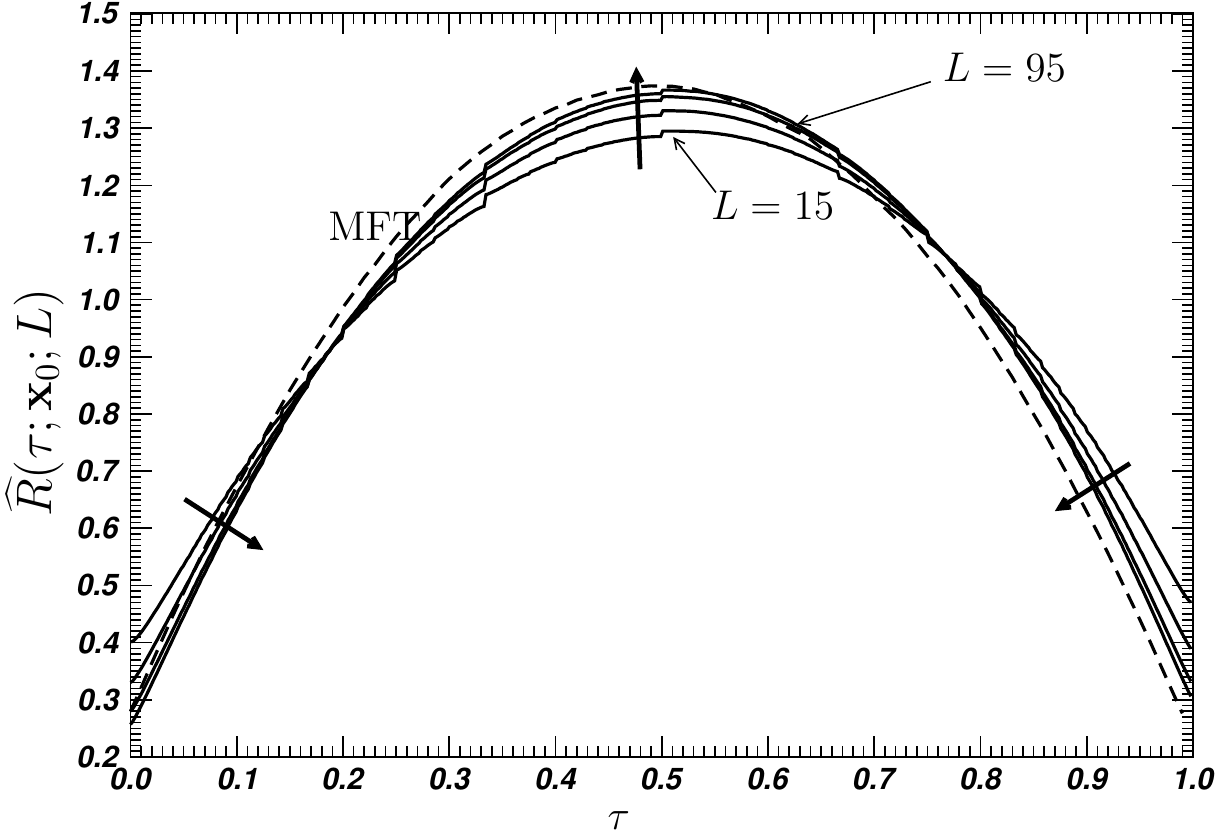}
\caption{\flabel{ava_rescaled_5D} The rescaled activity
$\ShapeAvaRel(\tau; \xvec_0,L)$ in five-dimensional systems of size
$L=15,31,63,95$ driven at the centre, $x_0=(L+1)/2$, to be compared to
the behaviour in one (\Fref{ava_rescaled}) and three dimensions
(\Fref{ava_rescaled_3D}).  The thick arrows point in the direction of
increasing system size.  The rescaling of time maps the spatially integrated
activity to the interval $\tau\in[0,1]$. Normalisation is applied so
that the integral under the curve is unity.  The numerical data for five
dimensions is somewhat ``rugged'' (mostly for small system sizes)
because of the many small avalanches which produce stepped activity
profiles.
 The  mean-field theory is
shown as a dashed line, ($L=255$, see \Fref{ava_rescaled}).
}
\end{figure}

\begin{figure}[b]
\includegraphics[width=0.95\linewidth]{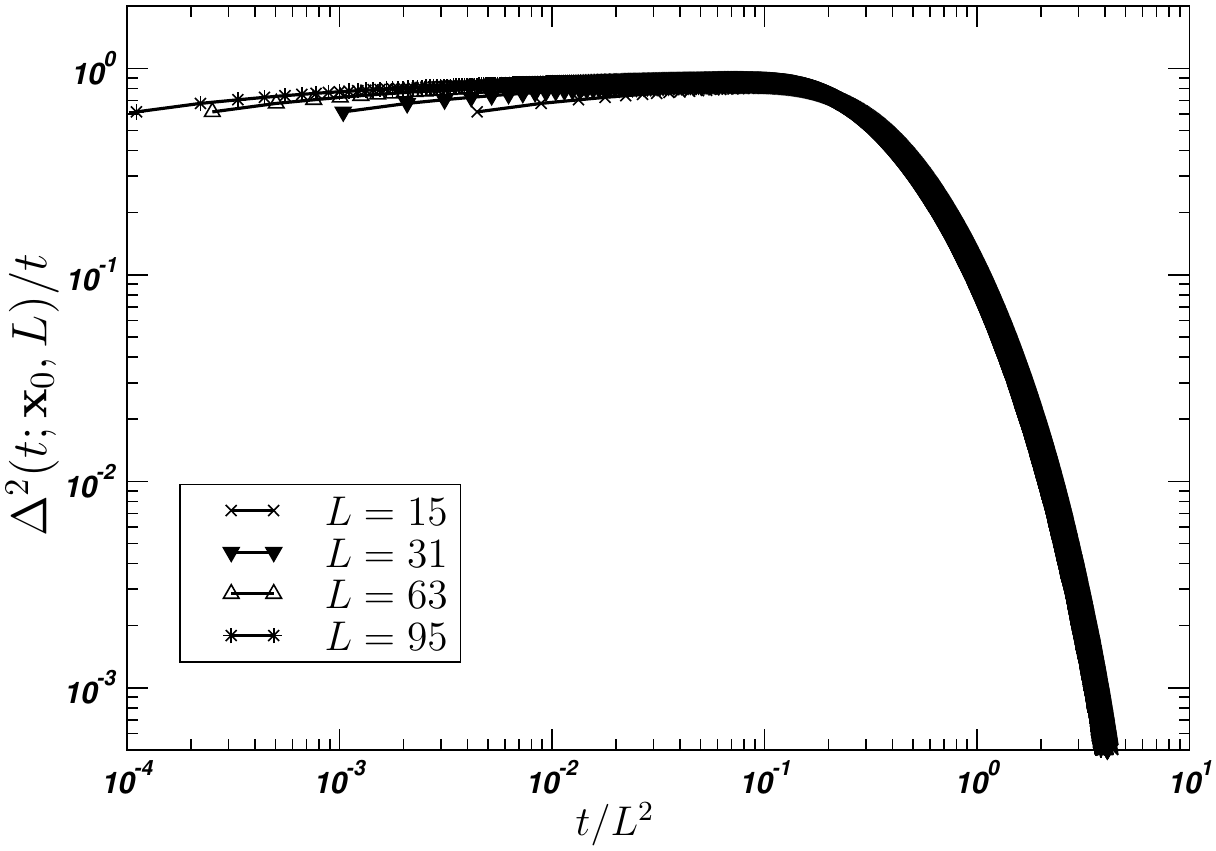}
\caption{\flabel{msd_5D}
Collapse of the width $\msd(t;\xvec_0,L)$ (\Eref{def_msd}) in the five-dimensional, centre driven Manna Model. Data is binned.
}
\end{figure}

\begin{figure*}
\subfigure[Time and sheet-averaged unconnected spatial activity-activity correlation function $\ActActTimeAveSpave_u(x_2,x_0;x_0,L)$, \Eref{def_ActActTimeAveSpave_u}, collapsing for different system sizes $L$ according to \eref{actact_scaling}.]{\includegraphics*[width=0.45\linewidth]{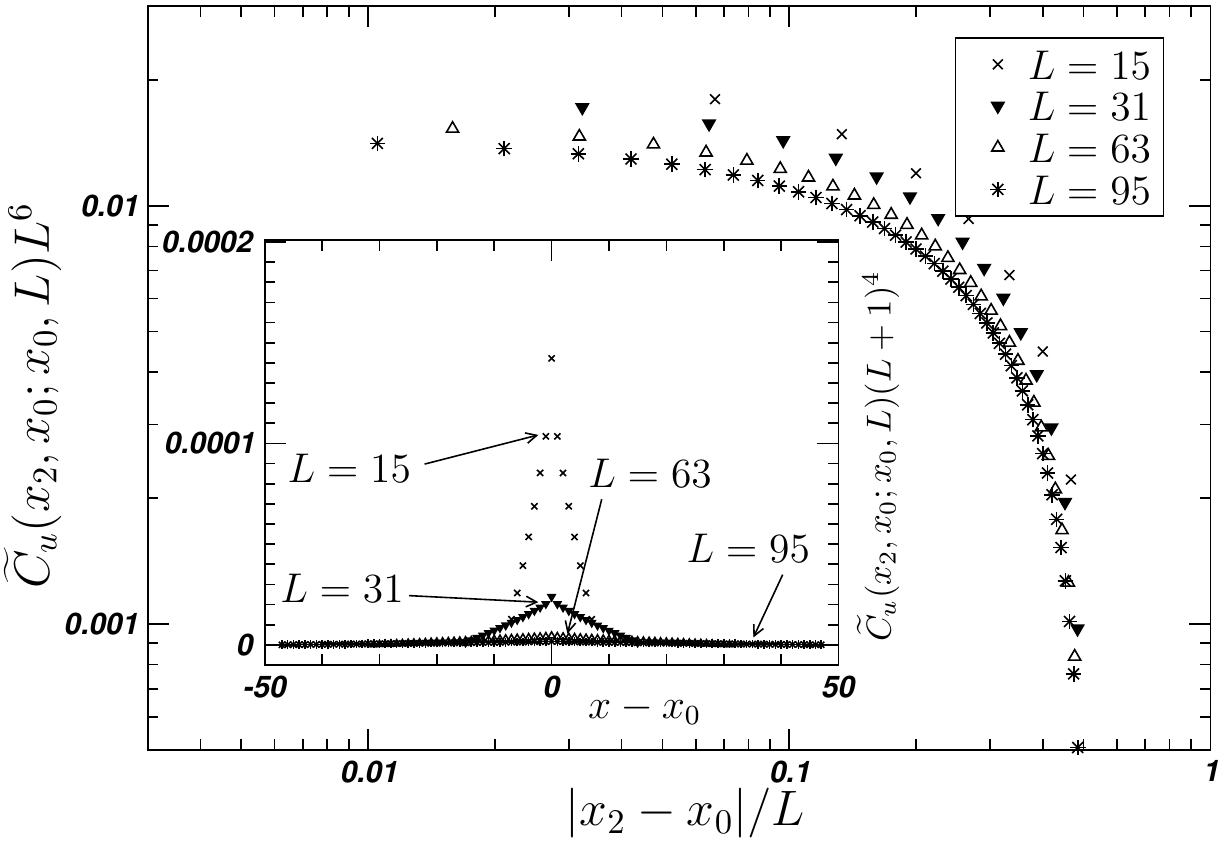}\flabel{actact_u_collapse_5D}}
\subfigure[Time and sheet-averaged connected spatial activity-activity correlation function $\ActActTimeAveSpave_c(x_2,x_0;x_0,L)$, \Eref{def_ActActTimeAveSpave_c}, collapsing for different system sizes $L$ according to \eref{actact_scaling}.]{\includegraphics*[width=0.45\linewidth]{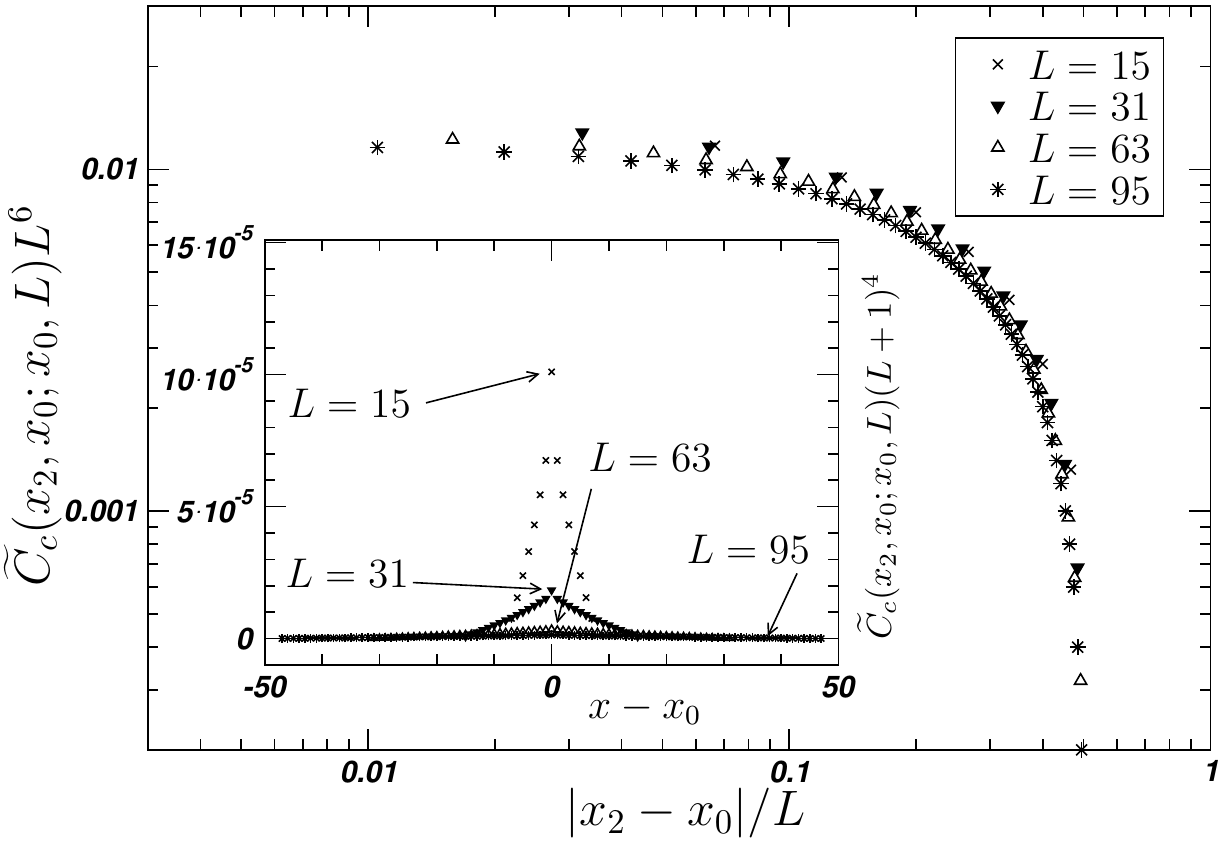}\flabel{actact_c_collapse_5D}}
\caption{\flabel{actact_collapses_5D} Collapses according to
\Eref{actact_scaling} of the two point spatial correlation function
$\ActActTimeAveSpave_u$ and $\ActActTimeAveSpave_c$ as defined in
\Erefs{def_ActActTimeAveSpave_u} and \eref{def_ActActTimeAveSpave_c}
respectively for different five-dimensional systems as indicated.  The
exponent $\mu=6$ is the theoretical value (see main text).
The insets
show the correlation function on linear scales, however rescaled by
the size of a sheet, $(L+1)^4$, so that different system sizes $L$ can
be shown simultaneously.  See also \Fref{actact} and
\Fref{actact_collapses_3D}.  }
\end{figure*}

Comparing the data in \Fref{ava_rescaled_5D} to that in
\Fref{ava_rescaled_3D} suggests that the mean-field theory works better
or applies at least as well in three dimensions as in five.  One
possible explanation is that system sizes in five dimensions are much
smaller and so the coincidence between mean field theory and data in
five dimensions would actually be significantly better, if bigger system
sizes were available. Inspecting the
data in \Fref{ava_rescaled_3D} suggests that the data may display
convergence away from the mean-field curve, whereas the data in
\Fref{ava_rescaled_5D} is still compatible with a convergence towards
it. However, even the mean-field data is based on a comparatively small
lattice ($L=255$, see \Sref{shape_of_ava_1D}) and might change slightly
as bigger lattices are considered. As in three dimensions, there is a slight
slant in opposite directions in the MFT data and the data from the Manna
Model.
Plotting $\ShapeAvaRel(1-\tau; \xvec_0,L)$ against the MFT
significantly improves the collapse. 

We expected the mean squared displacement $\msd(t;\xvec_0,L)$,
\Eref{def_msd}, to be almost perfectly linear in time $t$ in five
dimensions, corresponding essentially to the behaviour of a random
walker, so that the data collapses under rescaling of time by $L^2$ and
$\msd(t;\xvec_0,L)$ by $t$, \Fref{msd_5D}. The data in three dimensions,
\Fref{msd_3D}, displayed a near-perfect collapse and very clear
asymptotics. It turns out, however, that the data for the
five-dimensional system, \Fref{msd_5D}, is not as clear and clean as in
three dimensions, although it collapses nicely under the rescaling
expected. Two issues might play an important r{\^o}le: Firstly, the
linear extent of the lattices, which effectively limit the maximum
mean-square displacement from above and the finite size corrections from
below. These are small in five dimensions (maximum $L=95$) compared to three
(maximum $L=255$). Secondly, the exponents used in the collapse in five
dimensions are the analytical values expected. Allowing them to deviate
slightly from that (thus becoming effective exponents for the given
range of system sizes) improves the quality of the collapse by
compensating for some of the finite size corrections, something that was
not needed in three dimensions (possibly because of the larger linear
extents used there), which was based on the literature value.

Finally, \Fref{actact_collapses_5D} shows a collapse of the unconnected
and the connected activity-activity correlation functions,
\Eref{def_ActActTimeAveSpave_u} and \eref{def_ActActTimeAveSpave_c}
respectively. Again, the
collapse is based on the analytical value of the exponent $\mu=2(d+z-D)$
(see \Eref{actact_scaling} and the discussion thereafter), which gives
$\mu=6$ because $z=2$ and $D=4$ at $d=5$. This is not equivalent to
$\mu=d-\gamma'/\nu_\perp$ which gives $\mu=d$ as $\gamma'$ is expected
to vanish for all $d\ge d_c=4$ \cite{LuebeckHeger:2003b}. The reason for the mismatch is that in dimensions above
the upper critical dimension, the variance of the total activity no
longer follows $\Delta \rho_a \propto L^{\gamma'/\nu_\perp}$ with
$\gamma'=0$ as suggested in \Eref{def_gamma_dash}, but, rather, drops
off like $L^{4-d}$, so that $\Delta \rho_a L^{-d} \propto L^{-\mu}$
produces  $\mu=2d-4$ (see also the discussion around
\Eref{def_gamma_dash}). This is because activity is confined to a volume
of size $L^4$ whenever $d\ge4$, while the variance of the activity
density drops like $L^{-d}$, so the space integrated variance is $\Delta
\rho_a \propto L^{4-d}$.

However, the collapses in \Fref{actact_collapses_5D} are surprisingly
poor compared to the three-dimensional counterpart
(\Fref{actact_collapses_3D}). The same qualifications as above apply
(small system sizes in five dimensions and the exponents in five
dimensions not chosen to optimise the collapse as in $d=1$, \Fref{actact}, but not in $d=3$). In the present case
$\mu=6.2$ works best for $\ActActTimeAveSpave_u$ and $\mu=6.1$ works
best for $\ActActTimeAveSpave_c$.  Moreover, one may argue that space is
more accurately rescaled by $|x_2-x_0|/(L-1)$, because $|x_2-x_0|$
ranges from $0$ to $(L-1)/2$.  However, even at the small linear system
sizes used, this makes only a very small visible difference and does not
improve the collapse significantly.

Considering finally the avalanche-avalanche correlations on the
macroscopic time scale in five dimensions, it turns out that correlation
times are extremely long and the signal prohibitively noisy.  It is
virtually impossible to fit an exponential against the correlation
function, even when allowing for very large error bars. After some
window-averaging, the smallest system size $L=15$ gives a correlation
time (measured in number of attempted avalanches) of about $4000$, but
the next larger, $L=31$ has one at around $20000$ avalanches. Data for
$L=63$ does not produce any reasonable results even after taking window
averages. This result is all the more lamentable, as it would have been
interesting to verify the scaling $\propto L^{D-2}$ of the correlation time
above the upper critical dimension. 
This will require, however, many more than the $8\cdot 10^7$ avalanches
that we considered.

\section{Discussion and Conclusion}
\slabel{conclusion}
The key-objective of the present numerical study was to provide an extensive survey of
spatio-temporal correlations in SOC. They are frequently overlooked as the hallmark of criticality, as the vast majority of numerical
studies focuses on integrated observables, such as the space-time integral
of the activity during an avalanche (avalanche size) or its duration. 
As most clearly seen in one dimension, this is to large extent justifiable by the complications that arise from the additional independent variables (such as distances in time and space), the often noisier numerics and the less convincing scaling behaviour, even when modern computing resources can compensate for some of these disadvantages.

On the other hand, much of what ``self-organised criticality'' ought
to display, in particular self-organisation and spatio-temporal
self-similarity \cite{WatkinsETAL:2016}, is very well captured and in
fact confirmed by the correlation functions studied above. Among the
numerical results presented, the independence of the density profile of
the substrate from the driving position, further supported by the
calculations in the appendix, stands out as one of the clearest signs of
self-organisation \cite{GrassbergerDharMohanty:2016}.  
Interestingly, 
correlations in the substrate (\Fref{substrate_correlations_1D}) have a small
amplitude, which does not scale with the system size.
This is
very much in contrast to other models, such as the OFC Model
\cite{OlamiFederChristensen:1992}, where correlations are very visible
and part of the evolution towards a (supposedly) scale invariant steady
state \cite{MiddletonTang:1995,Lise:2002}. 
The lack of correlations is apparently also in sharp contrast to the fixed energy
version of the Manna Model, as Basu \etal refer to 
``the natural long-range correlations in the background''
\cite{BasuETAL:2012} and the hyperuniformity found by Hexner and Levine \cite{HexnerLevine:2015}, which we confirmed above \Sref{substrate_correlations}. In fact, Grassberger, Dhar and Mohanty \cite{GrassbergerDharMohanty:2016}
have successfully used \emph{periodic} initial conditions in the closely related
Oslo
Model to reduce 
transients. In other words, there are
clear indications of correlations, but they are not easily measured in the
density autocorrelation function studied above.

The effect of the boundary on
the substrate particle density further in the bulk is discernible only for
a
small, fixed number of sites (a number of sites that does not increase with
system size), because the amplitude of the deviation from the bulk density does not 
scale in $L$. Yet, there is a visible power law decay of the boundary effect, 
cut off by the system size, see \Frefs{substrate_density_1D_boundaries}, 
\ref{fig:substrate_density_3D_boundary} and \ref{fig:substrate_density_5D_boundary}.
As that data is rather noisy and the amplitude 
does scale up with system size, at a large scale the boundary
will barely
be noticeable in the density of immobile particles in the bulk. 
In contrast,
it is very clearly
visible in the activity, \Fref{TimeIntegratedActivity} and
\Eref{triangular_profile}, which is dissipated at the open boundaries of
the system \cite{HwaKardar:1989a,nonotePaczuskiBassler:2000}. The triangular
shape of the time-integrated activity is certainly not ``guided''
by the almost featureless density of particles in the substrate.  The
activity throughout would surely be affected by a change in substrate
density, but the latter self-organises such that the time integral of
the former displays the behaviour of mere random walkers, which at
stationarity have no sinks or sources in the bulk other than by driving.

That the environment organises visibly only as far as a scalar density is
concerned is different to the seemingly more complicated picture
emerging from experiments, say the ricepile \cite{FretteETAL:1996}, where particles
appear to arrange so a newly added one finds itself in a system that
mediates long range interaction. In the Manna Model, self-organisation
manifests primarily in a scalar density of substrate particles
and during avalanches, when large correlations of active particles arise whose amplitude scales with system size.
One may test this hypothesis further by destroying
the spatial correlations explicitly, while leaving the densities profile
unchanged \cite{Stapleton:2007}.  This is probably best done in
dimensions $d>1$, because of the uniformity of the density in the
periodic directions.

In summary, the self-organisation on the macroscopic time scale affects
almost exclusively the bulk density of immobile particles and does not
seem to result in them being significantly correlated. As far as
self-organisation towards the critical state is concerned, in the
present study we have not encountered significant effects in the system, other
than the critical scaling in the active state itself. Even, the algebraic scaling of the
density
away from the boundary,
as shown in \Frefs{substrate_density_1D_boundaries}, \ref{fig:substrate_density_3D_boundary}
and \ref{fig:substrate_density_5D_boundary}, has no discernible effect on
the bulk, even when the correlation length seems to be of order $L$. None
of the features
in the bulk of the substrate (can) have an amplitude that scales
with the system size and therefore may easily be overlooked in natural systems.
Although self-organisation to the
critical density takes place, at quiescence we were unable to detect an unambiguous
fingerprint of criticality.

The substrate particle density displays a noisy power law
decay of the (small) deviation from the bulk value away from the boundaries,
see \Frefs{substrate_density_1D}, \ref{fig:substrate_density_3D} and
\ref{fig:substrate_density_5D}.
The amplitude of that deviation cannot possibly increase indefinitely with
system size and thus may, again, be difficult to detect in real systems 
in particular without a well-defined lattice spacing.
Otherwise, the space-dependence of the substrate particle density
is rather trivial and almost
featureless. One may be tempted to say the same about the time-integrated
response function (\Fref{TimeIntegratedActivity}), although a
characteristic scale, \ie a correlation length, can be (analytically)
identified as the system size and the amplitude scales linearly in $L$. The time-resolved response has some significant spatial
structure (\Fref{response_sliced}, also \Fref{slices}) and the
characteristic length scale, a correlation length, is indeed the system size, as can be seen
vividly in \Fref{TimeIntegratedActivity}. However, without time-dependence,
the response function is that of a random walker with $\eta=0$ and no
renormalisation of the diffusion constant.

The scaling of the correlation function of the activity, on the other
hand, confirms the picture that spatial correlations build up over microscopic time
(such as $t\propto |x-x_0|^z$ in \Fref{full_collapse} and $t^*(L)$ in
\Fref{actact_time_sliced}) but are eventually curbed by the system size
$L$ (\eg inset of \Fref{actact_time_sliced}, \Fref{actact_collapse}). As
a result, the scaling that is ultimately (in the long-time limit)
displayed by the Manna Model is finite size scaling. For the correlation
functions, this result is far from trivial --- as the process takes place on
the lattice, there is no dimensional reason for the correlation length
not to be a non-trivial power of the system size. Nevertheless, the
correlation length (\ie the decay length of the correlation functions)
is indeed proportional to the system size (as demonstrated in
\Fref{actact}), so that
standard finite size scaling applies.

For quantities such as the time-integrated response
$\TimeIntegratedActivity(\xvec;\xvec_0,L)$,  the scaling with
$(x-x_0)/L'$ can be determined analytically,
\Erefs{triangular_profile} and \eref{triangular_profile_35}.  This
is a general theme: the scaling in space of the response is to
large extent determined by the particles' trajectories being random
walks. No such mapping exists for higher order correlation functions.
Scaling in time, on the other hand, is generally non-trivial. For
example, the mean squared displacement $\msd(t;\xvec_0,L)\propto
t^{2/z}$ (for $t\ll L^z$, see discussion after \Eref{def_eta_z})
displays strongly superdiffusive behaviour as activity ceases only at
low local (immobile) particle densities, \ie where it has not ceased in the
past. Only above the upper critical dimension, where $z=2$, does the mean
squared displacement slow down to reach $\msd(t;\xvec_0,L)\propto t$, \Fref{msd_5D}.
In that sense, the temporal behaviour is richer than the
spatial behaviour.

On the other hand, higher correlation functions (technically,
three-point correlation functions) display a highly non-trivial
spatial structure, \eg \Fref{actact}, \Fref{actact_collapses_3D} and
\Fref{actact_collapses_5D}. 
The presence of spatio-temporal scaling of the
activity (including its amplitude) is reassuring and has to be seen in contrast to the absence of
finite size
scaling of the 
amplitude of the
(weak) anti-correlations in the (immobile) substrate
particles, \Fref{substrate_correlations_1D}, which are bounded just as the
density is.
Summarising these
findings, one might be tempted to generalise that the substrate, the
``backdrop'' of the activity, is not the best place to search for
correlations and scaling in SOC. 
At least our direct measurements of the one-point and two-point correlation functions 
did not reveal any scaling of the amplitude with system size.  
In the Manna Model, interesting
behaviour is most easily found in the activity.

We have also considered temporal correlations on the macroscopic timescale,
not least in order to validate and generalise the mapping of SOC to
interfacial growth \cite{PaczuskiBoettcher:1996}. This seems to work
well in one and three dimensions, but in five dimensions was hampered by
the enormously long correlation times. If valid, the correlation time
between avalanche sizes on the macroscopic time scale scales like
$L^{D-2}$, \ie avalanches separated by $\propto L^{D-2}$ avalanche
attempts may be considered as independent. As the average number of
topplings scales like $L^2$ and determines the CPU-time needed to
perform the avalanches, the average amount of CPU-time for independent
events scales like $L^D$, to be considered when selecting observables
(possibly in the presence of self-averaging
\cite{FerrenbergLandauBinder:1991,Pruessner:2012:Book}).

The most concerning result above is the significant
deviation from the literature value of the dynamical exponent $z$ in one
dimension, as estimated in the collapses above.  Worse, some of the
exponents are inconsistent even for the same observable. The problem is
confined to one dimension, with data for $d=3$ producing all the
expected behaviour, while $d=5$ shows some deviations, likely to be
caused by finite size effects.  In
the following, we focus on the problems in one dimensions and discuss possible
explanations.

The dynamical exponents in one dimension that we found were $z=1.51$ for
the collapse of the spatially integrated activity (\Fref{ava_rel}), but
also $z=1.574$ from the slope in small $t$ in the same plot (which however assumes
constant $\FCtilde_0$, see discussion at the end of
\Sref{shape_of_ava_1D}). Further, we found $z=1.5625$ from the scaling
of the mean squared displacement (\Fref{msd}), $z=1.445$ to $z=1.59$ for
the scaling of the response propagator in time at fixed spatial
coordinates (\Fref{slices}), $z=1.59$ (but not $z=1.445$) for the
scaling of the response in space at short times (\Fref{full_collapse})
and finally $z=1.71$ if $\lambda^{-1}=z$ for the spatial scaling of the
unconnected correlation function in \Fref{actact_time_sliced}. The
scaling of the time averaged correlation function can be related to the
dynamical exponent through two different scaling relations which are not
fully consistent. Choosing the one based on SOC observables (see
discussion after \Eref{alternative_scaling_relation}) gives
$z=\mu/2+D-d=1.543(14)$ for the data shown in \Fref{actact_collapse} and
using $D=2.253(14)$ from \cite{HuynhPruessnerChew:2011}. The alternative
scaling relation $\mu=d-\gamma'/\nu_\perp$ produces $\mu=0.59(4)$
\cite{Luebeck:2004},
validating the scaling found in \Fref{actact_collapse}, but seems to
clash with $\mu=2(d+z-D)$.

To acknowledge the significance of the disparity of the dynamical
exponents in one dimension, it is worth comparing to three dimensions,
where results are quite consistent: The spatially integrated
activity collapses nicely with the literature value $z=1.777(4)$
(\Fref{ShapeAvaAbs_3D}) even for the slope in small $t$, as does the
mean squared displacement (\Fref{msd_3D}). The average  spatial
activity-activity correlation function shows some variability,
\Fref{actact_collapses_3D}, as $z=\mu/2+D-d=1.830(11)$ for the
unconnected correlation function (\Fref{actact_u_collapse_3D}, $\mu=2.92$) and
$z=\mu/2+D-d=1.765(11)$ for the connected on
(\Fref{actact_c_collapse_3D}, $\mu=2.79$), based on $D=3.370(11)$ in the literature
\cite{HuynhPruessner:2012b}. The collapse of the unconnected correlation
function is not as good as for the connected one and may generally be
expected to display slightly worse scaling. In all, it seems that the
numerics in three dimensions validate the methods and the scaling
proposed. Further support comes from the numerics in five dimensions, if
one accepts that the scaling was spoilt to some extent by finite size
corrections.

The results in one dimension (\ie the value of the dynamical exponent
$z$) are worrying firstly because they clash with the literature values
obtained in traditional SOC simulations. This is put in perspective by
the variety of results reported, ranging from 
$1.393(37)$ \cite{LuebeckHeger:2003a}
to
$1.66(7)$ \cite{DickmanETAL:2001}. 
However, all literature values known to us, except $z=1.445(10)$ in \cite{HuynhPruessnerChew:2011} were taken in 
variants of the Abelian Manna Model; 
$z=1.47(7)$ \cite{NakanishiSneppen:1997} was obtained in the non-Abelian version of the Manna Model,
$z=1.66(7)$ \cite{DickmanETAL:2001} in the FES version,
$z=1.45(3)$ and $z=1.54(5)$ \cite{DickmanTomedeOliveira:2002} in different versions of the height-restricted FES version,
$z=1.393(37)$ \cite{LuebeckHeger:2003a} in the non-Abelian version
and
$1.50(4)$ \cite{Dickman:2006} in the FES version with height restriction.
We would expect that as a matter of universality, the exponents found
for these models should coincide with the measurements of correlation
functions obtained above.
In fact, some traditional SOC observables such as the avalanche size are
``coarse grained'' variations of observables used above, such as the
spatial integral of the activity, $\ShapeAvaAbs(t;\xvec_0,L)$.

Secondly, as seen above, estimates for the dynamical exponent in one dimension found in the present work are inconsistent across different
observables.  This is particularly worrying in cases when the
exponent is not measured by finding a collapse whose quality is
difficult to assess, but when it is determined by measuring an
``obvious'' slope. Such a slope may, however, be obscured by the presence of a
lower cutoff and a scaling function
\cite{ClausetShaliziNewman:2009,DelucaCorral:2014,ChristensenETAL:2008}.
In case of $\ShapeAvaAbs(t;\xvec_0,L)$, shown in \Fref{ava_rel}, a good
collapse was obtained using $z=1.51$. A less good one was obtained
from $z=1.445$.  On the other hand, the initial slope ought to be
$(2-z)/z\approx0.325$ for $z=1.51$, but turns out to be much
closer to $0.27$, suggesting $z\approx1.574$. Although the exponents characterising the slope are small, close inspection
leaves little doubt about the latter. A similar clash of the dynamical
exponents used in the collapse versus the one used in the initial
time-dependence was \emph{not} found for the width of the propagator
\Fref{msd}.

To reconcile these results, one may revisit the scaling form
\Eref{def_eta_z} and relax it to
\begin{multline}
\Activity(\xvec, t; \xvec_0,L) = a |\xvec-\xvec_0|^{-(d-2+\eta+z')} \\
\times \FC\left(\frac{\xvec-\xvec_0}{L^{\sigma}},\frac{\xvec-\xvec_0}{b t^{1/z}}\right)
\ ,
\elabel{def_eta_z_relaxed}
\end{multline}
with additional exponents $z'$ and $\sigma$.
The time integral gives $\TimeIntegratedActivity(\xvec;\xvec_0,L)$,
whose scaling is known exactly, \Eref{triangular_profile}, $\propto
x^{2-d} \FCtilde(x/L)$ leading to
\Eref{time_integral_to_argue_vanishing_eta} with $\eta=0$.  The
corresponding result for the form \eref{def_eta_z_relaxed} is
$\eta+z'=z$ and $\sigma=1$, reproducing \Eref{def_eta_z}. The
pre-factor of
$L^{2-z}$ in \Eref{ShapeAvaAbs_collapse} then follows from the
spatial integral of \Eref{def_eta_z_relaxed} which gives
$-(d-2+\eta+z')+d\sigma=2-z$.  As for the scaling in small
$t\ll L^z$, integrating \Eref{def_eta_z_relaxed} in space to produce
$\ShapeAvaAbs(t;\xvec_0,L)$ gives necessarily the scaling form
\Eref{ShapeAvaAbs_scaling} with $\eta=0$,
\[
\ShapeAvaAbs(t;\xvec_0,L) = a \left(\frac{t}{b}\right)^{(2-z)/z}
\FCtilde_0\left( \frac{t}{b' L^z} \right) \ .
\]
The only possible reason why $\ShapeAvaAbs(t;\xvec_0,L)$ scales in small
$t$ not as $t^{(2-z)/z}$ is that $\FCtilde_0(\ldots)$ is itself a power law.
That, however, implies that $\lim_{L\to\infty}
\ShapeAvaAbs(t;\xvec_0,L)$ is not finite, \ie either diverges or
vanishes, which cannot possibly happen, because the total activity
across the entire system at fixed time must be monotonically increasing
in $L$, yet is bounded from above (for example by $2^t$).

The exponent describing the initial slope is rather small, we found
$(2-z)/z\approx 0.27$ as shown in \Fref{ava_rel}, whereas the collapse
was done with $z=1.51$, so that $(2-z)/z \approx 0.325$ as if
$\FCtilde_0(u) \propto u^{-0.055}$. However, the numerics giving
$(2-z)/z\approx 0.27$ is very reliable, suggesting that one should much
rather consider $z=2/1.27\approx1.575$ as the ``correct'' value of $z$
That value of $z$ produces a good collapse in small $t$, but a
comparatively poor one in the tail. The picture thus remains
dissatisfying: As far as the collapse \Fref{ava_rel} is concerned, the
powerlaw of the initial slope $\propto t^{(2-z)/z}$ cannot be made
consistent with the exponents needed for a satisfactory collapse.  Both
exponents are larger than what we think is the most reliable literature
value, $z=1.445(10)$, which is close to the conjectured value of $z=10/7=1.42857\ldots$ 
for the Oslo Model \cite{GrassbergerDharMohanty:2016}.

Based on our data we conclude that the scaling form \Eref{def_eta_z}
must be suffering from very significant corrections.  One might wonder
whether this is a matter of, say, the assumption of translational
invariance mentioned after \Eref{def_eta_z}.  This assumption could be
relaxed by allowing $x_0/L$ (in one dimension) as an argument of the
scaling function. However, $x_0/L=1/2+1/(2L)$ was essentially constant
for the different $L$ considered.

A more daunting explanation for the poor consistency of the exponents is
the definition of the time scale.  We have repeated some of the
simulations in one dimension using a Poissonian waiting time between
topplings and found
that the picture does not change significantly. While a collapse like
\Fref{ava_rel} works well with $z=1.48$ (\cf $1.51$ above), the initial slope still
suggests $z=1.59$ (\cf $1.575$ above).

Another observable that, in one dimension, displays unexpectedly poor
scaling with exponents from the literature is the width of the
propagator $\msd(t;\xvec_0,L)$, \Eref{def_msd}, as shown in
\Fref{msd_plots} (with $z=1.5625$).  Like the total activity
$\ShapeAvaAbs(t;\xvec_0,L)$, this is essentially a spatial integral of
the propagator.  In this case, the initial slope, $\propto t^{2/z}$,
which features very clearly, and the exponent to collapse can be chosen
consistently.  However, $z=1.5625$ is well away from the expected value
of $z=1.445(10)$. In fact, further inspection suggests that the collapse
and the match of the initial slope may possibly be further improved by taking $z$ as large as $z=1.59$.
An exponent of $z=1.445$ looks very poor in comparison, certainly for
the initial slope, which is more clearly visible for $\msd(t;\xvec_0,L)$
in \Fref{msd} than for $\ShapeAvaAbs(t;\xvec_0,L)$  in \Fref{ava_rel}. 

A possible explanation for the inconsistencies with the literature
values for the exponents are the corrections that were allowed for in
the latter, but are difficult to capture in a collapse.  According to
\Eref{msd_scaling} $\msd(t;\xvec_0,L) L^2$ collapses when plotted
against $t/L^z$. Instead of taking any specific $z$, one may use an
estimate of the characteristic time scale $T_c(L)$, which scales like
$L^z$ only to leading order, to rescale time by $1/T_c(L)$ rather than
$1/L^z$. The characteristic time scale is proportional to the moment
ratio $\ave{T^2}/\ave{T}$ (the second moment of the duration over the
first). A collapse is therefore expected by plotting $\msd(t;\xvec_0,L)
L^2$ against $t\ave{T}/\ave{T^2}$.

However, there is no improvement of the collapse in comparison to using
$L^z$ with $z=1.445$.  Even when considering only very small system
sizes, $\ave{T^2}/\ave{T}$ clearly scales with an exponent of less that
$z=1.5$, while the collapse clearly needs a dynamical exponent larger
than $z=1.55$.

In comparison with three dimensions, the one dimensional collapses
display many inconsistencies.  It was argued by L{\"u}beck and Heger
\cite{LuebeckHeger:2003a} that the Manna universality class splits in one
dimension into two distinct ones, at least as far as absorbing state
phase transitions are concerned. Basu \etal \cite{BasuETAL:2012}
suggested that such fixed energy sandpiles (FES) belonged ``generically to
[the] directed percolation'' universality class, although they studied
in fact only the Manna Model and only in one dimension.  Interestingly,
Lee \cite{Lee:2013} pointed out that the observations made by Basu \etal
are confined to one dimension. In two dimensions, the scaling of the FES
Manna Model clearly differs from that of directed percolation. The
original claim by Basu \etal was based on the
observation that 
under improved numerical conditions
five exponents studied ($\alpha$, $\beta$,
$\nu_{\perp}$, $\nu_{\parallel}$ and $z$) 
were closer to directed percolation than
previously reported in the literature \cite{Luebeck:2004}.\footnote{Closer inspection reveals, however, that both $\alpha$ and
$\nu_{\parallel}$ had already been reported \cite{LuebeckHeger:2003a} within one standard
deviation of
those by Basu \etal, \ie there has not been a claim that $\alpha$ or
$\nu_{\parallel}$
were much different from those in directed percolation.} Their finding of $z=1.51(5)$ is within the range
of some of the findings above and remarkably far from their own estimate
of $\nu_{\perp}=1.095(5)$ and $\nu_{\parallel}=1.75(5)$, which gives
$\nu_{\parallel}/\nu_{\perp}=1.60(5)$, supposedly equal to $z$.  One may
speculate whether their \emph{de facto} observation of an inconsistent
$z$ (in one dimension only) is linked to ours.  As far as the dynamical
exponent is concerned, in one dimension many of the findings above for
the Manna Model are not incompatible with directed percolation,
$z=1.580745(10)$ \cite{Jensen:1999}. However, many others are
incompatible. One may speculate whether this is due to an interplay of 
two microscopic timescales, the other one characterising avalanche durations, one
characterising correlations, or the two microscopic length scales,
namely finite distances on the lattice and its size.

Another surprise was the unexpectedly poor scaling in five dimensions.
While the collapses worked (mostly)
with exponents as expected from theory, their quality was not as good as most of
those in three dimensions.  In \Fref{ShapeAvaAbs_5D} the collapse worked
essentially as expected, but \Fref{ava_rescaled_5D} showed visible
artefacts and produced results seemingly further away from MFT than the
corresponding ones in three dimensions, \Fref{ava_rescaled_3D}.
Similarly, the collapse of the width, \Fref{msd_5D}, was good given the
(expected) MFT exponent, yet somewhat disappointing for early times
$t/L^2$. The worst behaviour was found for the correlation function,
\Fref{actact_collapses_5D}, which showed clear deviations from the
expected exponents. As mentioned above, the system sizes are bound to be
very small in five dimensions --- observables like the width are
necessarily bounded from above by the system size, while corrections are
bounded from below.  However, it seems somewhat inconsistent to accept
finite size corrections as an explanation for the poor behaviour in
$d=1$ and $d=5$ in the light of the very convincing results in three
dimensions (which, nevertheless, validates many of the scaling
assumptions).

To put the inconsistencies in perspective, one should keep in
mind that collapses are a comparatively poor tool to extract exponents
and as a result, to some extent also a poor test for scaling. Firstly,
there are the corrections alluded to above. 
While these are well understood for the finite size scaling of individual
moments \cite{Wegner:1972,Barber:1983}, we are not aware of a systematic way
of introducing
and
assessing them in  
a data collapse, say
\begin{multline}\elabel{def_eta_z_again_extended}
\Activity(\xvec, t; \xvec_0,L) = \\
a |\xvec-\xvec_0|^{-(d-2+z)} \FCtilde\left(\frac{t}{b' L^z},\frac{\xvec-\xvec_0}{b t^{1/z}}\right) \\
+ a_2 |\xvec-\xvec_0|^{-(d-2+z+\omega)}
\FCtilde_2\left(\frac{t}{b' L^z},\frac{\xvec-\xvec_0}{b t^{1/z}}\right)
\end{multline}
with $\omega>0$, \cf \Eref{def_eta_z_again}.  Secondly, the range within
which the collapse is supposed to ``work'' is cut off towards the
smaller scale, yet it is difficult to assert the value of that cutoff.
Thirdly, statistical errors are difficult to estimate, other than via
the range of exponents that seem to result in an acceptable collapse
(whatever that may be).  Finally, even a very simple form like
(\Eref{msd_scaling})
\begin{equation}
\msd(t;\xvec_0,L) = a' t^{2/z} \FCtilde_4 \left(\frac{t}{b' L^z}\right)
\ ,
\end{equation}
suggests a collapse of $\msd(t;\xvec_0,L) t^{q-2/z} L^{-qz}$ versus
$t/L^z$ for \emph{any} value of $q$. Technically, different $q$ should
make little difference, but a choice that makes the resulting range of
the ordinate large, will blur displacements in that direction.
Depending on the choice of $q$ some deviations are more readily
identified than others.  For example, $q=0$ (\Fref{msd_collapse}) is
more forgiving for deviations in the tail, whereas $q=2/z$ is more
forgiving for deviations in small $t$ (\cf \Fref{msd}). We used this
choice 
consciously when we collapsed $\ShapeAvaAbs(t;\xvec_0,L)$ in
\Fref{ava_rel},
\Fref{ShapeAvaAbs_3D} and \Fref{ShapeAvaAbs_5D} by rescaling it by a power of $L$ according to
\Eref{ShapeAvaAbs_collapse} (and ``read off''
the $t$-dependence according to \Eref{ShapeAvaAbs_scaling}) rather than collapsing by
\Eref{ShapeAvaAbs_scaling}.

In response to this ambiguity, and in light of the fact that a data collapse is
best regarded as a tool to illustrate and possibly test for scaling, our
initial decision was to plot all collapses with what is expected from
the SOC literature \cite{HuynhPruessnerChew:2011,HuynhPruessner:2012b}.
However, in one dimension, this led to very poor results, which could be
improved easily by using different exponents, namely those shown. In contrast, in three and
five dimensions, we mostly used literature values and the results were
mostly good or satisfactory (mild but clear deviations were visible in
the spatial activity-activity correlation function shown in
\Fref{actact_collapses_3D} and \Fref{actact_collapses_5D}). 

In summary, we see scaling of spatio-temporal correlations confirmed in
the Manna Model of SOC. As far as self-organisation and scaling of
activity and its correlations with system size go, the behaviour is as expected.
In contrast, scaling in the substrate has small, fixed amplitudes and is difficult to detect.
Some results
in one dimension are inconsistent, but this is in line with other findings
for the Manna Model in the FES mode
\cite{LuebeckHeger:2003a,BasuETAL:2012}.

\begin{acknowledgments}
The authors gratefully acknowledge interesting discussions with D. Dhar, N.
Huynh, P. Grassberger and N. Wei.
The authors would 
also 
like to thank Andy Thomas and Niall Adams for
computing support.
\end{acknowledgments}

\newcommand{\bibconferencename}[1]{\textit{#1}}
\bibliography{articles,books}

\appendix
\section{Markov matrices of the one-dimensional Manna Model}
\slabel{appendix_generate_matrices}
In the appendices we determine properties of the distribution of
quiescent or ``relaxed'' configurations of the one-dimensional Manna
Model in the stationary (or steady) state. A configuration is said to be quiescent or
relaxed if none of the sites carries more than one particle. Given that
the particle number at each site is non-negative, there are therefore
$N=2^L$ relaxed configurations in a system with $L$ sites. 

The distribution of quiescent configurations will be captured in a
(column) probability vector $\ket{p}\in\Rset^N$, with each component
$p_i\ge0$ corresponding to the probability to find the system in
configuration $i\in\{1,2,\ldots,N\}$. The evolution of this probability
vector is due to $N\times N$ Markov matrices $a_x$, ``charging'' the system at site
$x\in\{1,2,\ldots,L\}$ and fully
relaxing it. The elements $(a_x)_{ji}$ of these matrices are the
probability with which relaxed configuration $j$ goes over to relaxed
configuration $i$ after charging the system at $x$.  The distribution of
final configurations after charging an ensemble of systems given by
$\ket{p}$ on site $x$ is thus given by $a_x\ket{p}$. The steady state is
described by a distribution $\ket{p_{0,x}}$, not necessarily unique and 
initially expected to depend on $x$,
which is invariant under the application of $a_x$, \ie $a_x
\ket{p_{0,x}} = \ket{p_{0,x}}$.

Further below, we will make extensive use of the equation
$a_x^2=\quarter (a_{x-1}+a_{x+1})^2$ due to Dhar \cite{Dhar:1999a},
\Eref{a_squared_id}, which encapsulates the toppling rules of the Manna Model
by relating the charging of $x$ to
the charging
of its neighbours (subject to boundary conditions). For illustration
purposes, we will occasionally distinguish $a_{x_0}$, the charging at a
particular initial position $x_0$ and $a_x$, the charging at any other
$x$.

Much of the calculations in the present section are more straight forward in
the Oslo Model \cite{ChristensenETAL:1996}, where (in one dimension) two
particles are moved away from the toppling site by moving one particle
to each neighbour (corresponding to a downhill movement of
height-units). In contrast to the Manna Model, in the Oslo Model
avalanches cannot last indefinitely. As we will see, this is related to a
certain nil-potency of matrices (which is not found in the Manna Model)
and thus leads to a simplification of the Markov matrices $a_x$ which
have been determined in the Oslo Model for systems up to size $L=8$
\cite{Corral:2004c}. The nil-potency, however, also has the consequence
of certain configurations being inaccessible
\cite{ChuaChristensen:2002}.

In the following, we briefly outline how the Markov matrices $a_x$ can
be determined in closed form, using computer algebra and some simple
enumeration. As set out above, there are precisely $N=2^L$ relaxed
configurations in a system with $L$ sites. After charging site $x$ the
configuration may no longer be relaxed, \ie the number of particles
residing at site $x$ (and in the course of the avalanche at other sites)
may exceed $1$. We will refer to a configuration that \emph{may} still decay
into a relaxed configuration as an ``excited configuration''.  In a
slight abuse of terminology, the set of excited configurations contains
all relaxed configurations but not \latin{vice versa}.  While there is
in principle no ``height restriction'' in the Manna Model, given we
allow only single charges, there are at most $M=(L+2)^L$ excited
configurations,
because after being charged once, a system cannot
contain more than $L+1$ particles, which may be distributed among the
$L$ sites in all possible ways during the course of an avalanche. In
other words, the $M$ excited configurations are certainly closed under
the evolution of the Manna Model. In fact, $M$ is a rather generous
over-estimate, as many of those $(L+2)^L$ excited
configurations may not be accessible from any relaxed configuration,
because
they contain more than $L+1$ particles.

To construct the Markov matrices $a_x$ (for all $x$ at once),
we will first construct the rectangular $N\times M$ matrix $E$
which contains the probabilities with which each configuration (mostly,
however, non-relaxed ones) decays to a particular relaxed one. The
entries of $a_x$ are those $N$ \emph{columns} of $E$ (which has $M$
columns) which correspond to initial configurations that are relaxed
configurations except for one additional particle added at site $x$. For
example, to extract the resulting distribution of configurations after
charging the relaxed configuration $(1,0,1)$ at the first site, one has
to consult the entries for $(2,0,1)$ in $E$.

To construct $E$,
we first introduce the (column) vectors $\ket{e_i}\in\Rset^N$, with
$i\in\{1,2,\ldots,M\}$, whose entries are the probabilities for an
excited configuration $i$ to end up in a particular relaxed
configuration $k\in\{1,2,\ldots,N\}$. These $e_{ki}$ can be determined
implicitly as
\begin{equation}\elabel{implicit_e}
\ket{e_i} = \ket{r_i} + E \ket{\epsilon_i}
\end{equation}
where $E$ is the $N \times M$ matrix that maps each excited
configuration to a relaxed configuration, \ie 
\begin{equation}\elabel{E_by_e}
E=\Bigg( \ket{e_1} \ket{e_2} \ldots \ket{e_M} \Bigg)
\end{equation}
is made up of the $M$ column vector $\ket{e_i}\in\Rset^N$. The vector
$\ket{\epsilon_i}\in\Rset^M$ is a column vector where each entry
$\epsilon_{ji}$ is the probability with which the non-relaxed
configuration $i$ goes over into the non-relaxed configuration $j$, by
toppling, \emph{at least} one active site. The column vector
$\ket{r_i}\in\Rset^N$, $i=1,2,\ldots,M$ is the vector of probabilities
$r_{ji}$ that excited configuration $i$ goes over into relaxed
configuration $j$.  If $i$ is a relaxed configuration then $\ket{r_i}$
and $\ket{e_i}$ have a single non-vanishing entry and $\ket{\epsilon_i}$
vanishes everywhere.  Each relaxation (or ``decay channel'') is to be
accounted for exactly once in \Eref{implicit_e} either by $\ket{r_i}$ or
by $E \ket{\epsilon_i}$. No overcounting ought to take place, even when
$\ket{r_i}$ and $E \ket{\epsilon_i}$  both evolve excited configuration
$i$ to a relaxed configuration; $\ket{r_i}$ does it directly while $E
\ket{\epsilon_i}$ does it via non-relaxed configurations not considered
in $\ket{r_i}$.  To simplify accounting, $\ket{r_i}$ may account for
relaxations to a relaxed configuration in exactly one step (or none,
namely when $i$ is relaxed already) and $\ket{\epsilon_i}$ for the
transition to another non-relaxed configuration in exactly one step.

Combining \Eref{implicit_e} and \Eref{E_by_e}
gives the implicit equation
\begin{equation}\elabel{E_implicit}
E=R+E\EC
\end{equation} 
and thus
\begin{equation}\elabel{inverse_expression}
E=R(\ident - \EC)^{-1}
\end{equation}
with
\begin{equation}
R=\Bigg( \ket{r_1} \ket{r_2} \ldots \ket{r_M} \Bigg)
\end{equation}
the $N \times M$ matrix of relaxations of $M$ excited configurations to $N$
relaxed configurations directly (via relaxation of at least one site) and 
\begin{equation}
\EC=\Bigg( 
\ket{\epsilon_1} \ket{\epsilon_2} \ldots \ket{\epsilon_M}
\Bigg)
\end{equation}
the $M \times M$ matrix of transitions from one excited configuration to
another excited configuration (by relaxation of at least one site).  
One may read \Eref{E_implicit} as indicating that the decay of any
configuration into a relaxed one happens either within one step ($R$) or
by the decay of a configuration that has evolved by one step ($\EC$).

If
$R$ is easy to populate and all decay channels are considered at once,
then $\EC$ vanishes and $E=R$. In general, however, this is not the case
and both $R$ and $\EC$ contain single topplings (mostly, as $R$ may
contain entries corresponding to no toppling at all).  Because the
probabilities involving single relaxations are integer multiples of
$1/4$, the matrices $\EC$ and $R$, which are easily determined by
automated enumeration, can be represented as (sparse) matrices
containing integers, preceded by a factor $1/4$. Once $R$ and $\EC$ are
determined for a given system size in such an \emph{exact} enumeration
scheme, a
computer algebra system (such as Mathematica
\cite{Mathematica:10.0.2.0}) can derive $E$ in exact form and extract
all $a_x$.
Other successful methods to characterise the transition matrices, 
in particular for the Manna Model by Sadhu and Dhar \cite{SadhuDhar:2009}, can be found 
in the literature \cite{Corral:2004c}.

In the following we discuss a number of numerical implementation
details.  Given that memory requirements become a significant constraint,
the matrices are best kept small, which can be achieved by considering
only those excited configurations, which are actually encountered in the
decay channels of every singly charged (initially relaxed)
configuration.  In other words, in a numerical implementation, all $N$
relaxed configurations are generated first and charged at the $L$
different sites to generate precisely $LN/2$ excited configurations
(namely those $L$ times $N$ configurations which carried $1$ particle at
the site that is being charged, which is the case for precisely $N/2$ of
the configurations).  Those excited configurations must be part of the
following considerations, as are all excited configurations that appear
as intermediate configuration in the various decay channels. 

These initially excited configurations are placed on a stack, which in
the following contains one entry for each excited configuration not
considered yet.  In addition, a lookup-table of all possible excited
configurations is maintained.  That list contains an index for each
excited configuration that indicates their position in the matrices,
with $-1$ signalling that the configuration has not been considered yet.
The indices for the relaxed configurations are most easily determined by
interpreting their binary representation as the occupation; for example
the index $3$ indicates the first two sites occupied and the rest empty.
However, in general, configurations are best represented as an $L$ digit
number with base $L+2$ (namely up to $L+1$ particles per site). The
first two sites singly occupied therefore translates to $1+(L+2)$.
Because not all excited configurations will be generated, not all
indices are used and therefore maps are needed from indices to
configurations and vice versa.

For simplicity and to reduce memory requirements, the size of the
matrices is determined first by relaxing all $LN/2$ initially excited
states and all excited states appearing in their decay channels. Only
then the matrices are allocated and populated, by repeating this
process, as described in the following.

Taking an element off the stack of unprocessed excited states, it is
processed by, say, finding the leftmost excited site, generating three
(two for boundary sites) excited states, and putting them on the stack
if they have not been considered already. Repeating this process until the
stack is empty provides the total number $M'\le M$ of excited
configurations to be considered. 

After reserving memory for the $M'$ excited configurations (or, rather,
for the matrices of size $M'\times M'$ ), the matrices described above can be
generated. They are populated by repeating the process above (filling
the stack with singly excited configurations and updating those) using
the lookup-table described above to determine rows and columns of $R$
and $\EC$, which in an actual implementation may better be realised as
parts of a bigger, joint matrix, as they share the number and indexing of the columns. After determining and outputing those
two matrices, $E$ can be calculated in closed form using
\Eref{inverse_expression} in a computer algebra system. 

In principle, one could generate the relevant matrix at the time of
determining the indices (each configuration on the stack has been
assigned an index at the time it enters the stack, so in principle, at
the time of determining possible new configurations, all their indices
are known). However, the memory to be reserved for the matrix $M'\times
M'$ is significantly smaller than for $M\times M$, so $M'$ should be
determined first. This is not a computationally costly exercise.

Further code is needed to extract the correct data to compile the Markov
matrices $a_{x_0}$ for each of the driving sites $x_0$.  In line with
the notation above, each \emph{row} of the Markov matrix corresponds to
a particular target configuration, \ie the probabilities to make the
transition from $i$ to $j$ is stored in row $j$ and column $i$. This is
the same for $E$. The Markov matrix for a particular driving
site $x_0$ is compiled column by column or line by line,\footnote{Most
easily 
done so that each line corresponds to a particular
\emph{initial} state, which requires some transpose operations.} half of
which contain a single entry (unity) for a transition from one relaxed
configuration to another relaxed configuration, as the driven site $x_0$
is empty.
The other half contain the entries from
the matrix $E$ for those excited configurations, which actually feature
as those reached from driving any of the relaxed configurations at site
$x_0$.  All other elements of $E$ may be thought of as ``stepping
stones'' to compile the entries that make up the Markov matrix
$a_{x_0}$.

Finally, some more code may be needed to determine observables from the
eigenvectors of the Markov matrix, for example a suitable density
matrix, that translates each relaxed configuration to an occupation for
each site.

The procedure above may appear rather cumbersome, in particular in
comparison to similar procedures for, say, the Oslo Model
\cite{ChristensenETAL:1996,Dhar:2004,Corral:2004c}. The reason for the
extra complication, embodied in \Eref{inverse_expression} is the
appearance of decay channels of arbitrary duration in the Manna Model
for any $L>1$: In principle a pair of particles might move back and
forth indefinitely, therefore requiring the implicit determination of
decay probabilities.  This cannot possibly happen in the Oslo Model,
where particle transport is deterministic (even when the decision
whether or not it takes place is stochastic) and so each system size has
a finite maximum avalanche size.  In the Oslo Model, determining the
Markov matrices is therefore a ``mere'' counting exercise, which may be
tedious, but is finite. The Manna Model necessitates the solution of an
additional set of linear equations, \Eref{inverse_expression}.

To see the difference between Oslo and Manna Model mathematically
we note that the excited-excited relaxation matrix $\EC$ is
nilpotent in the Oslo Model, \ie a finite number of relaxations
produces a relaxed configuration. This is not the case in the Manna Model. The
problem is vividly expressed in \Eref{inverse_expression}, as 
\begin{equation}
(\ident - \EC)^{-1} = \sum_{i=0}^\infty \EC^i
\end{equation}
assuming convergence. In the Oslo model, the right hand side contains a
finite number of terms, because $\EC$ is nilpotent.

\section{Eigenstates}
\slabel{appendix_eigenstates}
In the following 
we carry on with the characterisation of operators and their eigenvectors in 
the one-dimensional Manna Model.
We will call an eigenvector $\ket{e}$ of $a_x$ with
eigenvalue unity, $a_x \ket{e} = \ket{e}$,
an \emph{eigenstate}. 
An eigenstate is thus a distribution of configurations that is invariant under
the action of $a_x$.
By normalisation, a row of unities is a left eigenvector with eigenvalue
unity of any Markov matrix, and a corresponding right eigenvector
exists. 
We will call a ``joint eigenstate'' any eigenstate that is common to all 
operators $a_x$.

As prominently pointed out by Grassberger, Dhar and Mohanty \cite{GrassbergerDharMohanty:2016}
(also \cite{SadhuDhar:2009}), 
the Abelianess (in particular the commutation property) of the $a_x$ guarantees
that a joint eigenstate exists.
For many applications, it may be enough to know of the existence of
a joint stationary state, that is the same eigenstate independent of the site
driven. However,
eigenstates may be degenerate and different ones reached depending on the
site driven and the initial condition.
Yet, by the Perron-Frobenius theorem, the eigenstate is \emph{unique}
provided all
final (recurrent) states are ``accessible'' from every initial state; otherwise the set of
all final states may be decomposable into disjoint subsets. As shown explicitly
in
\Aref{appendix_unique_estat} accessibility is 
easily demonstrated for any global, random drive operator $a$ \cite{Dhar:1999b},
driving randomly with finite probability on every site,
as defined
in \Eref{def_a}. As any such operators commute, their eigenstates are 
thus identical and unique.

However, accessibility is much more difficult to
demonstrate for an individual, single $a_{x_0}$, \ie if driving takes place
at only one site $x_0$. In \Aref{appendix_joint_eigenstates},
we show that all eigenstates of $a_1$ and $a_L$ are common to all operators,
including any random drive, which has a unique eigenstate. It follows that 
$a_1$ and $a_L$ have the same unique eigenstate as any random drive.
The proof does not hinge on accessibility by $a_1$ and $a_L$, but on the operator equation 
\Eref{a_squared_id}, due to Dhar \cite{Dhar:1999a} as mentioned above.

That accessibility is a non-trivial hurdle is shown in 
\Aref{appendix_degeneracy}, as the eigenstates of $a_x$ are \emph{not unique} 
if $L$ is odd and $x$ is even. The argument presented applies to other models, 
such as the Oslo Model \cite{ChristensenETAL:1996}. Consequently, the (degenerate) eigenstates of 
some operators do not coincide with those obtained for random drive, but depend
on initial conditions.

\subsection{Joint eigenstates}
\slabel{appendix_joint_eigenstates}
The key insight (to be proved in the following) is that every eigenstate
of $a_1$, that is the Markov matrix (or operator) controlling the
evolution of the system after charging the first site once (the left-most
site, $x_0=1$) is also an eigenstate of all other operators $a_{x_0}$ (which is
what we call a ``joint eigenstate'') 
\cite{Willis:2015}. The argument
can obviously be inverted to demonstrate the same for $a_L$ for driving
the right-most site. What makes these two sites, $x_0=1$ and $x_0=L$,
special are the boundary conditions. To show that every eigenstate of $a_1$
and $a_L$ is an eigenstate of all $a_{x}$ goes beyond demonstrating that there
is a common eigenstate of all $a_{x}$, because the uniqueness of the latter
implies the uniqueness and thus the identity of the eigenstates of
$a_1$, $a_L$ and the common one. In other words, the unique eigenstate that is common to all $a_x$ 
is the \emph{only} eigenstate of $a_1$ and $a_L$ and so boundary drive always
arrives at the same stationary state as any global drive, irrespective of
initial conditions.
As shown in \Aref{appendix_degeneracy}, in general, the same does not apply
to $a_{x_0}$ for ${x_0}\notin\{1,L\}$, 
\ie the eigenstates of, say, $a_2$ may be degenerate and thus may \emph{not} coincide with the unique, common one, even
when the numerics (\Fref{degeneracy_lifting}) suggests that \emph{asymptotically} the density profile at stationarity is independent of the driving position.
All that follows 
for the eigenstates of $a_x$
from their Abelianess and
the uniqueness of the eigenstate of random drive, is 
that one particular linear combination of their eigenstates coincides with
the unique, common one.

Squared Markov
matrices of the Manna Model have the important property \cite{Dhar:1999a}
\begin{equation}\elabel{a_squared_id}
a_x^2 = \quarter \left(
a_{x-1}+a_{x+1}
\right)^2 \ ,
\end{equation}
for all $x\in\{1,2,\ldots,L\}$ with boundary conditions
$a_0=\ident=a_{L+1}$, identities, as charging the system outside the boundary sites
$x=1$ and $x=L$ leaves it invariant. Abelianess of the Manna Model means
that $a_x$ and $a_y$ commute, so that the crossterm of
\Eref{a_squared_id} may be written as $2a_{x+1}a_{x-1}$.

In the following, we will show that if $\ket{e}$ is an eigenstate of
$a_{x-1}$ and $a_x$, \ie for charging the left two sites, then it is
also an eigenstate of $a_{x+1}$, the rightmost site of the three consecutive ones at
$x-1$, $x$ and $x+1$. A proof by induction then starts by considering an
eigenstate of $a_1$ and using $a_0=\ident$.

If $a_{x'} \ket{e} =\ket{e}$ for both  $x'=x-1$ and $x'=x$, then it
follows from $a_x^2 \ket{e} = \ket{e}$ and
\eref{a_squared_id} that
\begin{equation}
\ket{e} = \quarter \left( \ket{e} + 2 a_{x+1} \ket{e} + a_{x+1}^2 \ket{e}\right) \ .
\end{equation}
If we define $\ket{\delta}$ such that $a_{x+1} \ket{e} = \ket{e}
+\ket{\delta}$, \ie the deviation
of $a_{x+1} \ket{e}$ from $\ket{e}$, we have $a_{x+1}^2
\ket{e}=a_{x+1}(\ket{e}+\ket{\delta}) =
\ket{e}+\ket{\delta}+a_{x+1}\ket{\delta}$ and therefore
\begin{equation}
a_{x+1}\ket{\delta} = - 3 \ket{\delta} \ ,
\end{equation}
\ie $\ket{\delta}$ is either an eigenvector of $a_{x+1}$ with eigenvalue
$-3$ or it vanishes. Because $a_{x+1}$ is a Markov matrix, its spectrum
is bounded by the unit circle and it follows that
$\ket{\delta}$ vanishes and thus $a_{x+1} \ket{e} = \ket{e}$, \ie
$\ket{e}$ is an eigenstate of $a_{x+1}$.  In summary, if $a_{x'}
\ket{e}=\ket{e}$ for $x'=x-1$ and $x'=x$, then $a_{x+1}
\ket{e}=\ket{e}$. By induction it follows that if $a_{x'}
\ket{e}=\ket{e}$ for $x'=0$ and $x'=1$ then $a_{x'} \ket{e}=\ket{e}$ for
$x'\in {0,\ldots,L+1}$.  The base case is easily established for $x'=1$
and $\ket{e}$ an eigenstate of $a_1$, because $a_0\ket{e}=\ket{e}$
follows trivially from $a_0=\ident$. This concludes the proof.

The proof can obviously be applied ``in reverse'' to demonstrate that an
eigenstate $\ket{e}$ with $a_L\ket{e}=\ket{e}$ must be an eigenstate of
all $a_{x'}$. This is no longer possible if boundary conditions are
modified, for example to reflecting ones, $a_{L+1}=a_L$, so that
\begin{equation}\elabel{a_squared_id_reflecting_BC}
a_L^2 = \quarter \left(
a_{L-1}+a_{L-1}
\right)^2 = a_{L-1}^2 \ .
\end{equation}
One might think that means the proof above no longer
applies in the presence of reflecting boundaries, because they imply that
\Eref{a_squared_id} does not apply for $x=L$.  However, to prove that
$a_{x} \ket{e}=\ket{e}$ for $x\in\{0,\ldots,L\}$ (no longer including
the ``irrelevant'' site $x=L+1$) \Eref{a_squared_id} is only ever
invoked for $x\le L-1$, \ie the modification of \Eref{a_squared_id} for $a_L^2$,
\Eref{a_squared_id_reflecting_BC}, never enters. 
It follows that all eigenstates of $a_1$ are eigenstates of all operators.
The proof can even be
generalised to systems with anisotropy, but not to those with ballistic
motion, \ie when all particles moved during a toppling are moved to one
side only, which makes perfect sense, as that dynamics excludes
evolution of sites upstream, so charging there or charging downstream
produces a different sequence of configurations and thus a different
eigenstate.

A similar proof is available for the Oslo Model
\cite{ChristensenETAL:1996}
where
$a_x^3=a_{x+1}a_x a_{x-1}$ \cite{Dhar:2004} replaces \Eref{a_squared_id}
above, again in the presence of Abelianess. To arrive at a statement
about the independence of the steady state from the driving or the
uniqueness of the joint eigenstate \cite{GrassbergerDharMohanty:2016} 
by invoking Perron-Frobenius,
one has to
consider accessibility, as done below for the Manna Model.

\subsection{Uniqueness of eigenstate}
\slabel{appendix_unique_estat}
Above, we have shown that 
every eigenstate of $a_1$ and $a_L$ remains
invariant under the application of any of the
operators $a_x$ (for any $x\in\{1,2,\ldots,L\}$). 
However, we have also shown that \emph{any} eigenstate of $a_1$ and $a_L$ is an eigenstate of all $a_x$. 
This is specific to $a_1$ and $a_L$. We have no proof of that property for any other operator $a_x$. 
In the following, we show that the eigenstates of $a_1$ and $a_L$ are unique, which follows from them being joint eigenstates of all $a_x$.
In \Aref{appendix_degeneracy} we will demonstrate that eigenstates of $a_x$
for even $x$ are degenerate if $L$ is odd, so that not all of their eigenstates are also joint eigenstates.

In the following, we will consider the operators $a_{x}$ in a linear
combination to make them
amenable to the Perron-Frobenius theorem in its simplest form: if $a$ is
an irreducible Markov matrix, \ie there exists a power $k_{fi}>0$ for
each initial configuration $i$ and each final configuration $f$, such
that $(a^{k_{fi}})_{fi}>0$, then the eigenvector with eigenvalue $1$ is
unique \cite{GrimmettStirzaker:1992}.

If $\ket{e}$ is an eigenstate of all operators $a_{x_0}$, with
$x_0\in\{1,2,\ldots,L\}$, then 
\begin{equation}
\elabel{def_a}
a=\sum_{x_0=1}^L p_{x_0} a_{x_0} \ ,
\end{equation}
which is
the operator of ``random drive'' with probabilistic weights $p_{x_0}>0$ so that
$\sum_{x_0} p_{x_0}=1$, has obviously also that
eigenstate, $a\ket{e}=\ket{e}$. The Markov matrix $a$ is 
what is referred to above as random drive and
what is being
studied in the following. 
We will show that its eigenstate is unique, by demonstrating accessibility explicitly. Because every joint eigenstate is necessarily an eigenstate of $a$, it follows that joint eigenstates (simultaneous eigenstates of all $a_x$) are unique. Because all eigenstates of $a_1$ and $a_L$ are joint eigenstates, it follows that $a_1$, $a_L$ and $a$ have the same, unique eigenstate.

To determine the positivity of the entries of a positive power of $a$, all that matters is whether
a decay channel exists, that connects initial configuration $i$ and
final configuration $f$ within a finite number of relaxations, which
therefore occur with a finite probability.\footnote{If some final states
are accessible only from some particular initial states (and thus not
all are recurrent), it may happen
that several distinct stationary states exist, which may be reached
depending on initialisation or, given not all are recurrent, are accessed
depending on the initial (random) sequence of charges.}
By demonstrating that such a channel exists for each of the $N^2$ pairs
of initial and final states, we also demonstrate that all are recurrent.
In the following, we may
consider very unlikely decay channels, yet that suffices for the
argument. Unlike the Abelian Sandpile \cite{Pruessner:2012:Book}, not all
recurrent states appear with the same frequency.

The relevant decay channels are easily constructed explicitly: Any initial
quiescent configuration containing in total $n_i$ particles can be emptied by
applying $a$ repeatedly and choosing the decay channel whereby each site
containing a particle already is charged and the resulting pair of
particles moved to a boundary until it is dissipated by leaving the
system.  This procedure requires $n_i$ charges, \ie $a^{n_i}$ contains
entries indicating that the empty configuration is obtained 
with finite probability
starting
from a configuration containing $n_i$ particles.
To reach any configuration with $n_f$ particles from there, $a$ is
repeatedly applied and the ``decay'' channel is chosen whereby each site
to be occupied is charged, requiring $n_f$ further charges. It takes
therefore never more than $n_i+n_f$ charges to go from one quiescent
configuration to another, \ie $k_{fi}=n_i+n_f$ and
$(a^{n_i+n_f})_{fi}>0$. Given that $n_i=L$ is a unique state and $n_f=L$ from
$n_i=L$ therefore reached trivially,\footnote{Of course, we cannot allow
$k_{fi}=0$ as a valid number of applications of $a$ to go from $i$ to $f$, 
but from the completely full lattice $n_i=L$ the completely
full $n_f=L$ is accessed within two charges, namely by emptying one site and
refilling it. The same argument applies to to $n_i=0=n_f$.} the maximum number of charges to access a particular quiescent
configuration from another, given, quiescent configuration is $2L-1$ for $L>1$
(it is $2$ for $L=1$).

It follows that $a$ is irreducible and thus, by Perron-Frobenius, has a
unique eigenvector with eigenvalue unity.\footnote{In fact, one can
show, taking a route via the completely filled lattice,
that $a$ for $L>1$ is primitive (there is a single power $k=2(L+2)$ such
that $(a^k)_{fi}>0$ for all $i$ and $f$), so that periodic behaviour can
be excluded as well, \ie all eigenvalues other than $1$ have magnitude
strictly less than unity. For $L=1$ that is not the case, as is easily
seen by the periodic behaviour of that system.} By construction,
\Eref{def_a}, one eigenstate of $a$ is known, namely the 
eigenstate $\ket{e}$ of $x_0=1$ or $x_0=L$ 
studied in \Aref{appendix_joint_eigenstates}, which turned out to be a joint eigenstate 
such that $a_{x} \ket{e} = \ket{e}$ for all
$x\in\{1,2,\ldots,L\}$. 
With the present accessibility argument, we know that any such joint eigenstate $\ket{e}$ is unique
(there exists at most one joint eigenstate)
and that
it is also the eigenstate of random drive, in
particular uniform drive.  Because all $2^L$ states are accessible for
$a$ from all initial $2^L$ states, the eigenstate $\ket{e}$ has strictly positive elements (all states
recur with positive frequencies). 

This concludes the proof that there is exactly one stationary state $\ket{e}$ that is reached by either driving only at site $x_0=1$, or only at site $x_0=L$ or randomly throughout the lattice, \Eref{def_a}. In the next section we consider the question whether the same can be said about driving only at site $x_0=2,3,\ldots,L-1$. It turns out, that this is not the case. There are certain $x_0$ (namely even $x_0$ in lattices with odd $L$), that reach different stationary states depending on initial conditions. 
Their degenerate eigenstates form a subspace such that
the eigenstate $\ket{e}$ of $a_1$, $a_L$ and $a$ is only one particular linear combination of their eigenstates. In other words, driving at these sites $x_0$ generally leads to a different stationary state. However, as far as density profiles are concerned, in reasonably large systems, these different stationary states are numerically indiscernible, as shown below.

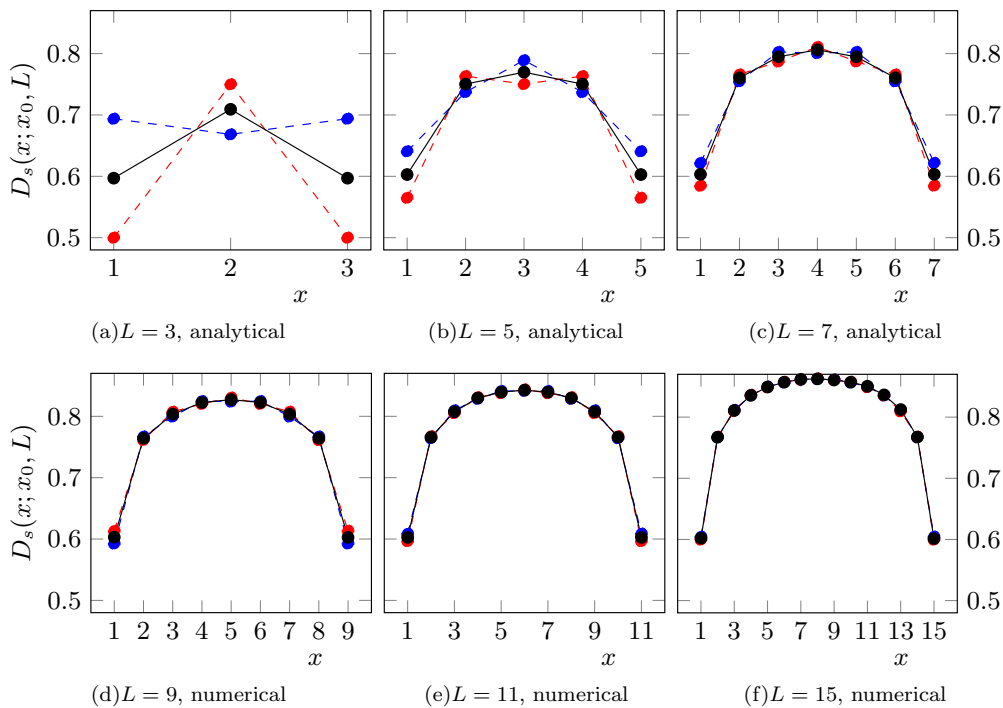
\begin{figure*}
\subfigure[$L=3$, analytical]{\resizebox{!}{4cm}{\begin{tikzpicture}[scale=2.5]
\begin{axis}[
xlabel={$x$},
ylabel={$\Ds(x;x_0,L)$},
xtick={1,...,3},
x label style={at={(axis description cs:0.75,0.05)},anchor=north},
y label style={at={(axis description cs:0.2,0.5)},anchor=south},
height=4.5cm,width=5cm,
ymin=0.48, ymax=0.87]
 \addplot[color = blue, mark = *,dashed] table[x=Pos,y=dens] {L3_DS1_EVEN.txt};
 \addplot[color = red, mark = *,dashed] table[x=Pos,y=dens] {L3_DS1_ODD.txt};
 \addplot[color = black, mark = *] table[x=Pos,y=dens] {L3_DS0.txt};
\end{axis}
\end{tikzpicture}}}
\subfigure[$L=5$, analytical]{\resizebox{!}{4cm}{\begin{tikzpicture}[scale=2.5]
\pgfplotsset{yticklabel style={draw=none}}
\begin{axis}[
xlabel={$x$},
yticklabel style={draw=none},
yticklabel style={xshift=5ex},
xtick={1,...,5},
yticklabels={,,},
x label style={at={(axis description cs:0.8,0.05)},anchor=north},
y label style={at={(axis description cs:0.2,0.5)},anchor=south},
height=4.5cm,width=5cm,
ymin=0.48, ymax=0.87]
 \addplot[color = blue, mark = *,dashed] table[x=Pos,y=dens] {L5_DS1_EVEN.txt};
 \addplot[color = red, mark = *,dashed] table[x=Pos,y=dens] {L5_DS1_ODD.txt};
 \addplot[color = black, mark = *] table[x=Pos,y=dens] {L5_DS0.txt};
\end{axis}
\end{tikzpicture}}}
\subfigure[$L=7$, analytical]{\resizebox{!}{4cm}{\begin{tikzpicture}[scale=2.5]
\begin{axis}[
xlabel={$x$},
yticklabel pos=right,
xtick={1,...,7},
x label style={at={(axis description cs:0.8,0.05)},anchor=north},
y label style={at={(axis description cs:0.2,0.5)},anchor=south},
height=4.5cm,width=5cm,
ymin=0.48, ymax=0.87]
 \addplot[color = blue, mark = *,dashed] table[x=Pos,y=dens] {L7_DS1_EVEN.txt};
 \addplot[color = red, mark = *,dashed] table[x=Pos,y=dens] {L7_DS1_ODD.txt};
 \addplot[color = black, mark = *] table[x=Pos,y=dens] {L7_DS0.txt};
\end{axis}
\end{tikzpicture}}}\\
\subfigure[$L=9$, numerical]{\resizebox{!}{4cm}{\begin{tikzpicture}[scale=2.5]
\begin{axis}[
xlabel={$x$},
ylabel={$\Ds(x;x_0,L)$},
xtick={1,...,9},
ytick={},
x label style={at={(axis description cs:0.8,0.05)},anchor=north},
y label style={at={(axis description cs:0.2,0.5)},anchor=south},
height=4.5cm,width=5cm,
ymin=0.48, ymax=0.87]
 \addplot[color = blue, mark = *,dashed] table[x=Pos,y=dens] {L9_DS1_EVEN.txt};
 \addplot[color = red, mark = *,dashed] table[x=Pos,y=dens] {L9_DS1_ODD.txt};
 \addplot[color = black, mark = *] table[x=Pos,y=dens] {L9_DS0.txt};
\end{axis}
\end{tikzpicture}}}
\subfigure[$L=11$, numerical]{\resizebox{!}{4cm}{\begin{tikzpicture}[scale=2.5]
\begin{axis}[
xlabel={$x$},
xtick={1,3,...,11},
yticklabels={,,},
yticklabel style={xshift=5ex},
x label style={at={(axis description cs:0.8,0.05)},anchor=north},
y label style={at={(axis description cs:0.2,0.5)},anchor=south},
height=4.5cm,width=5cm,
ymin=0.48, ymax=0.87]
 \addplot[color = blue, mark = *,dashed] table[x=Pos,y=dens] {L11_DS1_EVEN.txt};
 \addplot[color = red, mark = *,dashed] table[x=Pos,y=dens] {L11_DS1_ODD.txt};
 \addplot[color = black, mark = *] table[x=Pos,y=dens] {L11_DS0.txt};
\end{axis}
\end{tikzpicture}}}
\subfigure[$L=15$, numerical]{\resizebox{!}{4cm}{\begin{tikzpicture}[scale=2.5]
\begin{axis}[
xlabel={$x$},
yticklabel pos=right,
xtick={1,3,...,15},
x label style={at={(axis description cs:0.8,0.05)},anchor=north},
y label style={at={(axis description cs:0.2,0.5)},anchor=south},
height=4.5cm,width=5cm,
ymin=0.48, ymax=0.87]
 \addplot[color = blue, mark = *,dashed] table[x=Pos,y=dens] {L15_DS1_EVEN.txt};
 \addplot[color = red, mark = *,dashed] table[x=Pos,y=dens] {L15_DS1_ODD.txt};
 \addplot[color = black, mark = *] table[x=Pos,y=dens] {L15_DS0.txt};
\end{axis}
\end{tikzpicture}}\flabel{degeneracy_lifting15}}
\caption{\flabel{degeneracy_lifting}
Density profiles observed in a range of system sizes $L$, for driving at
$x_0=1$ (black, full line) and driving at $x_0=2$ (red and blue, dashed
lines), lines to guide the eye. The only degeneracy we observed was for
odd $L$ and driving at even $x_0$. Analytically, the profiles were
obtained for $L\le7$ by summing over the weights in the eigenstate
vector which correspond to configurations that had a given site
occupied.
Numerically, they were obtained for
$L\ge9$ by driving (typically $10^7$ times) at a given site, starting
from an empty lattice and, in a separate run, from a lattice occupied
initially only at $x_0+1$, thereby enforcing the other parity (see main
text).}
\end{figure*}

\subsection{Degeneracy}
\slabel{appendix_degeneracy}
In the proof above, the uniqueness of the stationary state hinges on the
positivity of the elements of the Markov matrix $a$ raised to some power
$k$. In contrast to random drive \Eref{def_a}, this positivity cannot be
shown for individual $a_{x_0}$.  In the following, we will demonstrate
that certain configurations are inaccessible from certain other
configurations for certain $a_{x_0}$, \ie that some of the $a_{x_0}$ are
decomposable \cite{vanKampen:1992} (the chain not irreducible
\cite{GrimmettStirzaker:1992}). As a result, these operators
have (at least) two distinct, \ie degenerate eigenstates, whose equally
weighted superposition is in fact the unique eigenstate $\ket{e}$
of $a_1$, $a_L$ and $a$, \Eref{def_a},
discussed above.

It is generally very difficult to demonstrate which set of states is
accessible by driving the system only at one particular site (but, as seen above,
very straight forward to show that all states are accessible when the
system is driven randomly at all sites). All we can demonstrate in the
following is that the eigenstate of some $a_{x_0}$ are \emph{at least}
two-fold degenerate (although we have convinced ourselves numerically
that no higher degeneracy occurs). From the proof in
\Aref{appendix_joint_eigenstates} eigenstates
of $a_1$ and $a_L$ are common to all $a_{x_0}$ and from
\Aref{appendix_unique_estat} we also know
that this eigenstate is unique, \ie we know that $a_1$ and $a_L$ have
unique eigenstates, so they they do
\emph{not} possess a degenerate eigenstate. We show now that the
$a_{x_0}$ with even $x_0$ have a degenerate eigenstate if $L$ is odd (we
know already that $a_L$ 
never has a degenerate eigenstate).

The degeneracy is due to parity conservation, namely the inaccessibility
of the set of configurations with an even number of particles on the odd
sublattice (which is the set of sites with odd coordinates $1\le x\le
L$) from those with an odd number of particles there and vice versa:
Focusing on bulk-dynamics and thus ignoring boundaries and driving site
for a moment, the number of particles transferred in a toppling from one
sublattice to the other is always even, \ie under bulk dynamics the
parity of the particle number on each sublattice is conserved. As far as
boundaries and driving is concerned, topplings at boundary sites break
that symmetry whenever one particle is lost. Driving at site $x_0$
changes the parity of its sublattice.

One boundary site is always odd, $x_0=1$, \ie the parity of the particle
number on the even sublattice is never conserved.  The second boundary
site is $x_0=L$.  If $L$ is even the parity on the odd sublattice is
also not conserved, \ie none of them are conserved.  If $L$ is odd, then
the parity on the odd sublattice is conserved, unless the driving site
itself if odd.\footnote{In fact, for odd $L$, driving on a site on the
odd sublattice flips the parity of the particle number of the odd
sublattice and because for odd $L$ this is the only mechanism by which
the parity can be changed, one may expect one eigenvalue of $-1$.} In
other words, when driving odd $L$ at even $x_0$, the phase space divides
into two mutually inaccessible regions. We conclude that for odd $L$ and
even $x_0$ degeneracy is at least two-fold.  With the findings above
(joint eigenstate of all operators and the lack of parity conservations
for even $L$ or odd $x_0$), we suspect that this is in fact the only
situation when the phase space becomes disjoint, \ie there is at most
two-fold degeneracy. Unfortunately, $L$ is odd and $x_0$ is even for the
centre driving used above.

We were able to verify exactly two-fold degeneracy up to the largest
system we could analyse analytically, $L=7$, driving at sites $x_0=2$
and at $x_0=4$. On each even site, the pair of eigenstates can obviously
be written so that they are orthogonal.  In fact, the two eigenstates on
each (even) site can always be chosen so that they correspond to the
stationary states where all recurrent configurations have either even or
odd parity of the odd sublattice, setting components corresponding to
states with the other parity to $0$.  
The eigenstates for a given parity found on different (even) sites must
be the same,
because the linear combination of
the two eigenstates must result in the (same) unique joint eigenstate.
The pair of eigenstates can, in fact, be constructed by
taking all components in the common, unique eigenstate corresponding to
one particular parity of the odd sublattice, setting the others to $0$.
In other words, if there is indeed only ever at most two-fold degeneracy at
even $x_0$ for odd $L$, then the pairs of eigenstates at those driving
sites $x_0$ can be chosen to be identical among different $x_0$.

To see that the unique joint eigenstate is an equally weighted sum of the two 
eigenstates containing only configurations of a single parity on the odd sublattice, 
we consider (any) global drive where even and odd parity of the odd sublattice 
occur with equal frequencies, because it changes only when odd sites are driven.
As a result, configurations with odd and even parity on the odd sublattice 
appear with equal frequencies in the unique joint eigenstate, which must therefore
be made up from the two eigenstates of opposite parity with equal weights.

As we were unable to identify any degeneracy in even $L$ or for driving
at odd $x_0$, the further analysis focuses on odd $L$ and even driving
$x_0$.  In $L\le7$, we were able to calculate matrices $a_{x_0}$ and
their eigenstates explicitly using the techniques discussed in
\Aref{appendix_generate_matrices} and the computer algebra system
Mathematica \cite{Mathematica:10.0.2.0}.  We confirmed that
the equally weighted sum of the two degenerate eigenstates (for even and odd parity) reproduces
the unique eigenstate $\ket{e}$ and that there was two-fold degeneracy
of the eigenstate only when driving even sites and not when driving odd
sites.  For larger $L$ we have determined numerically the density
profile (\ie the probability for a site to be occupied in the stationary
state) as a ``fingerprint'' of the stationary state,
\Fref{degeneracy_lifting}. In these systems, we could only ever see
two-fold degeneracy for driving at even $x_0$ in odd $L$ resulting in the same pair
of density profiles for all even $x_0$. With increasing system size the
(in total three) different density profiles become less and less
different.  \Fref{degeneracy_lifting15} shows the profiles obtained for
$L=15$ which are virtually indistinguishable from the unique density
profile of odd $x_0=1$. In reasonably large systems as those studied above, observing the
degeneracy in the density profile may be beyond numerical reach. 
We expect similar caveats to apply in the Oslo Model \cite{ChristensenETAL:1996,GrassbergerDharMohanty:2016}.

\end{document}